\documentclass[preprint]{aastex6}

\newcommand{\HI}{\mbox{H\footnotesize\,I}}
\newcommand{\HIs}{\mbox{{\scriptsize H}{\tiny\,I}}}
\newcommand{\WHI}{W_{\HIs}}
\newcommand{\cHI}{c_{\HIs}}
\newcommand{\yHI}{y_{\HIs}}
\newcommand{\THI}{\tau_{\HIs}}
\newcommand{\Htwo}{\mbox{H}_{2}}
\newcommand{\HtwoCO}{\mbox{H}_{\rm 2,CO}}
\fullcollaborationName{The Friends of AASTeX Collaboration}

\begin{document}


\title{Quantifying the Interstellar Medium and Cosmic Rays in the MBM 53, 54, and 55 Molecular Clouds
and the Pegasus loop using Fermi-LAT Gamma-ray Observations}


\author{
T.~Mizuno\altaffilmark{1}, 
S.~Abdollahi\altaffilmark{2}, 
Y.~Fukui\altaffilmark{3}, 
K.~Hayashi\altaffilmark{3}, 
A.~Okumura\altaffilmark{4}, 
H.~Tajima\altaffilmark{4,5}, 
and H.~Yamamoto\altaffilmark{3}
}
\altaffiltext{1}{Hiroshima Astrophysical Science Center, Hiroshima University, Higashi-Hiroshima, Hiroshima 739-8526, Japan; mizuno@hep01.hepl.hiroshima-u.ac.jp}
\altaffiltext{2}{Department of Physical Sciences, Hiroshima University, Higashi-Hiroshima, Hiroshima 739-8526, Japan}
\altaffiltext{3}{Department of Physics and Astrophysics, Nagoya University, Chikusa-ku Nagoya 464-8602, Japan}
\altaffiltext{4}{Nagoya University, Institute for Space-Earth Environmental Research, Furo-cho, Chikusa-ku, Nagoya 464-8601, Japan}
\altaffiltext{5}{W. W. Hansen Experimental Physics Laboratory, Kavli Institute for Particle Astrophysics and Cosmology, Department of Physics and SLAC National Accelerator Laboratory, Stanford University, Stanford, CA 94305, USA}










\begin{abstract}

A study of the interstellar medium (ISM) and cosmic rays (CRs)
using \textit{Fermi} Large Area Telescope (LAT) data,
in a region encompassing the nearby 
molecular clouds
MBM~53, 54, and 55 and a far-infrared loop-like structure in Pegasus,
is reported.
By comparing \textit{Planck} dust thermal emission model with \textit{Fermi}-LAT $\gamma$-ray data, 
it was found that neither 
the dust radiance $(R)$ nor 
the dust opacity at 353~GHz ($\tau_{353}$)
were proportional to the total gas column density $N({\rm H_{tot}})$
primarily because $N({\rm H_{tot}})/R$ and $N({\rm H_{tot}})/\tau_{353}$ depend on the 
dust temperature ($T_{\rm d}$).
The $N({\rm H_{tot}})$ distribution
was evaluated
using $\gamma$-ray data
by assuming the regions of high $T_{\rm d}$ to be dominated by
optically thin atomic hydrogen ($\HI$) and by employing an empirical linear relation
of $N({\rm H_{tot}})/R$ to $T_{\rm d}$. It was determined
that the mass of the gas not traced by the 21-cm or 2.6-mm surveys is
{$\sim$}25\% of the mass of $\HI$ in the optically thin case 
and is larger than the mass of the molecular gas traced by carbon monoxide by a factor of up to 5.
The measured $\gamma$-ray emissivity spectrum is 
consistent with a model based on CR spectra measured at the Earth
and the nuclear enhancement factor of ${\le}1.5$.
It is, however, lower than local $\HI$ emissivities reported by previous
\textit{Fermi}-LAT studies employing different analysis methods and
assumptions on ISM properties by 15\%--20\%
in energies below a few GeV, even if we take account of the statistical and systematic uncertainties.
The origin of the discrepancy is also discussed.
\end{abstract}




\section{Introduction} 


Interstellar space is permeated with ordinary matter (gas or dust), which is known as
the interstellar medium (ISM), high-energy charged particles known as cosmic rays (CRs),
interstellar radiation fields (ISRF), and magnetic fields.
These constituents have comparable pressures and are mutually interacting.
They play an important role in many physical and chemical processes (e.g., star formation)
that occur in the Milky Way and have been studied
in various wavebands---from radio to X-rays to 
$\gamma$ rays 
\citep[for a review, see, e.g.,][]{Ferriere2001}.
Of the multiwavelength observations,
cosmic $\gamma$-ray emission 
is known to be
a powerful probe to study the ISM and Galactic CRs.
High-energy CR protons and electrons interact with the interstellar gas 
or the ISRF
and produce 
$\gamma$ rays 
through nucleon--nucleon interactions, electron bremsstrahlung, 
and inverse Compton (IC) scattering.
Because the ISM is essentially transparent to these high-energy photons, we can 
study the ISM distribution via $\gamma$-ray observations.
Because the $\gamma$-ray production cross section is independent of the chemical or 
thermodynamic state of the interstellar gas, cosmic 
$\gamma$ rays 
have been recognized 
as a unique tracer of the total
gas column density regardless of its atomic or molecular state.
If the gas column densities are estimated with good accuracy using observations
in other wavebands such as radio, infrared, and optical, the CR spectrum and density distribution 
can be examined as well. In fact, the distributions of ISM and CRs obtained
are ambiguous because of the degeneracy;
therefore, $\gamma$-ray observations need to be complemented by using data from other wavebands.

Usually, 
the distribution of 
atomic hydrogen ($\HI$) is measured by 21-cm line surveys \citep[e.g.,][]{Dickey1990},
and 
the distribution of 
molecular hydrogen ($\Htwo$) is derived via 2.6-mm line observations of
carbon monoxide, CO \citep[e.g.,][]{Dame2001}.
The total gas column density can also be estimated from extinction, reddening, or emission by dust
\citep[e.g.,][]{Bohlin1978}.
These tracers have advantages and disadvantages.
$\HI$ 21-cm line surveys directly trace the distribution of atomic hydrogen and
provide us with the line velocity
along the line of sight inferred from the Doppler shift,
which, in turn, provides
distance information under the assumption
of the Galactic rotation curve \citep[e.g.,][]{Clemens1985}.
The obtained $\HI$ column density, however, suffers from uncertainties of the 21-cm line opacity
and self absorption.
CO 2.6-mm line surveys also provide us with velocity (and distance) information,
although it is an indirect tracer of $\Htwo$ and the derived molecular gas column density is affected 
by the assumption of the conversion factor (the so-called $X_{\rm CO}$).
This method may also miss CO-dark $\Htwo$ clouds due to, e.g., photodissociation \citep{Wolfire2010}.
Dust is expected to be well mixed with gas 
in the cold and warm phases of the ISM
and is a probe of the total gas column density,
although it lacks velocity (and distance) information.
Because dust is an indirect tracer of the interstellar gas 
(like CO is a tracer of $\Htwo$),
the derived gas column density is affected by 
assumptions of the dust-to-gas ratio 
and dust emissivity (or extinction).
Therefore, comparing $\HI$, CO, and dust observations is crucial to study the 
interstellar gas distribution,
and adding $\gamma$-ray data is important because it is another
independent tracer of the total gas column density.

Studies of the ISM (and CRs) have advanced significantly in last two decades.
The G236+39 cloud was found to have significant infrared emission from dust,
which was not accounted for by the $\HI$ 21-cm or CO 2.6-mm line observations,
suggesting the presence of an $\Htwo$ cloud with CO
emission below the detection threshold \citep{Reach1994}.
Combining the EGRET $\gamma$-ray data, $\HI$, CO, and dust extinction maps,
significant amount of gas not traced by the $\HI$ or CO surveys was revealed 
in the solar neighborhood
and has been referred to as ``dark gas'' \citep{Grenier2005}.
This work has been confirmed and improved in terms of significance and accuracy
by recent observations by \textit{Fermi}-LAT \citep[e.g.,][]{Fermi2ndQ,Fermi3rdQ,FermiCham}.
Taking account of the dark gas also makes it possible to obtain information
on Galactic CRs with unprecedented accuracy \citep[e.g.,][]{FermiHI2}.
The \textit{Planck} satellite provides 
an accurate dust thermal emission model,
which is crucial to study the ISM.
By comparing the \textit{Planck} dust emission 
model,
and the $\HI$ and CO data, 
the \citet{Planck2011} estimated the mass of
dark gas to be {$\sim$}30\% of the atomic gas and {$\sim$}120\% of the CO-bright molecular gas
in the solar neighborhood.
By comparing the \textit{Planck} dust optical depth map at 353~GHz ($\tau_{353}$),
and the $\HI$/CO data and assuming that the total gas column density was proportional to $\tau_{353}$, 
\citet{Fukui2014,Fukui2015} proposed that a significant amount of the atomic hydrogen was optically thick
in areas with low dust temperature ($T_{\rm d}$),
resulting in an excess mass comparable to the mass of $\HI$ in the optically thin case.
The \citet{Planck2014}, on the other hand, found that the dust radiance $R$ (bolometric luminosity)
was well correlated with the integrated $\HI$ 21-cm line intensity, $\WHI$, 
in wide range of $T_{\rm d}$ in the diffuse ISM,
and proposed that
it would be a better tracer of the dust (and the total gas) column density.
The \citet{Planck2015} combined the 
\textit{Fermi}-LAT data and \textit{Planck} dust emission model
to study the ISM in the Chamaeleon molecular cloud.
They employed a detailed model of the dust emission by \citet{Draine2007} and found a
good
correlation with $\gamma$-ray data. The obtained mass of the dark gas was
approximately twice that of
the CO-bright ${\rm H_{2}}$ and contributed {$\sim$}15\% of the total gas mass.

Here, we report an analysis of the \textit{Fermi}-LAT $\gamma$-ray data in the Galactic longitudes
$60\arcdeg \le l \le 120\arcdeg$ and the Galactic latitudes $-60\arcdeg \le b \le -28\arcdeg$.
Our region of interest (ROI) encompasses the MBM 53, 54, and 55 molecular cloud complexes 
(located at $l=84\arcdeg$ to $96\arcdeg$ and $b=-44\arcdeg$ to $-30\arcdeg$)
and an infrared loop-like structure in Pegasus (area of {$\sim$}$20\arcdeg{\times}20\arcdeg$ around
$(l,b) \sim (109\arcdeg, -45\arcdeg)$).
MBM 53, 54, and 55 are some of the nearest large molecular clouds \citep{Yamamoto2003},
located at a distance of {$\sim$}150~pc 
estimated by \citet{Welty1989}
based on measurements of interstellar NaI absorption toward stars associated with the clouds.
The loop-like structure in Pegasus (hereafter termed the ``Pegasus loop'') was identified
in IRAS 100~\micron \
maps
\citep{Kiss2004} and studied in CO 
using the NANTEN telescope
\citep{Yamamoto2006};
its distance has been estimated to be {$\sim$}100~pc
which is equal to the distance of the B2 star in the center of the loop.
The MBM 53, 54, and 55 clouds and the Pegasus loop are nearby ($100\mbox{--}150~{\rm pc}$) molecular
clouds located at high Galactic latitudes
(having small overlap with structures in the ISM at different distances),
and therefore are expected to have uniform ISM and CR properties 
(e.g., dust--to--gas ratio and CR density).

This paper is organized as follows. 
We describe the properties of the 
ISM tracers in the complexes studied
in Section ~2, 
and the $\gamma$-ray observations, data selection, and modeling in Section~3. 
The results of the data analysis are presented in Section~4, where we find that
neither $R$ nor $\tau_{353}$ are good measures of the total gas column density [$N({\rm H_{tot}})$].
We use the \textit{Fermi}-LAT $\gamma$-ray data to compensate for the observed $T_{\rm d}$ dependence
and evaluate $N({\rm H_{tot}})$ (also shown in Section~4).
We discuss the ISM and CR properties of the studied region in Section~5. 
A summary of this study and future prospects are presented in Section~6.

Before describing the analysis and results of the study,
we note the difference of our approach from that of preceding studies.
Most previous \textit{Fermi}-LAT studies of diffuse $\gamma$-ray emission used
$\HI$, CO, and dust data to prepare template 
maps of the neutral gas distribution in the atomic phase, the molecular phase, and the dark gas phase, respectively,
and analyzed $\gamma$-ray data 
to study the ISM and CRs
(the method is hereafter called a ``conventional template-fitting method'').
Motivated by our finding 
that the ratio of the $\gamma$-ray intensity associated with the ISM gas
[i.e., a tracer of $N({\rm H_{tot}})$] to dust tracers ($R$ or $\tau_{353}$)
depends on $T_{\rm d}$,
we took a different approach:
we focus on evaluating $N({\rm H_{tot}})$ using the \textit{Planck} dust map
by applying the correction based on $T_{\rm d}$
in the $\gamma$-ray data analysis (Section~4.3), and then discuss the relation 
of the obtained $N({\rm H_{tot}})$ distribution
with $\HI$ 21-cm and CO 2.6-mm line intensities (Section~5).
A comparison with a conventional template-fitting method is also given in Section~5 and Appendix~D.

\clearpage

\section{Properties of the ISM Tracers}

We analyzed the $\gamma$-ray data in a region with 
Galactic longitude $60\arcdeg \le l \le 120\arcdeg$ and
Galactic latitude $-60\arcdeg \le b \le -28\arcdeg$, which encompasses
the MBM 53, 54, and 55 cloud complexes and the Pegasus loop.
Because preparing good templates of the interstellar gas is crucial for
$\gamma$-ray data analysis, we first investigated the properties of the ISM tracers.
We prepared dust maps, a $\WHI$ map, and an integrated CO 2.6-mm line intensity
($W_{\rm CO}$) map, all stored in a HEALPix \citep{Gorski2005} equal-area sky map of order 9
(pixel size is ${\sim}0.013~{\rm deg^{2}}$).
We used
the \textit{Planck} dust maps 
(of $R$, $\tau_{353}$, $T_{\rm d}$, and dust spectral index $\beta$)
of the public data release 1 (the version R1.20)\footnote{\url{http://irsa.ipac.caltech.edu/data/Planck/release_1/all-sky-maps/}}
described by \citet{Planck2014},
since the latest release (public data release 2) does not include the dust radiance map.
Assuming a uniform dust temperature along the line of sight,
they have modeled the dust thermal emission with a single modified black-body, and constructed those maps
\citep[for details of the procedure, see][]{Planck2014}.
As described in \citet{Planck2014},
the dust optical depth is the product of the dust opacity (cross section) 
per H atom and the total gas column density. 
Therefore if the dust cross section is uniform $\tau_{353}$ is proportional to $N({\rm H_{tot}})$. 
The dust radiance $R$ is also expected to trace the total gas column density, 
since it is proportional to $N({\rm H_{tot}})$ under the assumption of a uniform dust--to--gas ratio, 
dust emissivity, and ISRF (see also Section 4.2).

To construct the $\WHI$ map, we referred to the Leiden/Argentine/Bonn (LAB) survey \citep{Kalberla2005}
integrated over the velocity range from $-450$ to $400~{\rm km~s^{-1}}$.
\footnote{
According to
\citet{Kalberla2005}, the velocity range of the survey spanned $850~{\rm km~s^{-1}}$ at a resolution of $1.3~{\rm km~s^{-1}}$
with a root-mean-square (RMS) noise per channel of $0.07\mbox{--}0.09~{\rm K}$. Therefore, the RMS noise in the integrated intensity
over the entire velocity range is estimated to be ${\sim}0.08~{\rm K} \times \sqrt{850/1.3} \times 1.3~{\rm km~s^{-1}} 
\sim 2.7~{\rm K~km~s^{-1}}$, much smaller than the values of $\WHI$ in our ROI.
}
We used a $W_{\rm CO}$ map internally available to the LAT team, which combines 
the work by \citet{Dame2001} and new data at high Galactic latitudes sampled in $0\fdg25$.
The new CO data includes most of the high-latitude CO clouds in the region studied here.
The CO spectra were filtered to suppress the noise and integrated over velocities \citep{Dame2011}.
We converted the $\WHI$ map into the column density $N(\HI)$ using the optically thin approximation
[$N(\HI_{\rm thin})({\rm cm^{-2}}) = 1.82 \times 10^{18} \cdot \WHI({\rm K~km~s^{-1}})$].
The obtained $N({\HI})$ model map, 
the ${\rm W_{CO}}$ map (${\rm K~km~s^{-1}}$), and the $T_{\rm d}$ map (K) in our ROI are shown in Figure~1.
In the \textit{Planck} dust maps, we identified several areas with high $T_{\rm d}$
indicating localized heating by stars. We refilled these areas 
(in the $R$, $\tau_{353}$, and $T_{\rm d}$ maps),
with the average of the peripheral pixels. Details of this procedure are described in Appendix~A.

The correlations
between $\WHI$ and $R$, and those between $\WHI$ and $\tau_{353}$,
are shown in Figure~2,
in which the colors represent different dust temperatures.
We masked areas with $W_{\rm CO}$ intensity greater 
than 1.1~${\rm K~km~s^{-1}}$ in order to match the
procedure of \citet{Fukui2014}, who analyzed the ISM in and around the MBM 53, 54, and 55 clouds.
Therefore the regions of $\Htwo$ associated with appreciable $W_{\rm CO}$ are not included in the figure.
We can confirm the trends of the dust--gas relation found by previous studies described in Section~1 as
(1) we observe in Figure~2a a good correlation between the $\WHI$ and $R$
in a wide range of $T_{\rm d}$ \citep{Planck2014}
\footnote{
They reported a good correlation up to column densities of (at least)
$5 \times 10^{20}~{\rm cm^{-2}}$
(Figure~20 of the reference), 
which corresponds to $\WHI$ of ${\rm {\sim}280~K~km~s^{-1}}$.
}
and
(2) we observe in Figure~2b a strong $T_{\rm d}$ dependence of the $\WHI\mbox{--}\tau_{353}$ relation,
which \citet{Fukui2014} interpreted to be primarily due to optically thick $\HI$ in low-$T_{\rm d}$ 
areas.

Although the region studied is dominated by the local ISM, contamination from clouds with different
velocities (and therefore likely having different distances) is inevitable.
We identified ISM clouds with velocities in $-80$ to $-30~{\rm km~s^{-1}}$
[reported by \citet{Wakker2001} as some of intermediate-velocity clouds (IVCs) in the southern sky],
while the main clouds have velocities in $-30$ to $+20~{\rm km~s^{-1}}$ (see Appendix~B for details).
We masked the areas shown in Figure~B1b to eliminate the contribution from the IVCs and confirmed the same trends
as described above.
We also examined the 
dust--$\WHI$ relation in sub-regions:
one is in $80\arcdeg \le l \le 100\arcdeg$ and $-44\arcdeg \le b \le -28\arcdeg$ which covers
the MBM 53, 54, and 55 clouds, and the other is in
$100\arcdeg \le l \le 120\arcdeg$ and $-55\arcdeg \le b \le -35\arcdeg$ which covers
the Pegasus loop. Again, we confirmed the same trends as described 
above; the difference 
seen among sub-regions is smaller than the difference seen between two tracers ($R$ and $\tau_{353}$).

The 
correlation
between the dust tracers and $\WHI$ alone is not sufficient to distinguish
which ($R$ or $\tau_{353}$) is the better tracer of the total dust (and gas) column density.
We therefore prepared two types of $N({\rm H_{tot}})$ model maps
based on $R$ and $\tau_{\rm 353}$ and tested them against the \textit{Fermi}-LAT $\gamma$-ray data.
We started with a single $N(\rm H_{tot})$ map (Section~4.1) and then employed
multiple $N(\rm H_{tot})$ maps sorted by $T_{\rm d}$ (Section~4.2).
We finally came back to a single $N(\rm H_{tot})$ map with a 
$T_{\rm d}$-dependent correction
applied
in order to better represent the $\gamma$-ray data (Section~4.3).

\begin{figure}[ht!]
\figurenum{1}
\gridline{
\fig{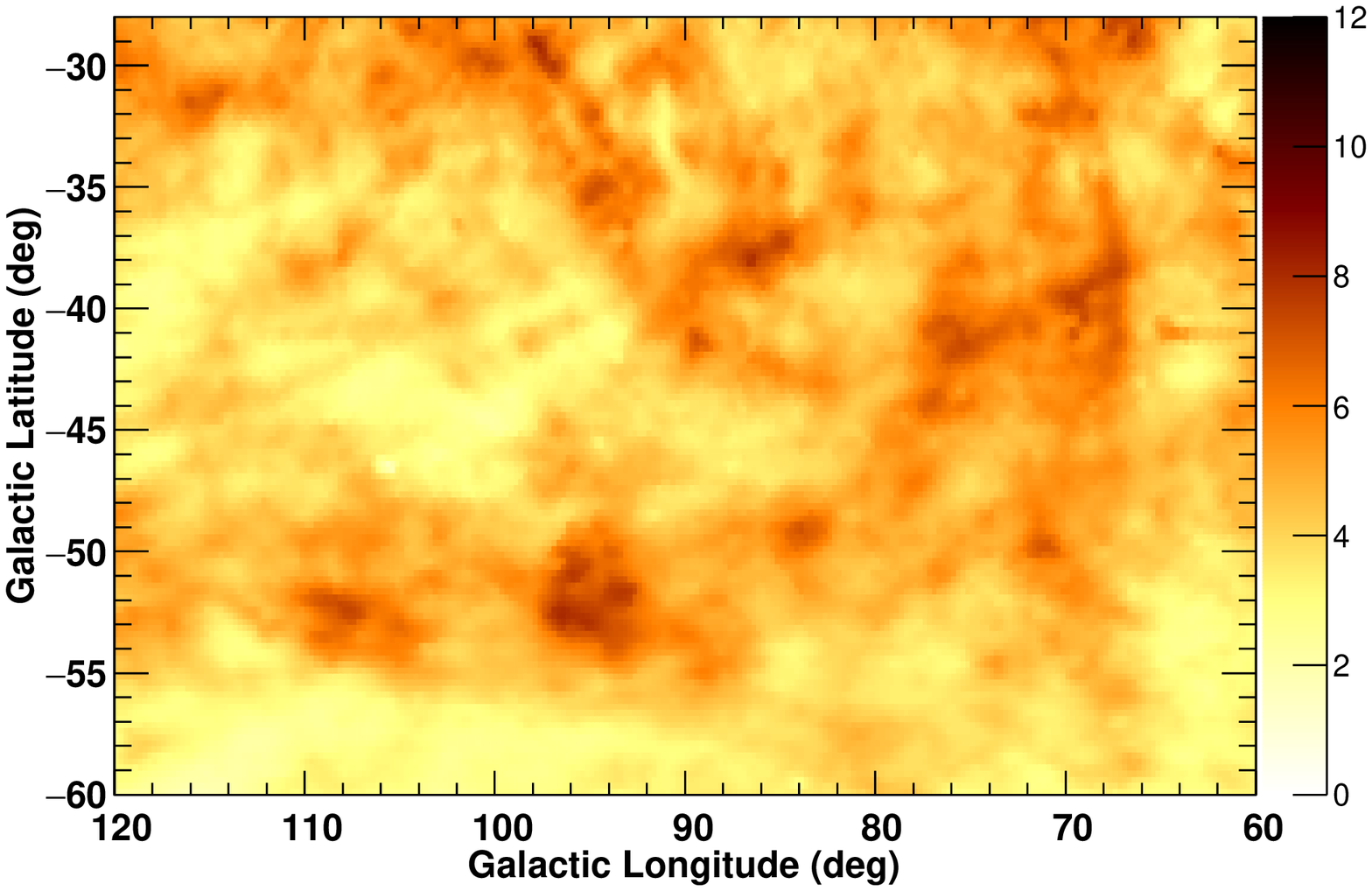}
{0.5\textwidth}{(a)}
\fig{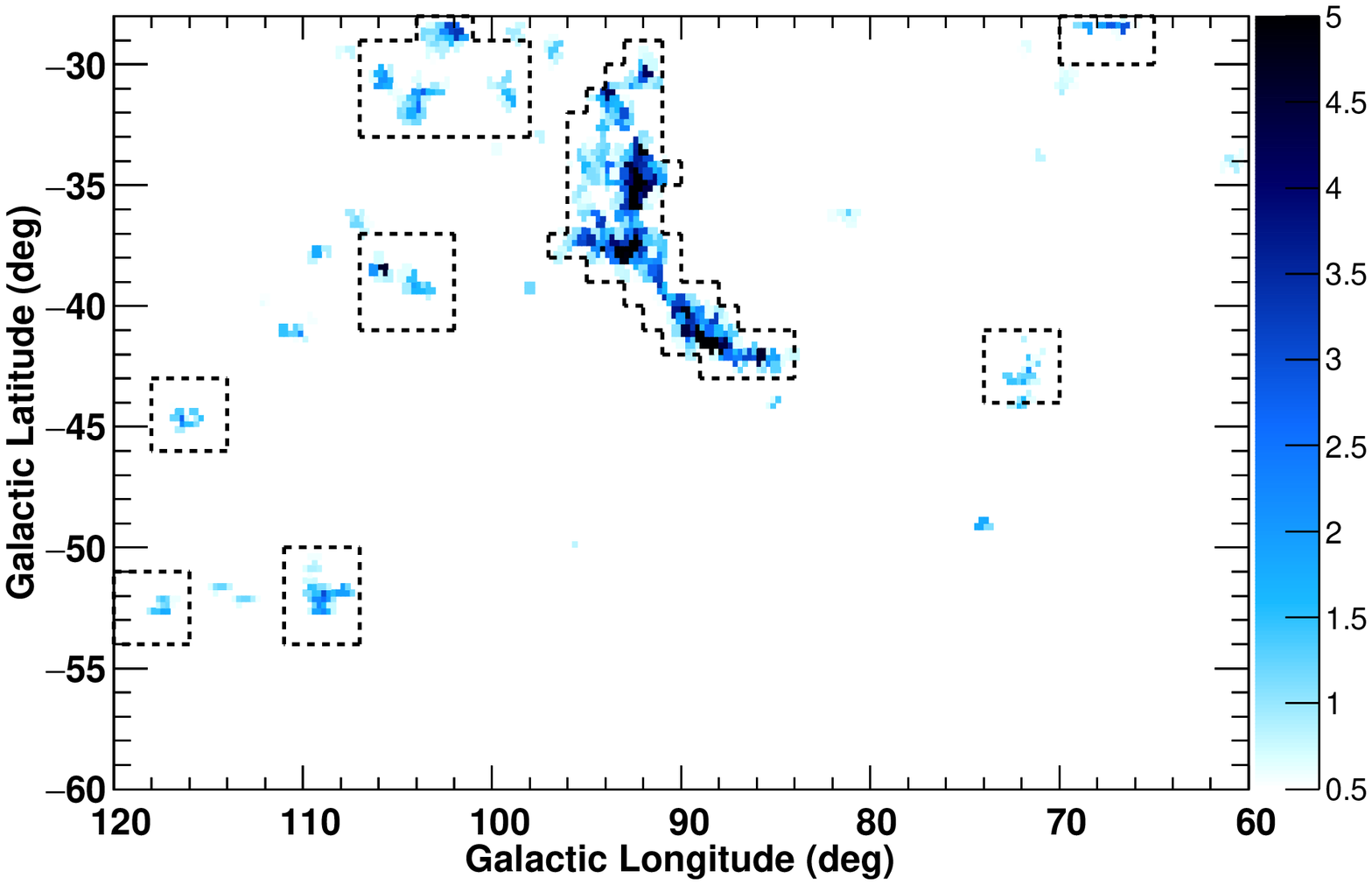}
{0.5\textwidth}{(b)}
}
\gridline{
\fig{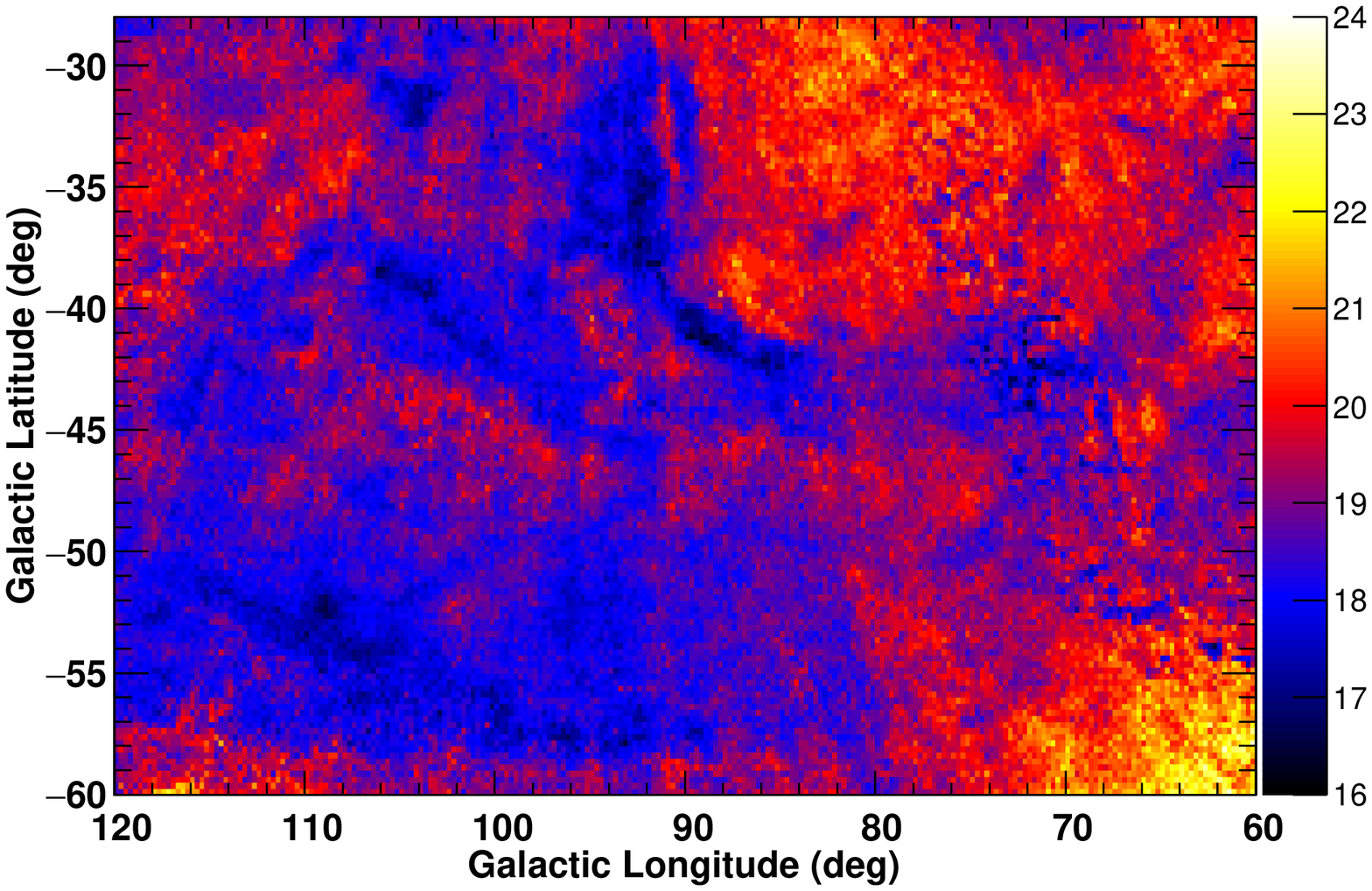}
{0.5\textwidth}{(c)}
}
\caption{
(a) The $\WHI$ map converted into $N(\HI)$ with the optically-thin approximation
($N(\HI_{\rm thin})({\rm cm^{-2}}) = 1.82 \times 10^{18} \cdot \WHI({\rm K~km~s^{-1}})$),
shown in units of $10^{20}~{\rm cm^{-2}}$;
(b) the ${\rm W_{CO}}$ map (${\rm K~km~s^{-1}}$);
and (c) the $T_{\rm d}$ map (K).
The dotted lines in panel~(b) indicate the areas to be masked in Section~4.2.
\label{fig:f1}
}
\end{figure}

\begin{figure}[ht!]
\figurenum{2}
\gridline{
\fig{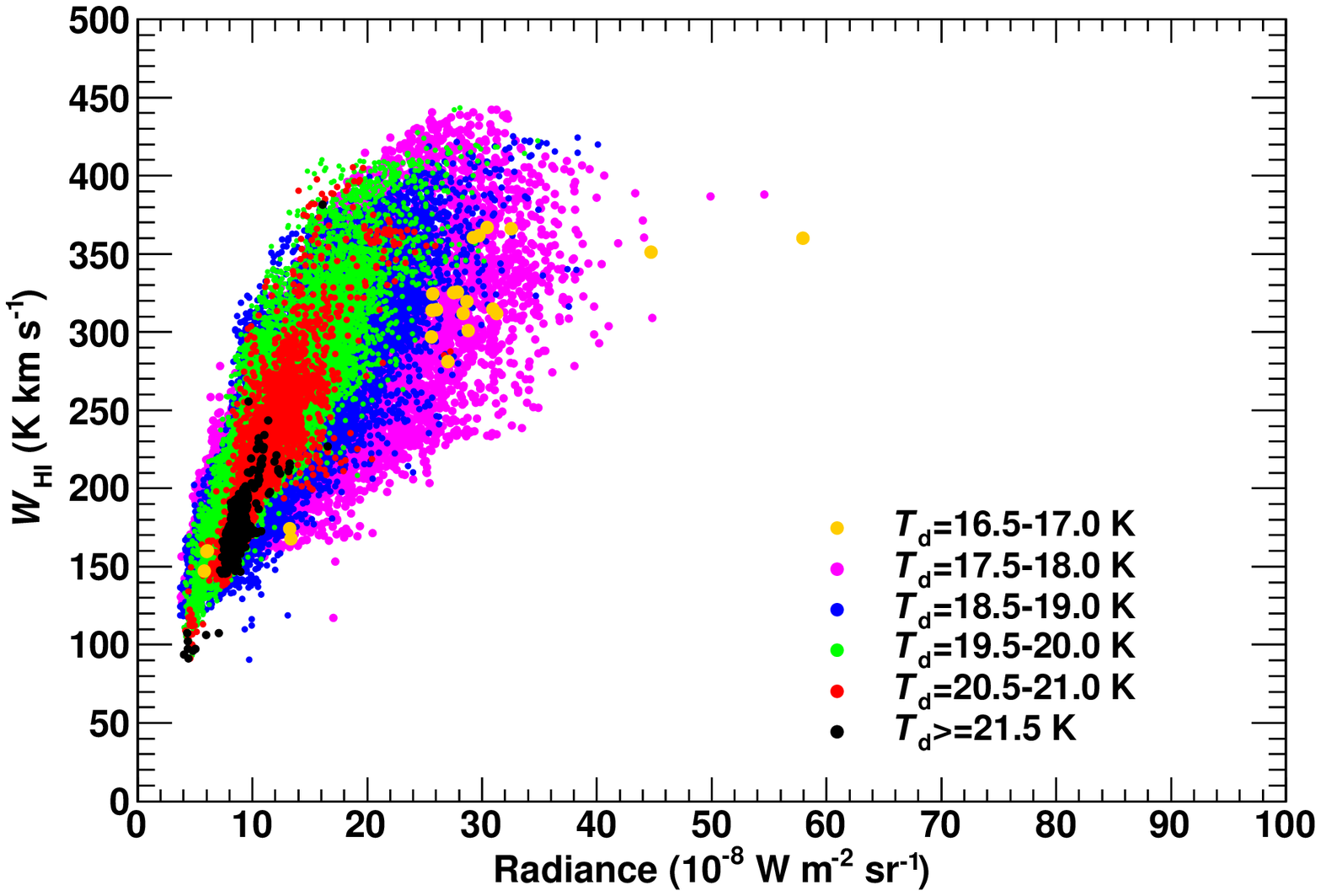}
{0.5\textwidth}{(a)}
\fig{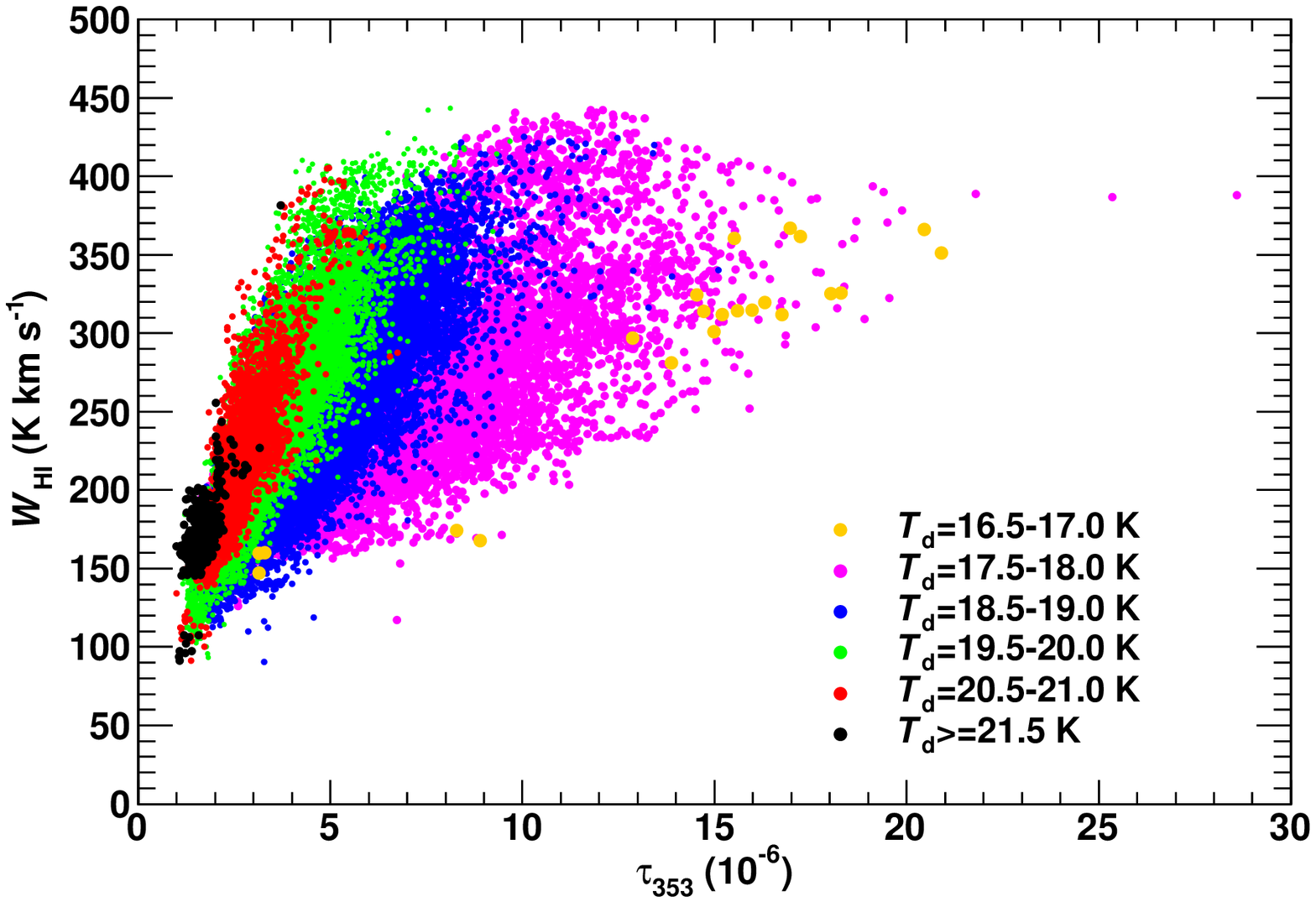}
{0.5\textwidth}{(b)}
}
\caption{
The correlations between $\WHI$ and dust tracers:
(a) scatter plot of $\WHI$ versus $R$ and
(b) scatter plot of $\WHI$ versus $\tau_{353}$. 
Data are shown in 0.5~K ranges of $T_{\rm d}$ with 0.5~K gaps between intervals for clarity.
Each point represents each pixel of our HEALPix map (order 9; pixel size is 
${\sim}0.013~{\rm deg^{2}}$).
\label{fig:f2}
}
\end{figure}

\clearpage

\section{Gamma-ray Data and Modeling}
\subsection{Gamma-ray Observations and Data Selection}


The LAT on board the \textit{Fermi Gamma-ray Space Telescope}, launched in 
2008 June,
is a pair-tracking $\gamma$-ray telescope, detecting photons in the range of {$\sim$}20~MeV to more than
300~GeV. Details of the LAT instrument and the pre-launch performance expectations
can be found in \citet{Atwood2009}, and the on-orbit calibration is described in \citet{Abdo2009}.
Thanks to its wide field of view, the \textit{Fermi}-LAT is an ideal telescope to study
Galactic diffuse 
$\gamma$ rays.
Past studies of Galactic diffuse emission by \textit{Fermi}-LAT
can be found in, e.g., \citet{FermiPaper2}, and \citet{FermiHI2}.

Routine science operations with the LAT started on 
2008 August 4.
We have accumulated events 
from 2008 August 4 to 2015 August 2 
(i.e., 7 years) to study
diffuse 
$\gamma$ rays 
in our ROI. During 
most of 
this time interval,
the LAT was operated in sky survey mode, obtaining complete sky coverage every
two orbits and relatively uniform exposure over time.
We used the Pass~8 event data, and
used the standard LAT analysis software, \textit{Fermi} Science Tools
\footnote{\url{http://fermi.gsfc.nasa.gov/ssc/data/analysis/software/}}
version v10r00p05,
to select
events satisfying the so-called \textit{Clean} class 
in order to
obtain low-background events. We also required that the reconstructed zenith angles of the
arrival direction of the photons be less than $100\arcdeg$ 
to reduce contamination by photons from the Earth 
atmosphere.
In addition, we excluded the periods of time during which the LAT
detected bright 
$\gamma$-ray bursts
or solar flares. 
(The integrated 
period of time excluded 
in this procedure is negligible
compared to that excluded to remove data with flares of 3C~454.3, described below)
We also referred to the 
Monitored Source List 
light curves
\footnote{\url{http://fermi.gsfc.nasa.gov/ssc/data/access/lat/msl_lc/}},
and excluded the 
periods 
of time (300~days in total) during which the LAT detected flares of 3C~454.3.
This reduced the contamination from the bright active galactic 
nucleus
in diffuse emission modeling
while keeping {$\sim$}90\% of the good time interval.
We used the latest response 
functions that match
our dataset and event selection, 
P8R2\_CLEAN\_V6, in the following analysis. Although we did not take into account the energy dispersion
\footnote{\url{http://fermi.gsfc.nasa.gov/ssc/data/analysis/documentation/Pass8_edisp_usage.html}}
in the analysis,
the impact on the results
is expected to be small 
since we analyzed the data above 0.3~GeV as described in Section~4.1.

\subsection{Model to Represent the Gamma-ray Emission}

We modeled the $\gamma$-ray emission observed by \textit{Fermi}-LAT
as a linear combination of the gas column density model map(s) constructed from 
the \textit{Planck} dust map,
IC emission, isotropic component and $\gamma$-ray point sources.
The use of the gas column density maps as a template is based on the assumption that 
$\gamma$ rays 
are generated
via interactions between the CRs and ISM gas and that CR densities do not vary significantly
over the scale of the interstellar complexes in this study.
This assumption is simple
but very plausible, particularly in 
high-Galactic latitude 
regions,
such as the one studied here. We 
started with a single $N({\rm H_{tot}})$ map 
based on \textit{Planck} dust model maps ($R$ or $\tau_{353}$)
in Section~4.1 and
employed 
multiple $N({\rm H_{tot}})$ maps sorted by $T_{\rm d}$ in Section~4.2.
We note that our $N({\rm H_{tot}})$ model map traces not only atomic but also molecular hydrogen
since dust is expected to be well mixed with the ISM gas in both phases.
We also 
included 
an IC model map and models for point sources.
To model the 
$\gamma$ rays 
produced via IC scattering, we used GALPROP 
\footnote{\url{http://galprop.stanford.edu}}
\citep[e.g.,][]{Galprop1,Galprop2},
a numerical code that solves the CR transport equation within the Galaxy and predicts
the $\gamma$-ray emission produced via the interactions of CRs with interstellar matter
and low-energy photons (IC scattering). The IC emission is calculated from the distribution
of propagated electrons and the interstellar radiation field developed by \citet{Porter2008}.
Here, we adopted the IC model map produced in the GALPROP
run 54\_77Xvarh7S,
which was used in 
LAT collaboration publications
\footnote{\url{https://www-glast.stanford.edu/cgi-bin/pubpub}}
such as \citet{Fermi3rdQ},
as our baseline model. Since the integrated intensity (over the solid angle) of the IC emission is lower 
than that of the isotropic component
(see below) and the gas-related diffuse $\gamma$ rays in the region studied (see Figure~7), the specific choice of the IC model
does not affect the obtained results significantly.
The effect of the IC model uncertainty is examined in Section~4.2.
To model the individual $\gamma$-ray sources, we referred to the third
\textit{Fermi}-LAT catalog (3FGL) described in \citet{Fermi3FGL}, 
which is based on the first four years of the science phase of the mission and 
includes more than 3000 sources detected at a significance of {$\ge$}4$\sigma$. 
For our analysis we considered 57 3FGL sources (detected at a significance of {$\ge$}5$\sigma$
\footnote{
As described in Section~4.1, we iteratively included sources in several groups at a time 
in the order of decreasing significance,
and confirmed that including sources with significance from 5$\sigma$ to 6$\sigma$
did not affect the gas-related scale factors significantly.
We therefore did not include sources with lower 3FGL significance.
}
)
in our ROI, and 17 bright sources ({$\ge$}20$\sigma$) just outside it (within $10\arcdeg$)
to take account of their possible contamination.
We also included 3FGL~J2338.7+0251 (4.8$\sigma$ detection in 3FGL) 
and a source located at $(l,b)=(71\fdg75, -53\fdg5)$
which became brighter after the period of time investigated in 3FGL.
The position of the latter source was determined by visual inspection of the $\gamma$-ray map and was fixed in the analysis.
We also added an isotropic component 
to represent the extragalactic diffuse emission
and the residual instrumental background from misclassified CR interactions
in the LAT detector. 
Another possible source of diffuse $\gamma$-ray emission is CR interactions with ionized gas.
In order to estimate its contribution, we referred to
\citet{FermiHI2} and used the free-free intensity map at a frequency of 22.7~GHz
extracted from 9 year of \textit{WMAP} observations \citep{Bennett2013} 
as a template for the $\gamma$-ray emission
correlated with ionized hydrogen.
We used the scaling factor adopted by \citet{FermiHI2} and found that the estimated column density is
at most ${\sim}10^{20}~{\rm cm^{-2}}$ at three spots in our ROI.
Two of them are positionally coincident with the two brightest $\gamma$-ray sources in our ROI, 
3C~454.3 and 3FGL~2232.5+1143.
The third spot is positionally coincident with localized residuals
seen in our $\gamma$-ray count map
[Figure~8a; $(l,b) \sim (63\fdg7, -34\fdg2)$].
Therefore we can securely expect that the impact of the ionized gas
on the determination of the neutral gas component is minimal
and we did not take the ionized gas into account in our analysis.

Then, $\gamma$-ray intensities $I_{\gamma}(l, b, E)~{\rm(ph~s^{-1}~cm^{-2}~sr^{-1}~MeV^{-1})}$
can be modeled as

\begin{equation}
I_{\gamma}(l, b, E) = 
\sum_{i} c_{1,i}(E) \cdot q_{\gamma}(E) \cdot N({\rm H_{tot}})_{i}(l, b)  + 
c_{2}(E) \cdot I_{\rm IC}(l, b, E) +
I_{\rm iso}(E) + \sum_{j} {\rm PS}_{j}(l, b, E)~~,
\end{equation}
where $N({\rm H_{tot}})_{i}$ is the total gas column density model (${\rm cm^{-2}}$) map(s)
in either atomic or molecular phase,
$q_{\gamma}(E)$ (${\rm ph~s^{-1}~sr^{-1}~MeV^{-1}}$) is the differential $\gamma$-ray yield 
or $\gamma$-ray emissivity per H atom,
$I_{\rm IC}(l, b, E)$ and $I_{\rm iso}(E)$ are the IC model and isotropic background intensities
(${\rm ph~s^{-1}~cm^{-2}~sr^{-1}~MeV^{-1}}$), respectively, and
${\rm PS}_{j}(l, b, E)$ represents the point source contributions.
The subscript $i$ allows for the separation of $N({\rm H_{tot}})$ maps by $T_{\rm d}$ (Section~4.2).
We applied the $\gamma$-ray emissivity model for
the local interstellar spectrum (LIS) of CRs and the so-called
nuclear enhancement factor $\epsilon_{\rm M}$ 
(a scale factor to take account of the effect of heavy nuclei in both CRs and the target matter)
of 1.84 \citep{Mori2009} adopted by \citet{FermiHI}.
To accommodate the uncertainties in the LIS and $\epsilon_{\rm M}$, 
we included 
scale factors
[$c_{1,i}(E)$ in Equation~(1)] as 
free parameters.
It will be 1 if the measured $\gamma$-ray emissivity agrees with the LIS and $\epsilon_{\rm M}$ we adopted.
The IC emission model (see above) also is uncertainly known, and we included another
scale factor [$c_{2}(E)$ in Equation~(1)] as a free parameter.
The isotropic component $I_{\rm iso}$ and the point source contributions were also taken 
to be free parameters
as a function of energy.
The positions of sources were fixed to the values in 3FGL.
We divided $\gamma$-ray data into several energy ranges and fit Equation~(1)
to $\gamma$-rays in each energy range using the binned likelihood method implemented in
\textit{Fermi} Science Tools. When using multiple $N({\rm H_{tot}})$ maps (Section~4.2 and Appendix~C), 
we used wider energy ranges and modeled $c_{2}(E)$ with a power law function as
$c_{2}(E)=c_{\rm 2n} \cdot (E/E_{0})^{c_{\rm 2i}}$ where $E_{0}$ is a reference energy.

\clearpage

\section{Data Analysis}
\subsection{
Initial Modeling with a Single Gas Map
}


We started our data analysis using a single total gas column density model maps based on $R$ or $\tau_{353}$.
To construct the $N({\rm H_{tot}})$ model maps,
we assumed a proportionality between $N({\rm H_{tot}})$ and $R$ (or $\tau_{353}$)
and that $\HI$ is optically thin and well represents the total gas column density
at least for regions with the high-temperature areas ($T_{\rm d} \ge 21.5~{\rm K}$).
First we made least-squares fit to the $T_{\rm d} \ge 21.5~{\rm K}$
\footnote{
Later we confirmed that the coefficient for $R$ was unchanged when we broadened the
temperature range to $T_{\rm d} \ge 20.5~{\rm K}$ in Section~4.3.
We also evaluated the systematic uncertainty of the coefficient and its effect on the $\HI$ emissivity spectrum
in Section~5.}
dust--$\WHI$ relation in Figure~2 with a linear function
with an intercept fixed at 0
\footnote{
Whether we allow the intercept to be free to vary or hold it fixed
at 0 (or a small value) is not expected to affect the results
significantly, since we determined the coefficient in regions with high $T_{\rm d}$
where the scatter is narrow, and then applied the correction based on $T_{\rm d}$ [Equation~(3)] to match the $\gamma$-ray data.
}
and obtained coefficients of 
$(19.8 \pm 1.5) \times 10^{8}~{\rm K~km~s^{-1}~(W~m^{-2}~sr^{-1})^{-1}}$
and 
$(102 \pm 13) \times 10^{6}~{\rm K~km~s^{-1}}$
for the $R$ and $\tau_{353}$, respectively,
where the errors are given as the RMS deviations.
We then converted $R$ (or $\tau_{353}$) into $N({\rm H_{tot}})$ maps,
using the coefficients obtained and multiplied by $1.82 \times 10^{18}~{\rm cm^{-2}~(K~km~s^{-1})^{-1}}$.
The obtained total gas column density template maps (proportional to the $R$ or $\tau_{353}$ maps)
are shown in Figure~3.
By comparing these maps to the $N(\HI_{\rm thin})$ map and the $W_{\rm CO}$ maps shown in Figure~1, 
we can recognize dense gas not accounted for by $\HI$ (the optically thin case)
in the MBM 53, 54, and 55 clouds and the Pegasus loop near the emission from CO. 
We can also see that the $\tau_{353}$-based map predicts a stronger contrast for
$N({\rm H_{tot}})$ distribution, and 
approximately a factor of two higher gas column density in dense clouds when compared to the $R$-based map.

Because good angular resolution is essential to examine the 
correlation between the $\gamma$ rays and the gas distribution, we restricted the energy 
to above 0.3~GeV.
The model described in Equation~1 was fitted to the data using
\textit{Fermi} Science Tools, which take into account the energy-dependent instrumental
point-spread function and the effective area. 
We analyzed the LAT data
from 0.3 to 72.9~GeV using the logarithmically equally spaced energy bands 0.3--0.52~GeV, 
0.52--0.9~GeV, 0.9--1.56~GeV, 1.56--2.7~GeV, 2.7--4.68~GeV, and 4.68--8.1~GeV. 
Above 8.1~GeV, we used wider energy
ranges of 8.1--24.3~GeV and 24.3--72.9~GeV to compensate for the low photon statistics.
We then have compared the data and model in each energy range using a
binned maximum-likelihood method with Poisson statistics in $0\fdg25 \times 0\fdg25$ bins.
Within each narrow energy range, we assumed constant spectra for the gas component and the IC emission
and assumed $c_{1}$ and $c_{2}$ to be free normalization parameters.
For $I_{\rm iso}$ and ${\rm PS}_{j}$ we assumed power-law spectra
with photon index fixed at 2.2 and free normalization.
In the highest energy range (24.3--72.9~GeV),
we found that the IC component (less intense than the isotropic component) was not well determined and 
fixed the scale factor to 1.
When modeling the point sources, we iteratively included them 
in several groups at a time in the order of decreasing significance.
We first included and fitted nine bright sources detected in 3FGL at more than 20$\sigma$;
then added and fit a second group (nine sources) detected at 13--20$\sigma$,
freezing the source parameters already included;
and added/fit a third group (10 sources) detected at
9--13$\sigma$ with the parameters of the already included sources again frozen.
In this way, we wound down to the sources detected at more than 5$\sigma$ in 3FGL.
In each step, the parameters of the diffuse emission model were always kept free to be varied.
We found that by including sources 
with significance from 5$\sigma$ to 6$\sigma$, the effects on gas-related scale factors were
1--2\% (comparable to or smaller than the statistical error) below 8.1~GeV, and
${\sim}4\%$ (about one-fourth of the statistical error) in the highest energy bin.
We therefore did not include sources with lower 3FGL significance.
Finally, the analysis was repeated with all the sources, letting only the parameters of
the diffuse model and those of the nine brightest sources vary freely.
To model the contamination from outside the ROI,
we took into account 17 point sources (with model parameters fixed to those of 3FGL) 
detected above 20$\sigma$ in 3FGL located 
at a distance ${\le}10\arcdeg$ from the region boundaries.
We also used $N({\rm H_{tot}})$ and $I_{\rm IC}$ maps
including peripheral regions.
The obtained log-likelihoods, $\ln{L}$
\footnote{
$L$ is conventionally calculated as $\ln{L}=\sum_{i}n_{i} \ln(\theta_{i})-\sum_{i}\theta_{i}$,
where $n_{i}$ and $\theta_{i}$ are the data and the model-predicted counts in each pixel
denoted by the subscript, respectively \citep[see, e.g.,][]{Mattox1996}
}
summed over individual energy ranges in 0.3--72.9~GeV
with the $R$-based and
$\tau_{353}$-based $N({\rm H_{tot}})$ maps are 757496.7 and 757452.3, respectively. 
Therefore, the $R$-based $N({\rm H_{tot}})$ map is preferred by the $\gamma$-ray data. The average of
the normalization for the gas component, $c_{1}$ in Equation~1, is $0.884 \pm 0.011$ and 
$0.391 \pm 0.005$ for the $R$-based and $\tau_{353}$-based maps, respectively.

\begin{figure}[ht!]
\figurenum{3}
\gridline{
\fig{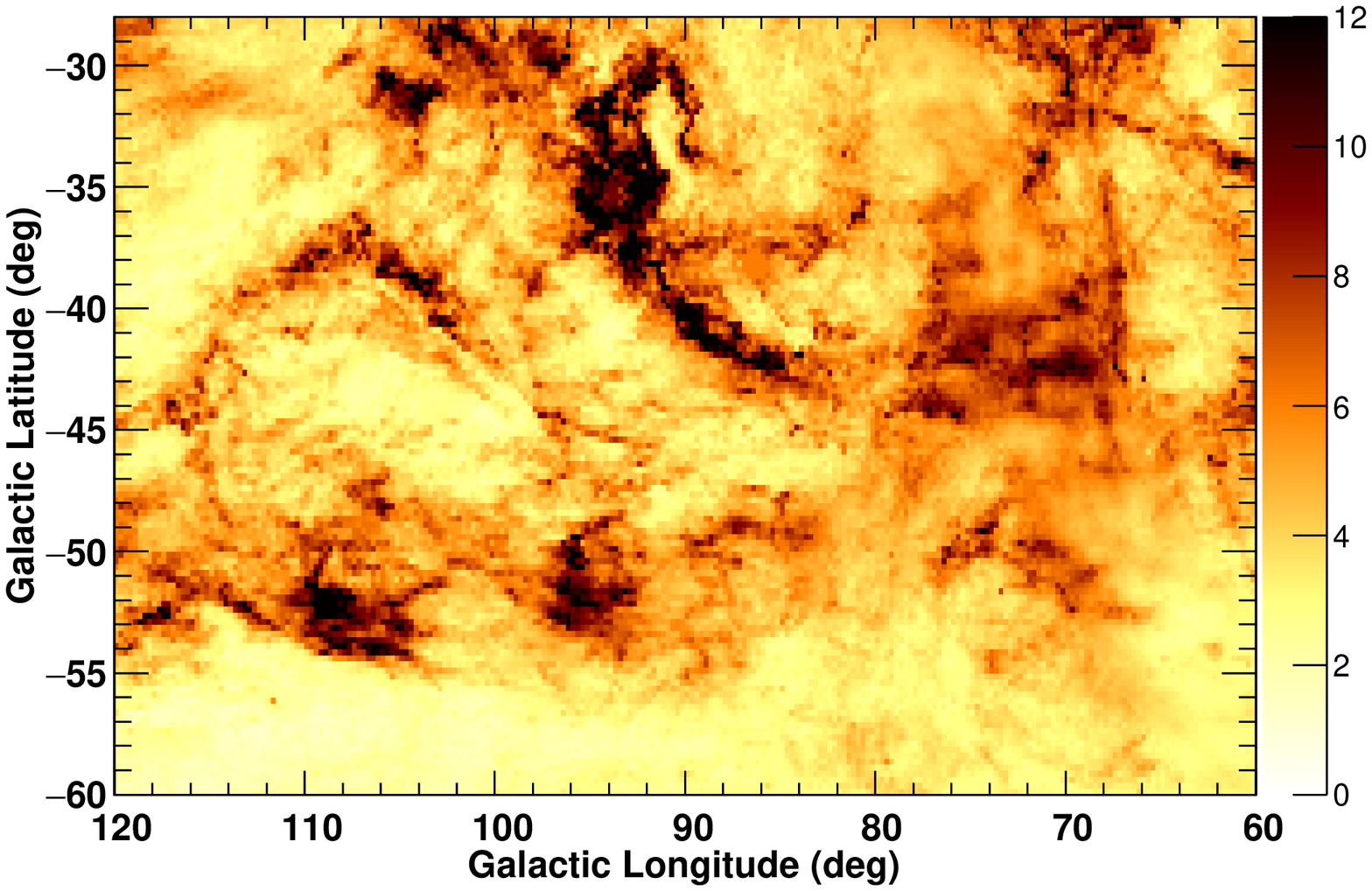}
{0.5\textwidth}{(a)}
\fig{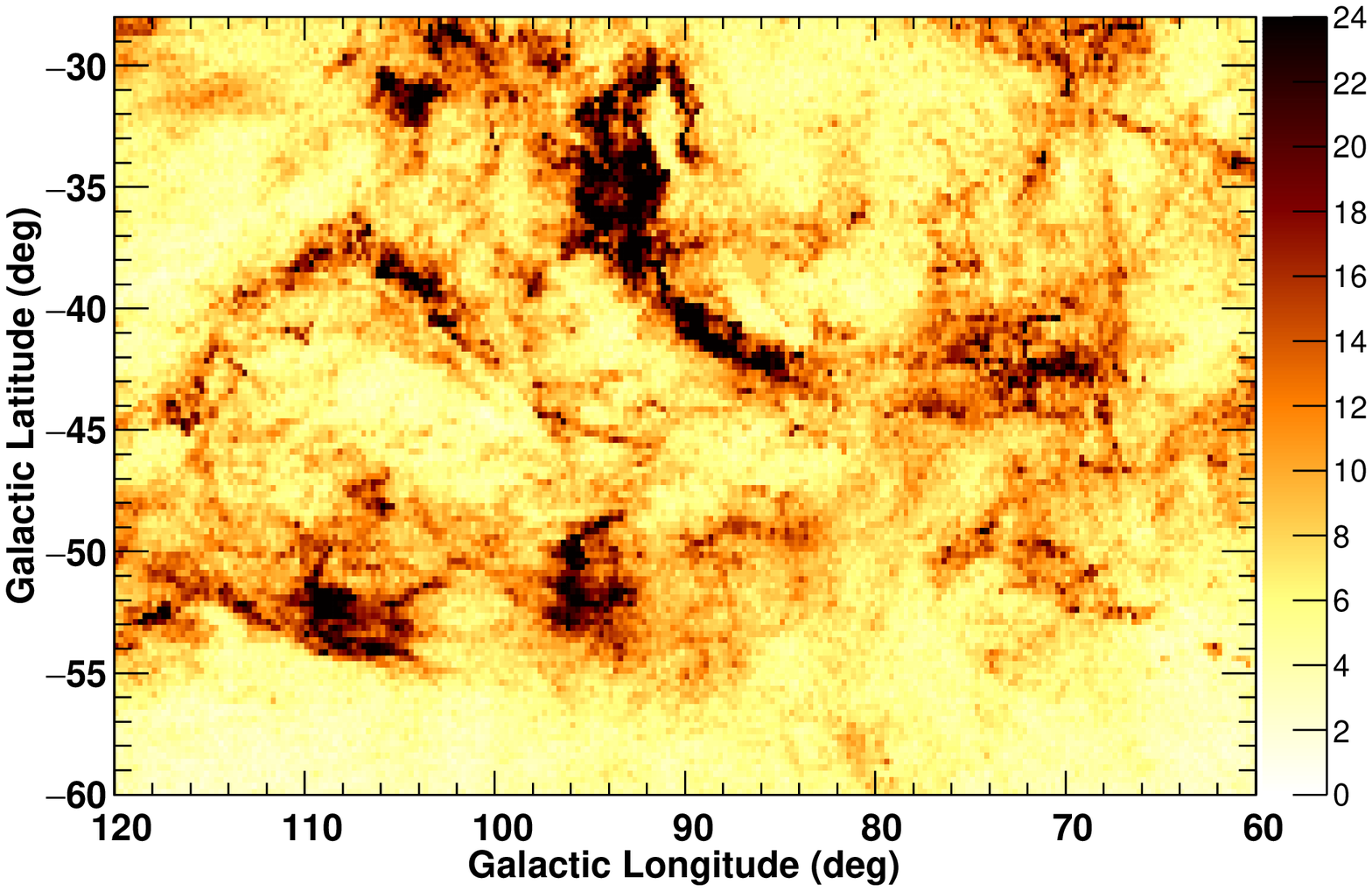}
{0.5\textwidth}{(b)}
}
\caption{
The total gas column density template maps
shown in units of $10^{20}~{\rm cm^{-2}}$:
(a) map based on $R$ ($N({\rm H_{tot}}) \propto R$) and
(b) map based on $\tau_{353}$ ($N({\rm H_{tot}}) \propto \tau_{353}$).
See the text in Section~4.1 for details of the procedure to construct those maps.
Note that the dynamic range is different in the two plots. 
\label{fig:f3}
}
\end{figure}

\clearpage

\subsection{
Dust Temperature-Sorted Modeling
}


As we saw in Section~2 (Figure~2), 
the correlations between $\WHI$ and dust tracers ($R$ or $\tau_{353}$) depend
on $T_{\rm d}$, and the temperature dependence is significantly different
in the cases of $R$ or $\tau_{353}$.
Even though the $R$-based $N({\rm H_{tot}})$ map is preferred to the $\tau_{353}$-based one
in terms of $\ln{L}$
by $\gamma$-ray data analysis as described in Section~4.1,
the true $N({\rm H_{tot}})$ distribution could be appreciably different from either of them.
To investigate the temperature dependence more quantitatively,
we proceeded to an analysis with $T_{\rm d}$-sorted template maps as described below.

We split the $N({\rm H_{tot}})$ template map (constructed from $R$ or $\tau_{353}$)
into four based on $T_{\rm d}$, for $T_{\rm d} \le 18~{\rm K}$, $T_{\rm d}={\rm 18\mbox{--}19~{K}}$,
$T_{\rm d}={\rm 19\mbox{--}20~{K}}$ and $T_{\rm d} \ge 20~{\rm K}$
\footnote{
The relative solid angles in our ROI are 9.5\%, 41.2\%, 36.7\%, and 12.6\% for
$T_{\rm d} \le 18~{\rm K}$, $T_{\rm d}={\rm 18\mbox{--}19~{K}}$,
$T_{\rm d}={\rm 19\mbox{--}20~{K}}$ and $T_{\rm d} \ge 20~{\rm K}$, respectively.
},
and fit $\gamma$-ray data with Equation~1
using the four template maps, with scaling factors ($c_{1,i}(E)$ for each of the four templates)
freely varying individually instead of using a single $N({\rm H_{tot}})$ map. 
Because our new model now had more
free parameters, the narrow energy ranges were no longer feasible 
and we combined two adjacent energy ranges
to make broader ranges:
0.3--0.9~GeV, 0.9--2.7~GeV, 2.7--8.1~GeV and 8.1--72.9~GeV. 
To accommodate these wider energy ranges,
we modeled the IC (scale factor $c_{2}(E)$) and isotropic ($I_{\rm iso}(E)$) 
intensities as power laws with both the normalization and photon index free to vary in each energy range. 
Two bright sources, 3C~454.3 and 3FGL~J2232.5+1143, were also fitted with power laws
with  both the normalization and photon index allowed to vary.
To test whether splitting the gas template map and fitting the scale factors individually 
yielded statistically significant improvement in the likelihood,
we first fit the $\gamma$-ray data using a single $N(\rm H_{tot})$ map in the wider energy range,
and then proceeded to the analysis with the four template gas maps.
The values of $\ln{L}$ obtained from a fit in four energy bands with a single $N(\rm H_{tot})$ map 
(summed over the individual energy ranges)
were 1262782.5 and 1262751.2
for the $R$-based and $\tau_{353}$-based maps, respectively, and the values of $\ln{L}$ obtained
from a fit with the four template maps were 1262821.1 and 1262802.8, respectively.
The improvement in the fit, the likelihood test statistic ${\rm TS} \equiv 2\Delta \ln{L}$,
was 77.2 and 103.2 with 12 more degrees of freedom (giving a statistical significance of 6.7$\sigma$ and 8.2$\sigma$)
for the $R$-based and $\tau_{353}$-based fits,
respectively. Therefore, the fit improvement was significant in both cases
\footnote{
TS for the null hypothesis is asymptotically distributed as
chi-square with the degrees of freedom. (\url{http://fermi.gsfc.nasa.gov/ssc/data/analysis/documentation/Cicerone/Cicerone_Likelihood/Likelihood_overview.html})};
however, the $R$-based analysis was still preferred.
Obtained fit parameters and spectra of each component are summarized in Appendix~C. 

As shown in Tables C2 and C3 in Appendix~C, we observed that
the scaling factors $c_{1,i}(E)$ depended on $T_{\rm d}$;
the averages over the entire energy range are summarized in Figure~4,
which shows a clear negative correlation (lower scaling factor at higher $T_{\rm d}$) 
and a positive correlation
for the $R$-based and $\tau_{353}$-based $N({\rm H_{tot}})$ maps, respectively.
These trends cannot be interpreted as being due to the properties of CRs, because the
physical environments that determine the $T_{\rm d}$ (e.g., the ISRF intensity and dust cross section)
do not affect the CR density.
The only possible explanation in terms of the CR properties is the exclusion of charged particles 
in dense clouds with large magnetic fields. 
However, CRs have been confirmed to penetrate into dense cloud cores
with $W_{\rm CO} \ge 10~{\rm K~km~s^{-1}}$ \citep[e.g.,][]{Fermi2ndQ,Fermi3rdQ,FermiCham}
which corresponds to densities much larger than those of clouds studied here.
Therefore, the main cause of $T_{\rm d}$ dependence found here is not attributable to the
properties of CRs.

Another possible cause of the apparent $T_{\rm d}$ dependence is the uncertainty of the IC model.
Even if we adjusted the IC spectrum 
by scaling it
in each energy range, the spatial distribution of our IC model 
might not be accurate.
This could affect the results shown in Figure~4 in two ways:
one by changing the slope of $T_{\rm d}$ dependence
[i.e., changing the measured $T_{\rm d}$ dependence of $N({\rm H_{tot}})/R$ and $N({\rm H_{tot}})/\tau_{353}$],
and the other by changing the values of the scaling factor
(i.e., measured $\gamma$-ray emissivity or CR density).
To investigate this possibility, we tested several other IC models. 
As described in Section~2, we used the IC model produced in the GALPROP run
54\_77Xvarh7S as our baseline model.
This configuration assumes a CR source distribution 
proportional to
\begin{equation}
f(r) = \left( \frac{r}{r_{\sun}} \right)^{1.25} \exp \left( -3.56 \cdot \frac{r-r_{\sun}}{r_{\sun}} \right)~~,
\end{equation}
where $r$ is the Galactocentric distance and
$r_{\sun}={\rm 8.5~kpc}$ is the distance from the Sun to the Galactic center.
In the baseline model, the
boundary of the cosmic-ray halo $z_{h}$ is chosen to be 4~kpc with 
uniform spatial diffusion coefficient $D_{xx} = \beta D_{0} (\rho/{\rm 4~GV})^{\delta}$ across the Galaxy,
where $\beta \equiv v/c$ is the velocity of a particle relative to the speed of light,
$\rho$ is the rigidity of the particle, and $D_{0}={\rm 5.8 \times 10^{28}~cm^{2}~s^{-1}}$
and $\delta=0.33$ (the Kolmogorov spectrum) were adopted. 
As described in \citet{Fermi3rdQ} and \citet{Palma2012},
of those parameters, the CR source distribution and the halo height 
typically most strongly affect the propagated
CR spatial distribution 
(in Galactocentric distance and the height from the Galactic plane)
and therefore the IC spatial distribution (in $l$ and $b$).
Therefore, we tested two more CR source distributions, the pulsar-based distribution and
the SNR (supernova remnant)-based distribution \citep[see figure~12 of][]{Fermi3rdQ}
in addition to the distribution in the baseline model,
and two more CR halo heights,
10~kpc and 20~kpc, 
in addition to the halo height (4~kpc) in the baseline model.
To match the direct CR measurements at Earth, $D_{0}={\rm 5.8 \times 10^{28}~cm^{2}~s^{-1}}$ was adjusted
when changing $z_{h}$ as described in \citet{Fermi3rdQ}.
The obtained IC maps show the smallest gradient in the Galactic longitude direction
with the SNR-based CR source distribution (the flattest distribution of our three choices), 
and the smallest gradient in the Galactic latitude direction with $z_{h}=20~{\rm kpc}$ 
(the largest halo height of our three choices). 
Our baseline model provides a reasonably good fit to the data and provides the
second highest value of $\ln{L}$ among all models considered.
All nine configurations show the same trend in the $T_{\rm d}$ dependence of the scale factor $c_{1,i}$
(negative/positive correlation with $T_{\rm d}$ in the $R$-based/$\tau_{353}$-based analysis),
and the scale factors are not affected significantly.
Therefore, we conclude that our finding concerning the $T_{\rm d}$ dependence is robust
against variations in the models of IC emission,
and that the primary cause of the dependence is the 
non-uniformity of 
$N({\rm H_{tot}})/R$ and $N({\rm H_{tot}})/\tau_{353}$.

One may also argue that $N({\rm H_{tot}})/R$ or $N({\rm H_{tot}})/\tau_{353}$
is appreciably different in dense molecular cloud complexes from that in translucent clouds,
affecting the slope of the $T_{\rm d}$ dependence 
(and the values of the scale factor if the best-fit isotropic component
and IC emission is affected) appreciably.
Therefore, we masked areas around the dense molecular clouds traced by CO,
as indicated by the dotted lines in Figure~1b, and repeated the same analysis
using our baseline IC model.
We again confirmed that the $T_{\rm d}$ dependence we found 
(negative/positive correlation with $T_{\rm d}$ in the $R$-based/$\tau_{353}$-based analyses) remained
and the scale factors were not affected significantly.
We also masked IVCs in our ROI (see Section~2 and Appendix~B)
as indicated by the dotted lines in Figure~B1b, since they could have different CR density 
and/or different properties of the ISM gas and dust,
and performed the same test. Again, the effect on the results shown in Figure~4 was found to be small.
To gauge systematic uncertainties, we bracketed the $T_{\rm d}$ dependence of our baseline model in Figure~4
with that obtained using the IC model of the pulsar-based CR source distribution and $z_{h}={\rm 20~kpc}$
(which shows the largest difference among the nine possibilities from our baseline IC model),
that obtained using our baseline IC model but with areas around the molecular clouds traced by CO masked, 
and that obtained using our baseline IC model but with areas of IVCs masked,
as shown by the shaded bands in Figure~4.

We thus found that the ratios of $N({\rm H_{tot}})$/$R$ and $N({\rm H_{tot}})$/$\tau_{353}$
(both $\propto c_{1, i}$) are not uniform and depend on $T_{\rm d}$.
Even though the variation is by only 30\%--40\% over the range of $T_{\rm d}$ in the region,
this shows both $R$ and $\tau_{353}$ are not accurate tracers of the total gas column densities
and we give possible explanations for this below.

\begin{description}

\item[{\boldmath $R$} (Radiance)]\mbox{}\\
Under the assumption of a uniform dust-to-gas ratio and dust emissivity,
$R$ (the dust bolometric luminosity) per H atom,
or the dust specific luminosity,
will be constant if the ISRF is uniform along the line of sight.
This is the basis of the claim by the \citet{Planck2014} that
$R$ is a good tracer of the total dust (and gas) column density.
However, even though the ISRF is uniform in the vicinity of the solar system,
the $R$ per H atom could decrease as the gas (and dust) density increases,
because the ISRF is more strongly absorbed by dust at higher density.
This will cause a correlated decrease 
in the $T_{\rm d}$ and dust specific luminosity.
This qualitative argument is supported by theoretical works, e.g., by \citet{Ysard2015}.
We note that results by \citet{Ysard2015} are for very diffuse ISM 
($N({\rm H_{tot})} \le 2.5 \times 10^{20}~{\rm cm^{-2}}$). Therefore direct comparison with our results 
is not appropriate.
We also note that a correlated decrease in the $T_{\rm d}$ and dust specific luminosity
was observed by \citet{Planck2015} in the Chamaeleon molecular cloud region (see Figure~13 of the reference).
Therefore, the trend we found is not specific to the region under study here but likely to be universal.

\item[{\boldmath $\tau_{353}$}]\mbox{}\\
In the optically thin limit, the specific intensity of the dust emission $I_{\nu}$ is given as
$I_{\nu}=\tau_{\nu} B_{\nu} (T_{\rm d}) = \sigma_{\nu} N({\rm H_{tot}}) B_{\nu} (T_{\rm d})$,
where $\tau_{\nu}$ and $\sigma_{\nu}$ are the optical depth and the dust opacity (cross section)
per hydrogen atom, respectively, and $B_{\nu}$ is the Planck function.
$\sigma_{\nu}$ depends on the frequency and is often described as
a power law, giving $I_{\nu}=\tau_{\nu_{0}} (\nu/\nu_{0})^{\beta} B_{\nu}(T_{\rm d})$ 
(i.e., a modified blackbody)
with $\beta = 1.5\mbox{--}2$ and $\nu_{0}$ is a reference frequency. Therefore, if the dust cross section is uniform
in the ROI, $\tau_{\nu}$ is proportional to $N({\rm H_{tot}})$ and we can measure the
total gas column density by measuring the dust optical depth at any frequency.
Even though this assumption has been adopted in multiple studies,
the dust spectral index $\beta$ and therefore the dust opacity are not uniform
but rather anti-correlate with $T_{\rm d}$ as reported by the \citet{Planck2014}.
We examined our map and confirmed this anti-correlation between $T_{\rm d}$ and $\beta$ in our ROI.
\end{description}

Therefore, neither $R$ nor $\tau_{353}$ is guaranteed to be a good tracer of $N({\rm H_{tot}})$. 
As described above, we used another gas tracer, the $\gamma$-ray data from the \textit{Fermi}-LAT, 
and found that both $N({\rm H_{tot}})$/$R$ and $N({\rm H_{tot}})$/$\tau_{353}$ 
(both $\propto c_{1, i}$) were not uniform in our ROI, whatever the physical reason may be.
Therefore, we propose to use the $\gamma$-ray data as a robust tracer of $N({\rm H_{tot}})$
to compensate for the $T_{\rm d}$ dependence, as described in Section~4.3 in detail.
We note that $\gamma$-ray observations suffer from low photon statistics, 
and contamination by non-gas-related components, such as IC, isotropic component,
and point sources, and therefore cannot determine the gas column density alone.
Combining the $\gamma$-ray data and other gas tracers 
(e.g., dust, $\HI$ 21-cm line, and CO 2.6-mm line)
is essential to quantify the true gas distribution.

\begin{figure}[ht!]
\figurenum{4}
\gridline{
\fig{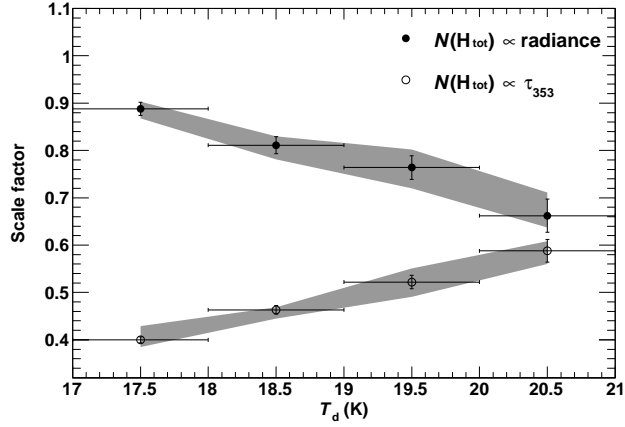}
{0.5\textwidth}{}
}
\caption{
Summary of the scale factors $c_{1,i}$ in Equation~(1) averaged over the entire energy range in
each range of $T_{\rm d}$.
The filled and open circles show the temperature dependence of the scale factors for the $R$-based 
and $\tau_{353}$-based $N({\rm H_{tot}})$ maps, respectively,
and gray bands show the systematic uncertainty (see the text in Section~4.2 for details).
Although a small fraction of the pixels show $T_{\rm d}$ below 17~K or above 21~K,
they are included in the first or last data points, respectively.
\label{fig:f4}
}
\end{figure}

\clearpage

\subsection{
Dust Temperature-Corrected Modeling
}


As described in Section~4.2, we found the $T_{\rm d}$ dependence of 
the scaling factors of the gas component
and concluded that it was primarily due to the dust properties.
We aim to use the $\gamma$-ray data as a robust tracer of $N({\rm H_{tot}})$
and apply a correction on the $N({\rm H_{tot}})$ model map to
compensate for $T_{\rm d}$ dependence.

We started from the $R$-based single $N({\rm H_{tot}})$ map (see Figure~3)
and modified the gas column density with an empirical function as below:
\begin{eqnarray}
N({\rm H_{tot, mod}})=
\left\{
\begin{array}{l}
N({\rm H_{tot, R}})~(T_{\rm d} > T_{\rm bk})~~, \\
(1+0.05 \cdot C \cdot \frac{T_{\rm bk}-T_{\rm d}}{\rm 1~K}) \cdot N({\rm H_{tot, R}})
~(T_{\rm d} \le T_{\rm bk})~~,
\end{array}
\right.
\end{eqnarray}
where $T_{\rm bk} = 20.5~{\rm K}$.
Above $T_{\rm bk}$ we retained the original $N({\rm H_{tot}})$ distribution. We re-examined the
correlation between $R$ and $\WHI$ (Figure~2a) and confirmed that the coefficient between
the two quantities, originally determined for $T_{\rm d}$ above 21.5~K, 
was unchanged above 20.5~K. 
Therefore, our procedure is self-consistent. 
We tested several choices of coefficients and summarized the value of $\ln{L}$ in Figure~5. 
We found that a coefficient $C$ of 2 (which corresponds to a 10\% apparent decrease 
in $R$ per H atom,
or 10\% required increase in the gas column density, as $T_{\rm d}$ decreases by 1~K) 
gave the best representation of the \textit{Fermi}-LAT data.
We also tested $T_{\rm bk}$ values of 20~K and 21~K instead of 20.5~K in Equation~3,
repeated the same analysis, 
and confirmed that the combination of $T_{\rm bk}={\rm 20.5~K}$ and $C=2$ gave the largest value of $\ln{L}$.

Having obtained our modified column density map $N({\rm H_{tot, mod}})$,
we returned to finer energy ranges to study the spectral shape of
each component in more detail. 
The $N({\rm H_{tot, mod}})$ map and the map of the excess gas column density above
$N(\HI_{\rm thin})$ are shown in Figure~6.
The fitting parameters and the obtained spectral components are summarized in Table~1 and Figure~7, respectively. 
The average of the scale factor for the gas component, $c_{1}$, is $0.675 \pm 0.008$ in 0.3--72.9~GeV.
The data count map, model count map and data/model ratio map are summarized in Figure~8.
By comparing Figure~6b with Figure~1, we can see that the excess gas column density 
above $N(\HI_{\rm thin})$ is greatest
in the MBM 53, 54, and 55 clouds (located at $l=84\arcdeg$ to $96\arcdeg$ and $b=-44\arcdeg$ to $-30\arcdeg$)
traced by CO emission.
We also see a significant amount of
excess gas in the Pegasus loop located at $l$ from $98\arcdeg$ to $118\arcdeg$ and
$b$ from $-55\arcdeg$ to $-35\arcdeg$, and an area of 
$l$ from $68\arcdeg$ to $78\arcdeg$ and
$b$ from $-44\arcdeg$ to $-40\arcdeg$.
They are regions with low $T_{\rm d}$ as indicated in Figure~1.

\floattable
\begin{deluxetable}{cccc}[ht!]
\tablecaption{Results of the fit with a single, corrected $N({\rm H_{tot}})$ map \label{tab:Final}}
\tablecolumns{4}
\tablenum{1}
\tablewidth{0pt}
\tablehead{
\colhead{Energy} & \colhead{$c_{1}$} & \colhead{$c_{2}$} & 
\colhead{Integ. $I_{\rm iso}$\tablenotemark{a}} \\
\colhead{(GeV)} & \colhead{} & \colhead{} & 
\colhead{}
}
\startdata
0.3--0.52 & $0.62\pm0.01$ & $0.80\pm0.10$ & $2.68\pm0.05$ \\
0.52--0.9 & $0.68\pm0.02$ & $0.93\pm0.12$ & $1.27\pm0.03$ \\
0.9--1.56 & $0.71\pm0.02$ & $0.92\pm0.15$ & $0.60\pm0.02$ \\
1.56--2.7 & $0.70\pm0.02$ & $0.62\pm0.19$ & $0.29\pm0.01$ \\
2.7--4.68 & $0.79\pm0.04$ & $0.74\pm0.25$ & $0.116\pm0.008$ \\
4.68--8.1 & $0.76\pm0.06$ & $0.93\pm0.35$ & $0.056\pm0.006$ \\
8.1--24.3 & $0.80\pm0.09$ & $0.45\pm0.41$ & $0.047\pm0.005$ \\
24.3--72.9 & $0.95\pm0.27$ & 1.0(fixed) & $0.008\pm0.001$ \\
\enddata
\tablenotetext{a}{The integrated intensity ($10^{-6}~{\rm ph~s^{-1}~cm^{-2}~sr^{-1}}$) in each band.}
\tablecomments{
The errors are 1-sigma statistical uncertainties.
In each energy bin,
$c_{1}$ gives the scale factor of the gas-related component,
IC is multiplied by a scaling factor $c_{2}$,
and
$I_{\rm iso}$ is modeled with a power law (photon index is fixed to 2.2) with the integrated intensity as a free parameter.
}
\end{deluxetable}

\begin{figure}[ht!]
\figurenum{5}
\gridline{
\fig{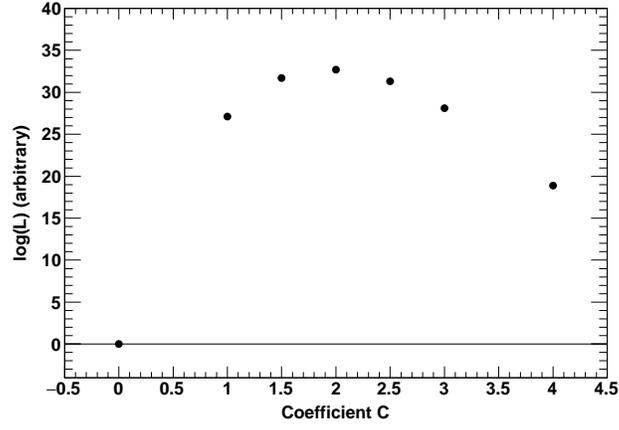}
{0.5\textwidth}{}
}
\caption{
Summary of $\ln{L}$ as a function of the coefficient $C$ in Equation~3 with
$T_{\rm bk}=20.5~{\rm K}$.
\label{fig:f5}
}
\end{figure}

\begin{figure}[ht!]
\figurenum{6}
\gridline{
\fig{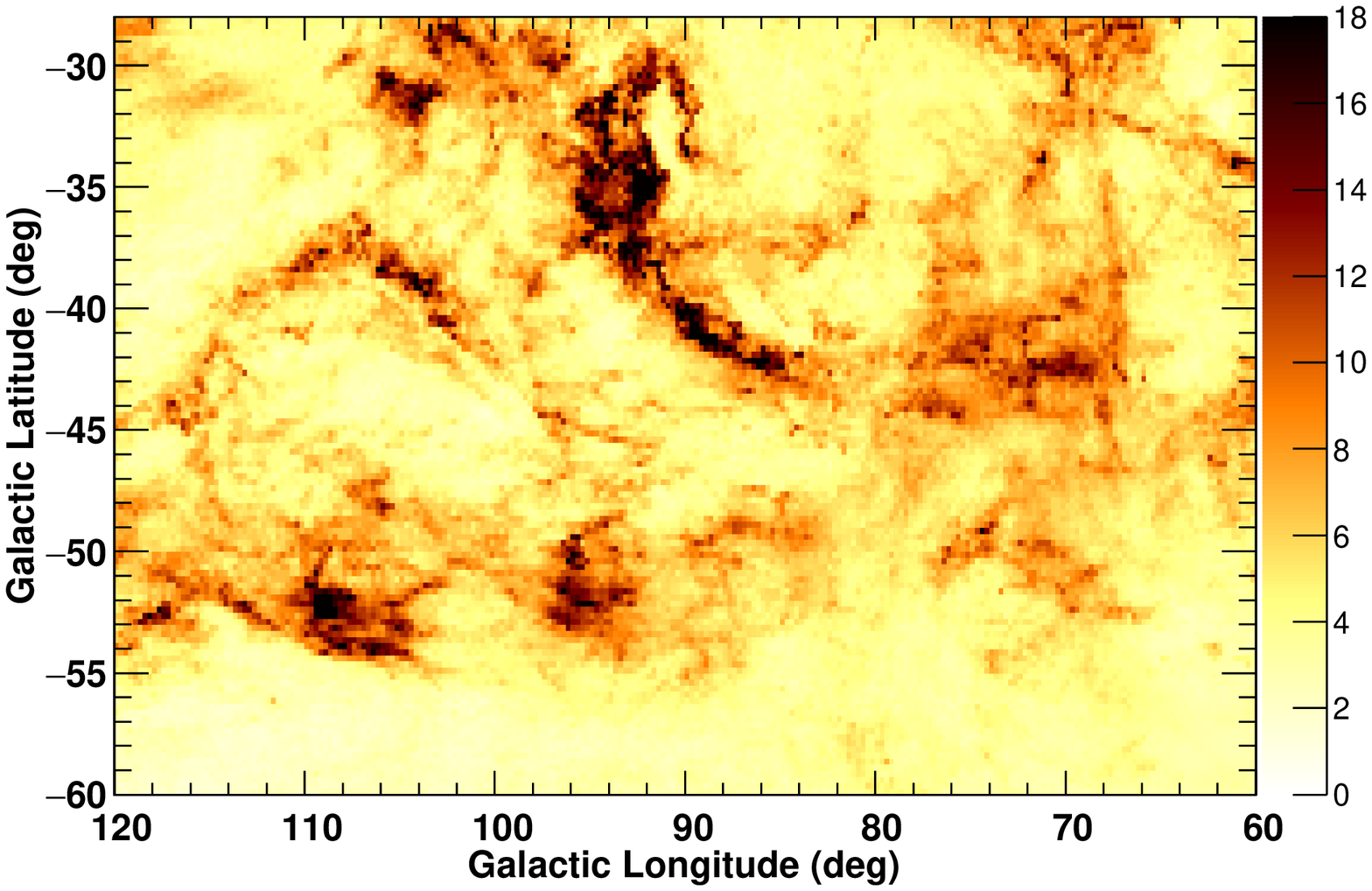}
{0.5\textwidth}{(a)}
\fig{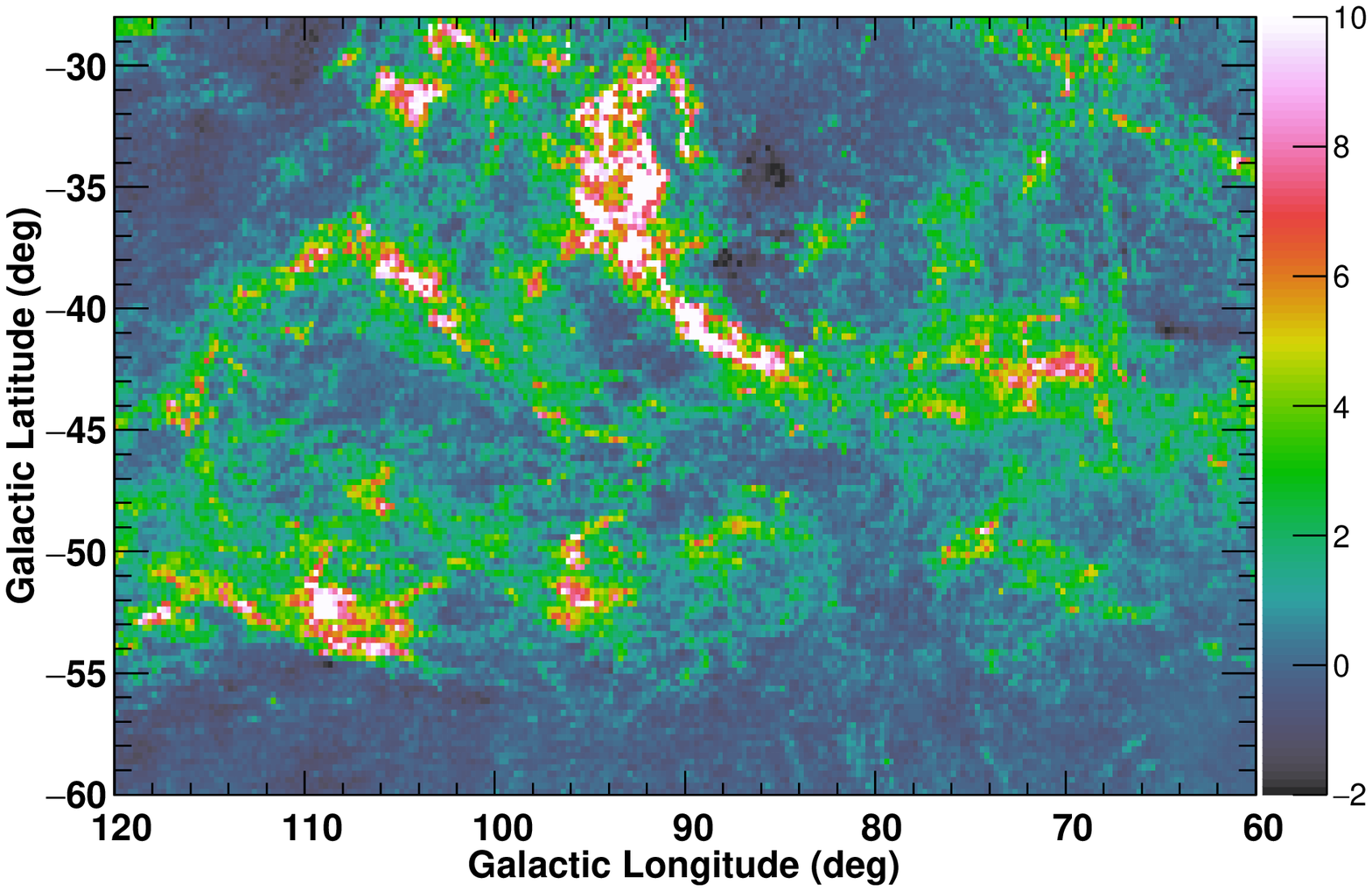}
{0.5\textwidth}{(b)}
}
\caption{
(a) The corrected $N({\rm H_{tot}})$ map and (b) the map of the excess gas column density above
$N(\HI_{\rm thin})$. Both panels are shown in units of $10^{20}~{\rm cm^{-2}}$.
\label{fig:f6}
}
\end{figure}

\begin{figure}[ht!]
\figurenum{7}
\gridline{
\fig{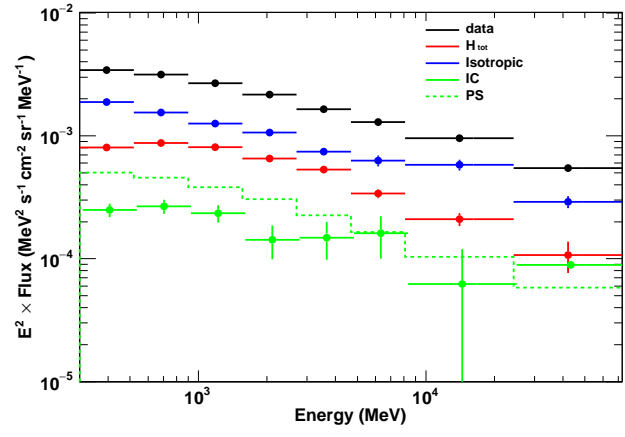}
{0.5\textwidth}{}
}
\caption{
Spectrum of each component obtained by the fit with the corrected $N({\rm H_{tot}})$ map.
\label{fig:f7}
}
\end{figure}

\begin{figure}[ht!]
\figurenum{8}
\gridline{
\fig{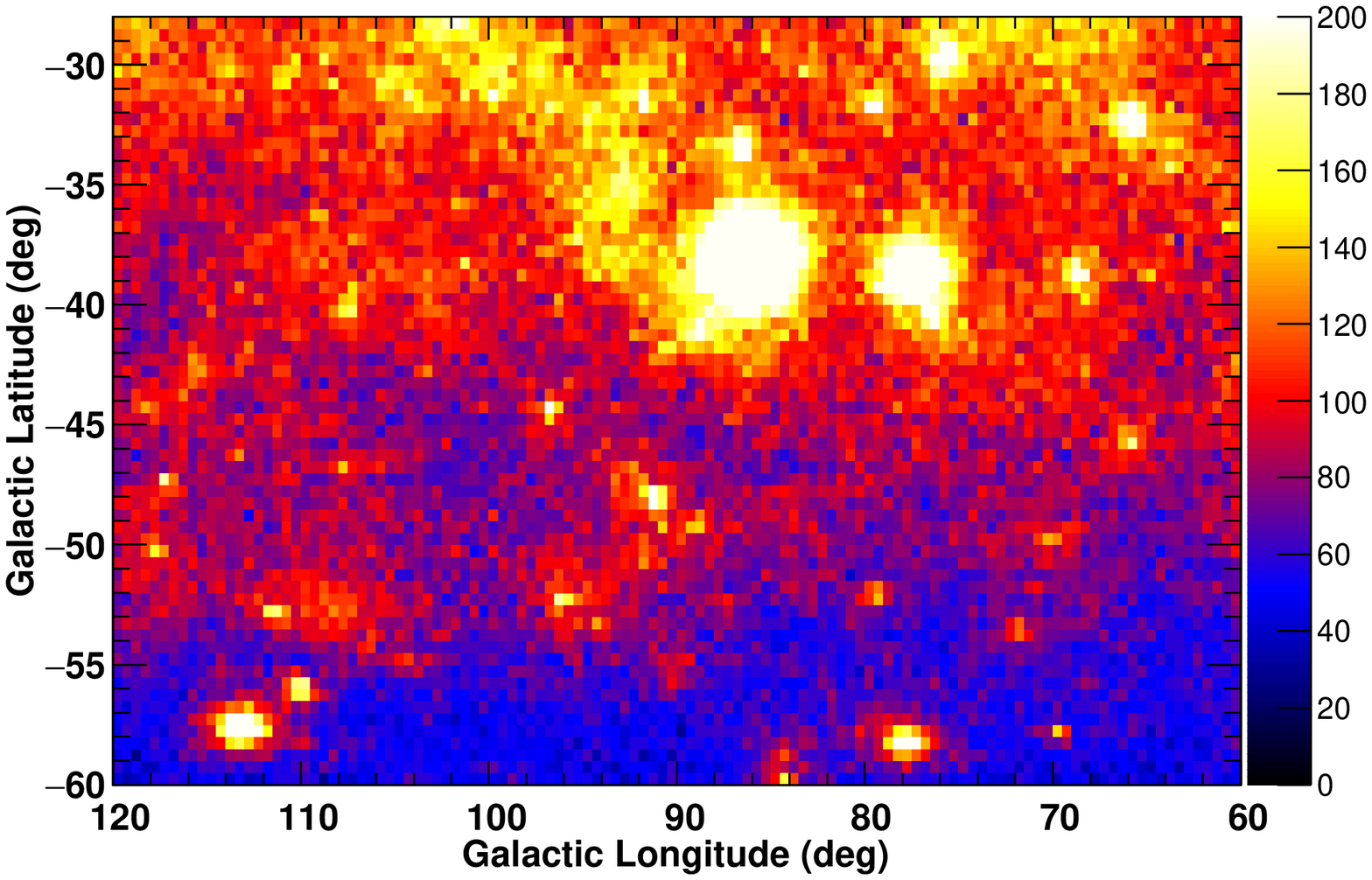}
{0.5\textwidth}{(a)}
\fig{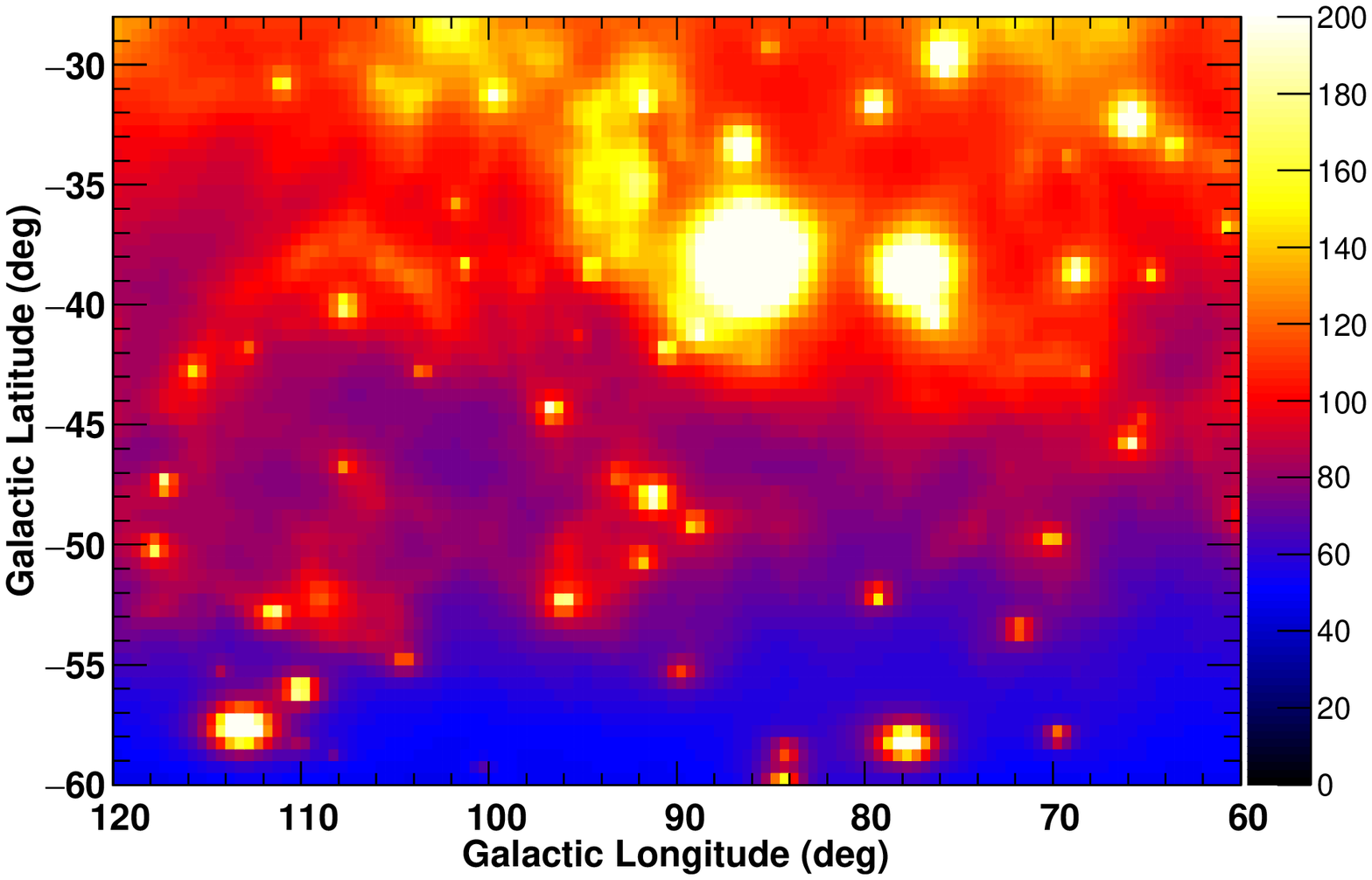}
{0.5\textwidth}{(b)}
}
\gridline{
\fig{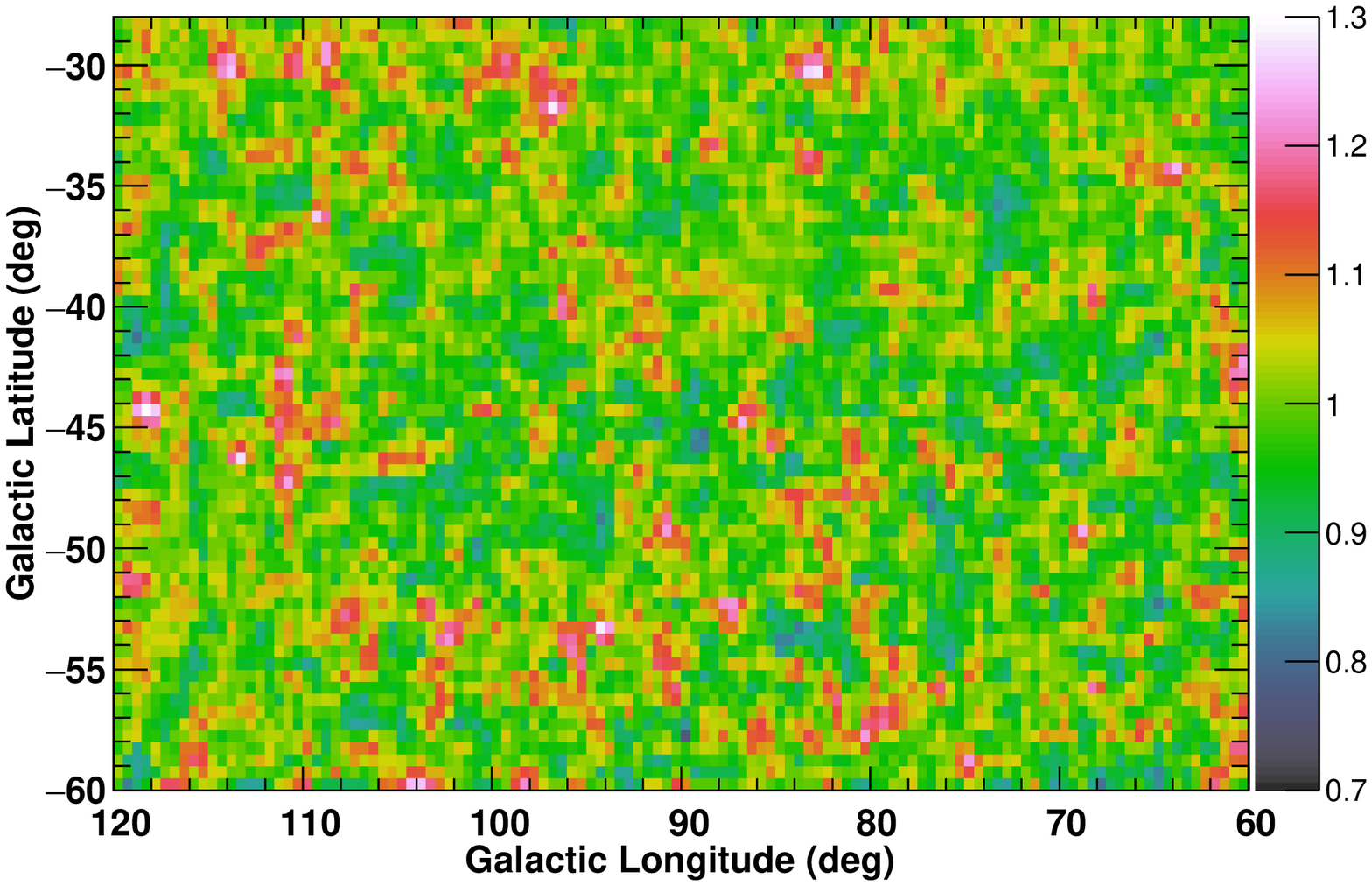}
{0.5\textwidth}{(c)}
}
\caption{
(a) The data count map, (b) the model count map, and (c) the data/model ratio map
obtained by the fit with the corrected $N({\rm H_{tot}})$ map. Although the fit has been performed
in $0\fdg25 \times 0\fdg25$ bins,
these maps have been rebinned in $0\fdg5 \times 0\fdg5$ pixels for display.
Data/model ratio map has been smoothed with a k3a kernel
(1-2-1 two-dimensional boxcar smoothing)
in ROOT framework (\url{https://root.cern.ch}).
\label{fig:f8}
}
\end{figure}

\clearpage

\section{Discussion}

In Section~4 we showed that neither $R$ nor $\tau_{353}$ were good representations of
the total gas column density 
in the ROI
and used the \textit{Fermi}-LAT $\gamma$-ray data to compensate 
for the 
$T_{\rm d}$ dependence of $N({\rm H_{tot}})/R$.
The correlation between $\WHI$ and the $N({\rm H_{tot}})$ 
inferred by the $\gamma$-ray data analysis 
obtained from Equation~3 with $T_{\rm bk}={\rm 20.5~K}$ and $C=2$ is shown in Figure~9a,
in which we observe moderate scatter.
We note that our corrected $N({\rm H_{tot}})$ model map is based on $R$, and therefore is expected
to include contributions from both atomic and molecular hydrogen.

Because the MBM 53, 54, and 55 clouds and the Pegasus loop are located at similar distances from the 
solar system and most of the $\HI$ clouds are expected to coexist with the $\Htwo$ clouds
(because they are located at high Galactic latitudes),
we can estimate the mass of gas from the column density $N({\rm H})$ as
\begin{equation}
M = \mu m_{\rm H} d^{2} \int N({\rm H}) \,d\Omega~~,
\end{equation}
where $d$ is the distance to the cloud, $m_{\rm H}$ is the mass of the hydrogen atom
and $\mu=1.41$ is the mean atomic mass per H atom \citep{AQ2000}.
Although our ROI includes IVCs, their contribution to the integral of $N({\HI})$
was at the 5\% level (see Appendix~B); therefore the impact on the discussion of the cloud mass distribution is small.
From Equation~(4),
$\int N({\rm H}) \,d\Omega=10^{22}~{\rm cm^{-2}~deg^{2}}$ corresponds to {$\sim$}740~${\rm M_{\sun}}$ for
$d=150~{\rm pc}$. 
The integrated $\HI$ column density for the optically thin case,
$\int N(\HI_{\rm thin}) \,d\Omega = 1.82 \times 10^{18} \cdot \int \WHI \,d\Omega$,
and the integrated column density of the excess gas ($\int (N({\rm H_{tot}})-N(\HI_{\rm thin})) \,d\Omega$)
are calculated
in 0.5-K step $T_{\rm d}$ bins and are summarized in Figure~9b.
We observed that the excess gas 
starts to appear below $T_{\rm d}=20~{\rm K}$ and 
contributes
{$\sim$}1/3 of the total amount of gas in the range of $T_{\rm d}=18\mbox{--}18.5~{\rm K}$.
Below $T_{\rm d}=18~{\rm K}$, the excess gas is as massive as that of
$\HI$ for the optically thin case.

Some fraction of this excess gas is molecular hydrogen traced by
CO (hereafter denoted as $\HtwoCO$), and we can calculate the mass as
\begin{equation}
M = \mu m_{\rm H} d^{2} \int 2N(\HtwoCO) \,d\Omega = \mu m_{\rm H} d^{2} \cdot 2 X_{\rm CO} \int W_{\rm CO} \,d\Omega~~.
\end{equation}
To estimate the value of $X_{\rm CO}$, we examined the correlation between
$N({\rm H_{tot}})-N(\HI_{\rm thin})$ and (moment-masked) $W_{\rm CO}$ as shown
in Figure~9c.
There, we observe large scatter, particularly in the low $W_{\rm CO}$ area, 
likely due to dark gas
(gas not traced by the $\HI$ 21-cm line or the CO 2.6-mm line).
If we restrict $T_{\rm d}$ below 17~K, where the $\Htwo$ gas traced by CO is expected to be dominant,
we see that the correlation becomes better and $W_{\rm CO}$ starts to appear above 
${\sim}5 \times 10^{20}~{\rm cm^{-2}}$.
The average of $N({\rm H_{tot}})-N(\HI_{\rm thin})$ in $T_{\rm d} \le 17~{\rm K}$ and
$W_{\rm CO} \le 0.1~{\rm K~km~s^{-1}}$ is $5.4 \times 10^{20}~{\rm cm^{-2}}$, which can be interpreted
as an offset due to the $\Htwo$ gas not being traced by CO or 
$\HI$ gas in the optically thin condition.
We then fit the other data points in $T_{\rm d} \le 17~{\rm K}$ using a linear function
with its intercept in the horizontal axis fixed at this average, and obtained a slope
of $0.40~{\rm K~km~s^{-1}~(10^{20}~cm^{-2})^{-1}}$. 
This translates into 
$X_{\rm CO} = 1.25 \times 10^{20}~{\rm cm^{-2}~(K~km~s^{-1})^{-1}}$ which is a typical value
obtained via the $\gamma$-ray data analysis of nearby molecular cloud complexes
by \textit{Fermi}-LAT \citep[see Figure~9 in][]{CRreview}.
We note that, given the larger scatter in the $W_{\rm CO}$ versus $N({\rm H_{tot}})-N(\HI_{\rm thin})$ relation
(at least partially due to the dark gas as described above), the uncertainty of
our estimated $X_{\rm CO}$ is large and 
possibly by a factor as large as two.
Although a precise determination of $X_{\rm CO}$ is important,
it is beyond the scope of this study. Using the $X_{\rm CO}$ estimated above,
we calculated $2 X_{\rm CO} \int W_{\rm CO} \,d\Omega$
which is a measure of the CO-bright $\Htwo$ mass (Equation~5), and 
plotted the distribution in Figure~9b.
The integral of $N(\HI_{\rm thin})$ and $N({\rm H_{tot}})-N(\HI_{\rm thin})$
is 60.9 and 16.5 in units of $10^{22}~{\rm cm^{-2}~deg^{2}}$, respectively,
and that of $W_{\rm CO}$ multiplied by $2X_{\rm CO}$ is 2.6 in the same units.
By comparing the excess mass distribution (red line) and the CO-bright $\Htwo$ mass distribution (blue dotted line) in Figure~9b,
we can see that most of the excess mass can be attributed to CO-bright $\Htwo$ gas
at $T_{\rm d} \le 17~{\rm K}$, as expected. 
However, above 17.5~K, the contribution of the CO-bright
molecular mass is {$\le$}10\% and cannot explain the excess mass
even if we assume $X_{\rm CO}$ is uncertain by a factor of 2.
This is the so-called ``dark gas'' 
and it contributes $(16.5-2.6)/60.9 {\sim}25\%$ of the mass in
$\HI$ in the optically thin case.

The ratio of the mass of dark gas to that of the $\Htwo$ traced by CO is
$(16.5-2.6)/2.6 {\sim}5$. 
(The value should be considered as an upper limit since we assumed that $\HI$ is optically thin.)
This is significantly higher than the values reported by
the \citet{Planck2011} ({$\sim$}120\% on average at high Galactic latitude) and that by
the \citet{Planck2015} ({$\sim$}200\% in the Chamaeleon clouds).
In other words, the region studied here is very dark-gas-rich when compared to $\Htwo$ traced by CO.
To explain the dark gas as being primarily due to CO-dark $\Htwo$, the ratio of
the CO-dark $\Htwo$ and the CO-bright $\Htwo$ should be {$\le$}5 in our case.
This is about a factor of 10 higher than model predictions,
e.g., by \citet{Wolfire2010} and \citet{Smith2014}.
A possible cause of this difference is the assumed physical conditions;
for example, the nominal cloud modeled in \citet{Wolfire2010} is relatively large,
with a total column density of $1.5 \times 10^{22}~{\rm cm^{-2}}$ which is larger than
the largest $N({\rm H_{tot}})$ that we found by a factor of 10. 
They also assumed simplified geometries for the clouds.
The applicability of their results to
more translucent clouds of complicated geometries
is not clear and theoretical investigations are needed.
Another possibility to explain the large dark-gas fraction (compared with the $\Htwo$ traced by CO)
and the scatter in the $\WHI$--$N({\rm H_{tot}})$ relation,
both seen in Figure~9, is the optical thickness of the $\HI$ 21-cm line
\citep[e.g.,][]{Fukui2014,Fukui2015}.
Then, $\WHI$ can be correlated with $N({\rm H_{tot}})$ as a function of the spin temperature $T_{\rm s}$ as
\begin{equation}
\WHI ({\rm K~km~s^{-1}}) = [T_{\rm s}({\rm K})-T_{\rm bg}({\rm K})]
\cdot \Delta V_{\HIs}({\rm km~s^{-1}}) \cdot [1-\exp(-\THI)]~~,
\end{equation}
and 
\begin{equation}
\THI = \frac{N_{\rm H_{tot}}({\rm cm^{-2}})}{1.82 \times 10^{18}}
\cdot \frac{1}{T_{\rm s}({\rm K})}
\cdot \frac{1}{\Delta V_{\HIs} ({\rm km~s^{-1}})}~~,
\end{equation}
where $\Delta V_{\HIs}$ is the $\HI$ line width
[defined as $\WHI$/(peak $\HI$ brightness temperature)],
$T_{\rm bg}$ is the background continuum
radiation temperature, and $\THI$ is the $\HI$ optical depth.
In Figure~9a, 
making an approximation by assuming that all of the gas is atomic and associated with
MBM 53, 54, and 55 clouds, we overlay
the model curves for several choices of $T_{\rm s}$
with $\Delta V_{\HIs}={\rm 14~km~s^{-1}}$
\citep[the median velocity dispersion in the vicinity of the MBM 53--55 complexes; see][]{Fukui2014}
and $T_{\rm bg}=2.7~{\rm K}$ (the cosmic microwave background radiation).
To illustrate the uncertainty of model curves due to the spread of $\Delta V_{\HIs}$, we also plot the curves of
$\Delta V_{\HIs}=11$ and ${\rm 22~km~s^{-1}}$ 
(which covers the half width of the $\Delta V_{\HIs}$ distribution of MBM~53-55)
for $T_{\rm s}=20$ and 100~K as dotted lines.
As inferred from the figure, the region with $T_{\rm d} \ge 20.5~{\rm K}$ is 
almost optically thin.
As the dust temperature decreases, $\HI$ becomes optically thicker with
$T_{\rm s} {\sim}100~{\rm K}$ for $T_{\rm d}=19.5\mbox{--}20~{\rm K}$,
$T_{\rm s} {\sim}60~{\rm K}$ for $T_{\rm d}=18.5\mbox{--}19~{\rm K}$, and
$T_{\rm s} {\sim}40~{\rm K}$ for $T_{\rm d} \le 18~{\rm K}$.

We note that the current data considered in this study
cannot distinguish between the two scenarios
for the primary origin of the dark gas
(CO-dark molecular gas, or optically thick atomic gas, or a mix of both contributions).
Therefore systematic and large surveys of background radio sources 
for direct measurements of the $\HI$ optical depth are important
(although such large surveys may not be feasible). We also look forward to the
progress in theoretical work (e.g., detailed modeling of translucent clouds)
for more detailed discussions of CO-dark $\Htwo$ hypothesis.
The main achievement of this study is quantification of the distribution of $N({\rm H_{tot}})$ and 
the dark gas
by combining the \textit{Fermi}-LAT $\gamma$-ray data and the \textit{Planck} dust model.
If we use the (uncorrected) $R$-based $N({\rm H_{tot}})$ and $\tau_{353}$-based
$N({\rm H_{tot}})$ maps, the integral of $N({\rm H_{tot}})-N(\HI_{\rm thin})$ 
is 6.2 and 56.8 in units of $10^{22}~{\rm cm^{-2}~deg^{2}}$, respectively,
whereas we obtained 16.5 in the same units through $\gamma$-ray data analysis.
Then, the dark gas contribution (obtained by subtracting $2 X_{\rm CO} \int W_{\rm CO} \,d\Omega$
estimated to be ${\sim}2.6$) 
based on uncorrected $R$-based and $\tau_{353}$-based $N({\rm H_{tot}})$ maps will be
3.6 and 54.2, respectively, and is
a factor of {$\sim$}4 lower/higher than what we found ($16.5-2.6=13.9$);
therefore, the correction based on the $\gamma$-ray analysis is crucial.
We also note that the required correction we found
({$\sim$}10\% increase of $N({\rm H_{tot}})$ as $T_{\rm d}$ decreases by 1~K) is the average of the studied region.
A systematic study of other high-latitude regions 
by \textit{Fermi}-LAT with the latest event selections and response functions (Pass~8)
in combination with the \textit{Planck} dust model and careful examinations
of the properties of ISM tracers
is required to examine the uniformity/variation of the ISM properties in the solar neighborhood.
Investigating sub-regions of this study with more LAT data 
also would be worthwhile.

Finally we discuss the $\HI$ emissivity spectrum obtained in this study summarized in Figure~10.
To examine the systematic uncertainty, we repeated the same analysis
in Section~4.3 
[first searched for the coefficient of Equation~(3) and then fit $\gamma$-ray data
with narrow energy bins] using the IC model of pulsar-based CR source distribution and
$z_{\rm h}=4~{\rm kpc}$, 
our baseline IC model but with areas of molecular clouds
traced by CO masked,
and our baseline IC model but with areas of IVCs masked (see Section~4.2).
Another source of systematic uncertainty of the $\HI$ emissivity spectrum is the
$\WHI$--to--$R$ ratio evaluated in Section~4.1. To examine this uncertainty, 
we divided the region with $T_{\rm d} \ge 20.5~{\rm K}$ into six sub-regions, 
collectively spanning the region
while requiring that the area of each sub-region has more than 10\% of the whole area.
The best-fit values of $\WHI/R$ of each sub-region were found to be within $+3.6\%/{-8.4\%}$ from
the average.
Although this uncertainty does not affect the slope of $T_{\rm d}$ dependence shown in Figure~4,
it changes the normalization of the emissivity spectrum independent of energy, and
we added this uncertainty to that due to the modeling of $\gamma$-ray data
(choice of the IC model and masking areas of clouds traced by CO or IVCs) as a linear sum.
The obtained systematic uncertainty is shown by the shaded band in Figure~10.
For comparison, we plotted the model curve for
the LIS that we adopted and $\epsilon_{\rm M}$ of 1.84 in the same figure.
To gauge the uncertainty in the emissivity model
[mainly due to the uncertainty of the elemental composition of CRs and the cross sections
other than proton-proton (p--p) collisions], 
we also plotted the model curve for
$\epsilon_{\rm M}=1.45$ 
\citep[the lowest value referred to in][]{Mori2009}
which gives 15\%-20\% lower emissivity.
We also plotted the emissivity spectrum of the local $\HI$ clouds (in different regions of the sky)
measured by \citet{FermiHI} and \citet{FermiHI2} for comparison,
in the analysis for which
different LAT event selections and response functions
were employed
(the so-called Pass~6 and Pass~7 by \citet{FermiHI} and \citet{FermiHI2}, respectively).
Most recent studies of high-latitude regions by \textit{Fermi}-LAT,
e.g., local $\HI$ emissivities in \citet{Tibaldo2015} and
$\HI$ emissivity of the Chamaeleon complex in \citet{Planck2015}
find $\HI$ emissivity spectra similar
to that of \citet{FermiHI2}.
One can recognize that our emissivity spectrum is apparently harder than the model curves and those from previous
\textit{Fermi}-LAT studies. However, within the systematic and statistical uncertainties 
the slope of our spectrum
is consistent with those of the models (and previous \textit{Fermi}-LAT results) 
above 1~GeV in which the index of the emissivity spectrum 
follows that of CR protons \citep[e.g.,][]{Aharonian2000}. We thus do not claim nor deny a harder emissivity spectrum
than that inferred from the CR spectra measured at the Earth.
A more significant difference can be seen in the normalization.
Our result agrees with 
the model with $\epsilon_{\rm M}=1.45$
except for the lowest energy bin,
in which the degeneracy among model components are most severe due to the 
breadth of the point-spread function
and for which the LIS is uncertain at the ${\sim}20\%$ level due to the solar modulation
\citep[e.g., see discussions by][]{FermiHI}.
On the other hand,
relevant studies by \textit{Fermi}-LAT favor the model with $\epsilon_{\rm M}=1.85$.
The difference is larger than the statistical and systematic uncertainties for energies below a few GeV.
This difference
cannot be fully explained by the 
uncertainty in the LAT effective area ({$\sim$}5\%)
\footnote{\url{http://fermi.gsfc.nasa.gov/ssc/data/analysis/LAT_caveats.html}}.
It can be understood, at least partially,
due to the assumption of the $\HI$ optical thickness.
\citet{FermiHI} and \citet{FermiHI2} assumed a uniform $T_{\rm s}$ of 125~K and 140~K, respectively,
and \citet{Tibaldo2015} and \citet{Planck2015} assumed the optically thin case.
If uniform $T_{\rm s}$ of such values (greater than or equal to 125~K) is
applied to our region, we will have smaller $N({\rm H_{tot}})$ on average than that we obtained in Figure~9,
and larger $\HI$ emissivity.

More specifically, the difference comes from different assumptions on gas and dust properties.
The method presented in this study is based on several assumptions:
(A1) Optically-thin $\HI$ dominates the ISM gas in areas with high $T_{\rm d}$.
\footnote{
We recall that $T_{\rm d}$ was obtained under the assumption that the dust temperature is uniform.
}
(A2) $N({\rm H_{tot}})/R$ is constant for the same value of $T_{\rm d}$.
We also assume that, through the $\gamma$-ray data analysis (Figure~4),
(A3) the $T_{\rm d}$-dependence of $N({\rm H_{tot}})/R$ can be compensated for 
by employing the
empirical function of Equation~(3). 
On the other hand, (leaving aside small differences in the analysis procedure and detailed assumptions)
a conventional template-fitting method \citep[e.g.,][]{Fermi2ndQ,Fermi3rdQ,FermiCham,FermiHI2,Tibaldo2015,Planck2015} 
is based on the following
assumptions on gas and dust properties:
(B1) The ISM gas can be divided into the atomic gas, the molecular gas associated with CO emission, and the dark-gas.
(B2) In each phase (the atomic, molecular and dark-gas), the gas and dust properties are uniform
across the ROI (or sub-regions).
More specifically, the statement~(B2) can be broken down as follows.
(B2a) Atomic hydrogen has uniform $T_{\rm s}$ and uniform $N(\HI)/D$, where $D$ is the employed dust map
(e.g., $R$ and $\tau_{353}$).
(B2b) Molecular hydrogen has uniform $X_{\rm CO}$ and uniform $N(\Htwo)/D$.
(B2c) Dark gas has uniform $N({\rm H_{DG}})/D$ where $N({\rm H_{DG}})$ is the column density of the dark gas.
In short, the method presented here relies on the uniformity of $N({\rm H_{tot}})/R$ for the same value of $T_{\rm d}$
regardless of the gas phases, while
a conventional template-fitting method relies on the uniformity of $T_{\rm s}$, $X_{\rm CO}$, and $N({\rm H})/D$
in each gas phase across the ROI (or sub-regions).
Therefore using a good dust tracer is crucial for the conventional method, 
and several alternative tracers of dust need to be compared to properly trace the ISM and CRs \citep[e.g.,][]{Planck2015}.
The method developed here, on the other hand, has more flexibility to adjust $N({\rm H_{tot}})/D$.
The obtained emissivity, however, depends on the calibration of the gas--to--dust ratio and the
$\gamma$-ray fit in high $T_{\rm d}$ regions; therefore careful examination
of the systematic uncertainty as we have done
is required.

To discuss the difference from previous relevant studies by \textit{Fermi}-LAT more quantitatively,
we also employed a conventional template-fitting.
As shown in Figure~2, $R$ shows better correlation with $\WHI$ than $\tau_{353}$ does,
and therefore we used the $R$ map as a dust tracer and performed a fit to $\gamma$-ray data.
Details of the template preparation and obtained results are given in Appendix~D,
in which the analysis using $N({\rm H_{tot,mod}})$ with $T_{\rm bk}=20.5~{\rm K}$ and $C=2$ [Equation~(3)]
is also shown for comparison.
As shown by Figure~D5, the two methods show similar data/model ratio, and their $\gamma$-ray model maps 
agree to within
${\le}5\%$ with each other in most of the ROI. 
\footnote{
Small differences are seen, e.g., in the upper-left corner
[$(l,b) \sim (106\arcdeg, -33\arcdeg)$] where optically-thin $\HIs$ dominates the ISM gas,
the conventional method slightly overpredicts the data.
}
Nevertheless, the scaling factor for the atomic gas phase obtained by the conventional 
analysis is $0.924\pm0.022$, which agrees (within ${\le}10\%$) with the model of $\epsilon=1.84$ like the other \textit{Fermi}-LAT studies
employing the similar method but is ${\sim}35\%$ larger than what we obtained
using the $N({\rm H_{tot,mod}})$ map.
A significant fraction of the difference is attributable to the inferred gas column density.
As shown by Figure~D6, while the two analyses show similar $N({\rm H_{tot}})$ in $T_{\rm d} \ge 19~{\rm K}$,
they give distinct $N({\rm H_{tot}})$ below 19~K 
where the conventional method has less flexibility to adjust $N({\rm H_{tot}})$.
The difference is ${\sim}12\%$ in $T_{\rm d}=18.5\mbox{--}19~{\rm K}$ and
becomes gradually larger as $T_{\rm d}$ decreases;
in $T_{\rm d}=17\mbox{--}17.5~{\rm K}$, the difference is
more than 25\%. 
We can also see that the conventional method requires significantly smaller IC emission 
as shown by Tables~D4 and D5 and Figure~D4, attributing more $\gamma$ rays to the gas-related component 
than our analysis does.
As a result of these two effects, most of the difference of scale factors ($\HI$ emissivity) can be explained.
Given the similarity of data/model ratios and $\gamma$-ray model maps, 
we do not rule out the conventional template-fitting analysis.
However, the underlying assumption of the method,
the uniformity of $N({\rm H})/D$ in each gas phase across the ROI, 
apparently does not agree with
the result of the $\gamma$-ray analysis shown in Figure~4.
[If the values of $X_{\rm CO}$ agree between the dust-fit and $\gamma$-ray analysis
and $N({\rm H_{DG}})/R$ is similar to $N({\HI_{\rm thin}})/R$,
as inferred in Appendix~D, the conventional method predicts a uniform $N({\rm H_{tot}})/R$ independent of $T_{\rm d}$.]
We therefore consider that the analysis we have developed is more accurate in this region of the sky,
implying lower fraction of heavy CR nuclei in the solar neighborhood and/or 
smaller cross sections other than for p--p collisions than previously inferred from gas emissivities in $\gamma$ rays.
Finally we note that the conventional template-fitting method is well established
and has long been applied for
many previous \textit{Fermi}-LAT studies, and has improved our understanding of the ISM and CRs.

Although our analysis was based on 
a few basic assumptions on gas and dust properties
(optically-thin $\HI$ in regions with high $T_{\rm d}$ and a uniform $N(\rm H_{\rm tot})/R$ in each $T_{\rm d}$ range)
and the plausible assumption that the $\gamma$-ray data can be used as a robust tracer
of $N({\rm H_{tot}})$, and
we have carefully examined the systematic uncertainty,
the method is still in an early phase of development
and should be tested and improved by applying to other regions of the sky. 
In such future analyses, a comparison with the conventional method should be done at the same time
to examine the merits and demerits of each method and the validity of the underlying assumptions,
and to better understand the properties of the ISM and CRs in the solar neighborhood.

\begin{figure}[ht!]
\figurenum{9}
\gridline{
\fig{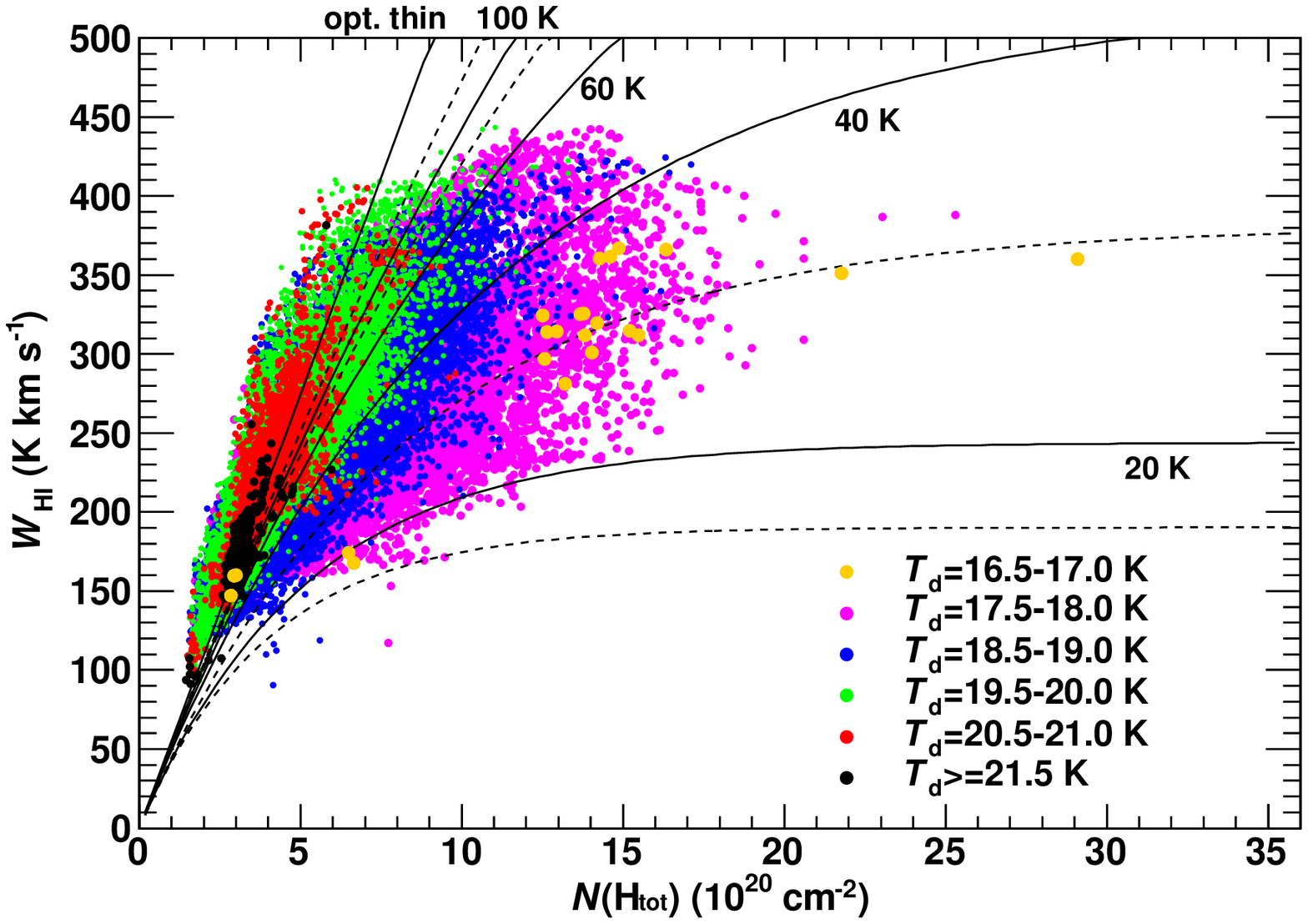}
{0.5\textwidth}{(a)}
\fig{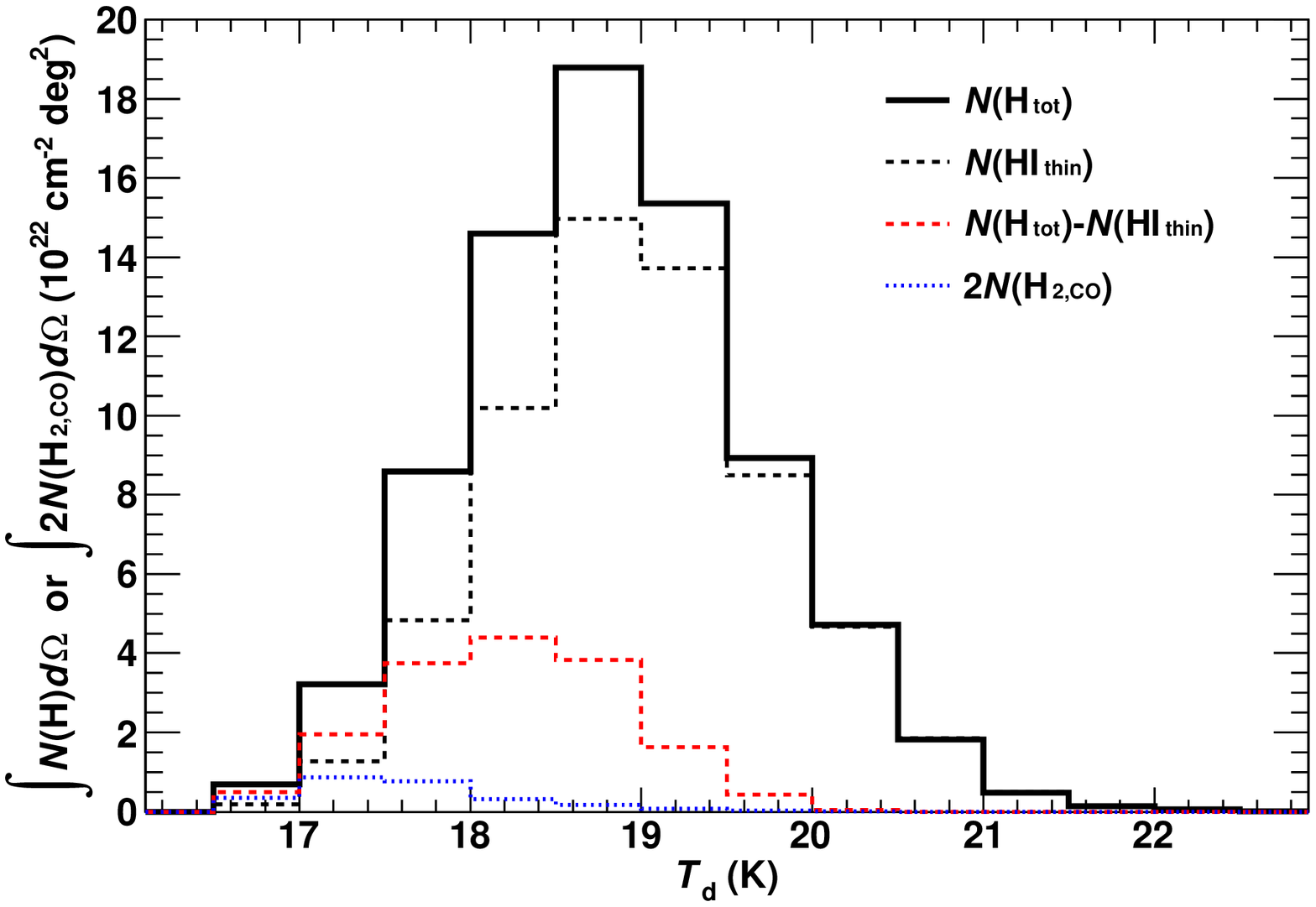}
{0.5\textwidth}{(b)}
}
\gridline{
\fig{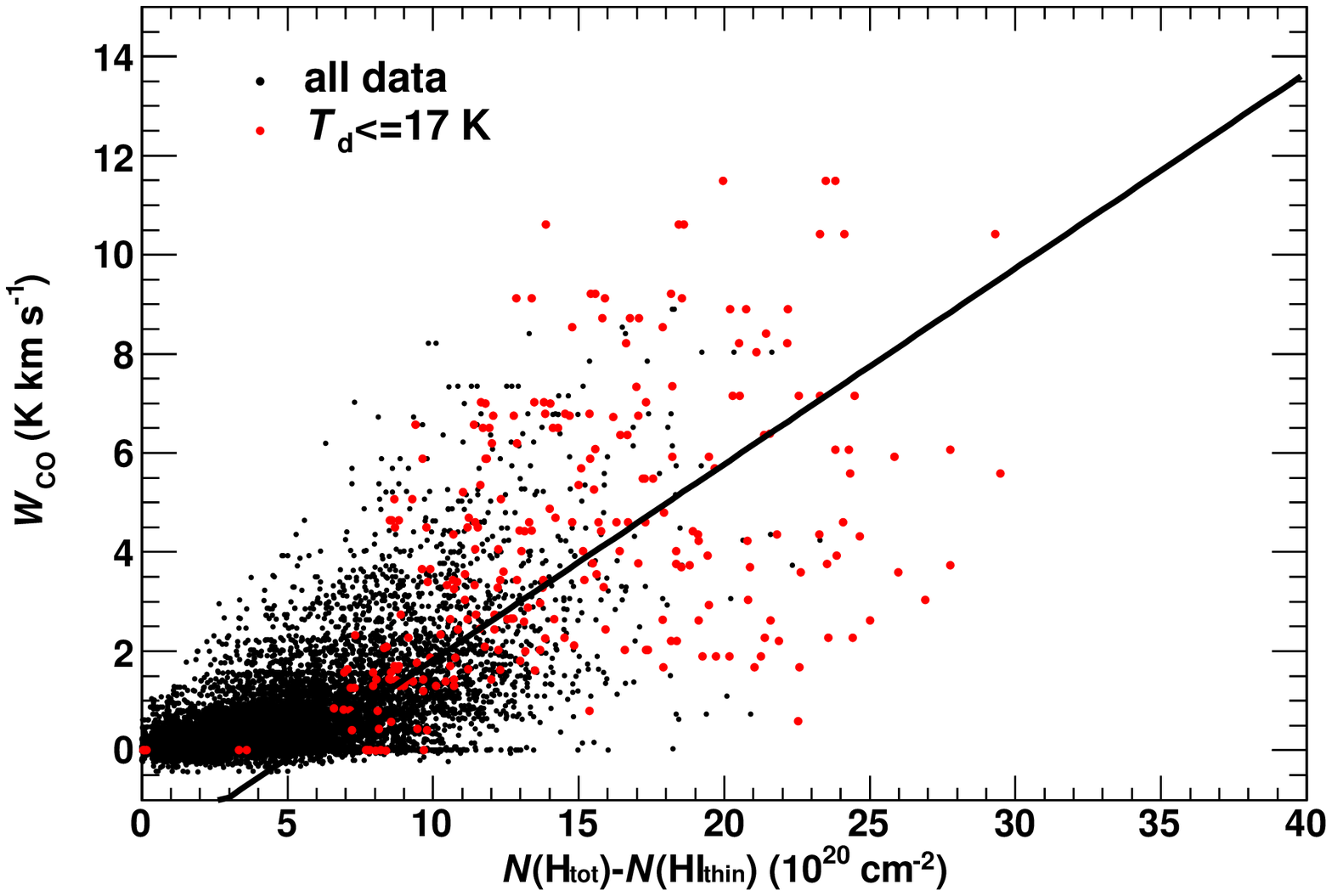}
{0.5\textwidth}{(c)}
}
\caption{
(a) The correlation between $\WHI$ and the $N({\rm H_{tot}})$ 
inferred by the $\gamma$-ray data analysis
obtained from Equation~(3) with $T_{\rm bk}={\rm 20.5~K}$ and $C=2$.
Making an approximation by assuming that all of the gas is atomic,
the model curves for several choices of $T_{\rm s}$ are overlaid.
Dotted lines illustrate the uncertainty of model curves for $T_{\rm s}=20$ and 100~K. 
See the text for details.
(b) The distribution of the integrated $\HI$ column density for the optically thin case,
$\int N(\HI_{\rm thin}) d\Omega = 1.82 \times 10^{18} \cdot \int \WHI d\Omega$,
and that of the integrated excess gas column density above $N(\HI_{\rm thin})$
calculated in 0.5-K step $T_{\rm d}$ bins.
The integrated $W_{\rm CO}$ as a measure of the CO-bright $\Htwo$ is also plotted.
(c) Scatter plot of $N({\rm H_{tot}})-N(\HI_{\rm thin})$ vs. $W_{\rm CO}$ in the studied region.
\label{fig:f9}
}
\end{figure}

\begin{figure}[ht!]
\figurenum{10}
\gridline{
\fig{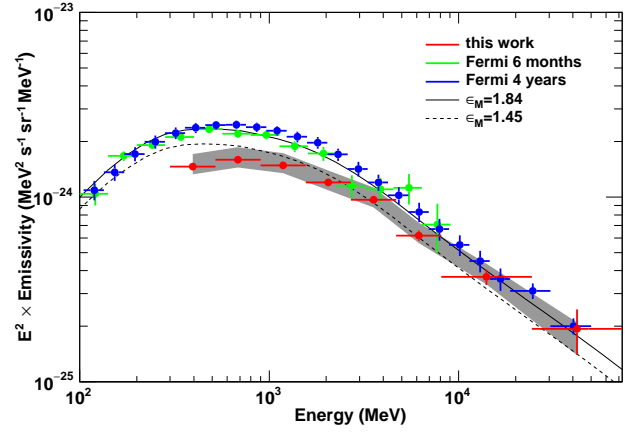}
{0.5\textwidth}{}
}
\caption{
Summary of the $\HI$ emissivity spectrum obtained in this study compared with 
the model curves based on the LIS for
$\epsilon_{\rm M}=$ 1.45 and 1.84, and the results of the relevant studies by \textit{Fermi}-LAT,
based on 6 months of observation and 4 years of observation by \citet{FermiHI} and \citet{FermiHI2}, respectively.
The shaded band shows the systematic uncertainty of our $\HI$ emissivity spectrum
(see the text in Section~5 for details).
\label{fig:f10}
}
\end{figure}

\clearpage 
\section{Summary and Future Prospects}

We carried out a study of the ISM and CRs
using the \textit{Fermi}-LAT data in the 0.3--72.9~GeV range and other 
interstellar gas tracers,
in particular the \textit{Planck} dust model,
at Galactic longitudes from $60\arcdeg$ to $120\arcdeg$ and Galactic latitudes from 
$-60\arcdeg$ to $-28\arcdeg$.
The region encompasses the nearby, 
high-Galactic latitude 
cloud complexes
MBM~53, 54, and 55 and a far-infrared loop-like structure in Pegasus.
We found that neither $R$ nor $\tau_{353}$
estimated from \textit{Planck} observations
were good representations of the total gas column density.
Instead, we found a systematic decrease of $N({\rm H_{tot}})/R$ or increase of
$N({\rm H_{tot}})/\tau_{353}$ as dust temperature increases.
We used the \textit{Fermi}-LAT $\gamma$-ray data to quantify the total gas column density,
obtaining the mass of the excess gas not traced by the $\HI$ 21-cm nor the CO 2.6-mm line surveys (dark gas) to be
{$\sim$}25\% of that from the $\HI$ gas in the optically thin case.
The ratio of the mass of the excess gas to that of the $\Htwo$ gas traced by CO is {$\le$}5.
This ratio is about a factor of 10 larger than model predictions in the CO-dark $\Htwo$ scenario,
requiring a better modeling of the CO-dark $\Htwo$ which is applicable to
the translucent clouds studied here.
Another possibility is 
that the $\HI$ has appreciable optical depth
which implies a
spin temperature of less than 100~K
in regions with dust temperatures below 19~K.
The correlation of gas templates based on $\gamma$-ray data and dust temperature is crucial,
since the dark gas contribution calculated from $R$ and $\tau_{353}$ without the correction
is a factor of {$\sim$}4 lower/higher than what we obtained.
We measured the $\gamma$-ray emissivity spectrum and found it 
agrees with the model for the LIS with $\epsilon_{\rm M}=1.45$,
while most of relevant \textit{Fermi}-LAT studies based on different analysis methods and assumptions
on ISM properties
show 15\%--20\% higher emissivity normalizations in energies below a few GeV.
The difference can be understood as due to the different gas column density
inferred in areas with low $T_{\rm d}$.
Although we do not rule out the conventional template-fitting analysis, 
its underlying assumption does not coincide with the $T_{\rm d}$ dependence
of $N({\rm H_{tot}})/R$ we found; therefore we regard our analysis is more accurate in this region of the sky,
implying lower fraction of heavy nuclei in local CRs and/or smaller cross sections
other than for p--p interactions than previously inferred from gas emissivities in $\gamma$ rays.
The method is still in early development phase, however, and needs to be tested and improved
by applying to other high-latitude regions.
In such studies, a comparison with the conventional template-fitting should also be carried out
to examine the limitations of each method and to better understand the properties of the ISM and CRs
in the solar neighborhood.
\\

The \textit{Fermi} LAT Collaboration acknowledges generous ongoing support
from a number of agencies and institutes that have supported both the
development and the operation of the LAT as well as scientific data analysis.
These include the National Aeronautics and Space Administration and the
Department of Energy in the United States, the Commissariat \`a l'Energie Atomique
and the Centre National de la Recherche Scientifique / Institut National de Physique
Nucl\'eaire et de Physique des Particules in France, the Agenzia Spaziale Italiana
and the Istituto Nazionale di Fisica Nucleare in Italy, the Ministry of Education,
Culture, Sports, Science and Technology (MEXT), High Energy Accelerator Research
Organization (KEK) and Japan Aerospace Exploration Agency (JAXA) in Japan, and
the K.~A.~Wallenberg Foundation, the Swedish Research Council and the
Swedish National Space Board in Sweden.
Additional support for science analysis during the operations phase is gratefully acknowledged 
from the Istituto Nazionale di Astrofisica in Italy and the Centre National d'\'Etudes Spatiales in France.

We thank T. M. Dame for providing the moment-masked CO data. 
Some of the results in this paper have been derived using the HEALPix
\citep{Gorski2005} package.

\vspace{5mm}
\facilities{\textit{Fermi} (LAT), \textit{Planck}}

\software{\textit{Fermi} Science Tools, GALPROP, HEALPix,  ROOT}

\clearpage

\appendix
\section{Treatment of the Infrared Sources}

In the \textit{Planck} dust model maps we identified several regions with high $T_{\rm d}$,
indicating localized heating by stars.
We refilled these areas in the $R$, $\tau_{353}$, and $T_{\rm d}$ maps
with the average of the peripheral pixels 
(since we used HEALPix maps of order 9, the pixel size is ${\sim}0.013~{\rm deg^{2}}$):
values in a circular region with radius of $r_{1}$ are filled with the average
of pixels in a ring with inner radius of $r_{1}$ and outer radius of $r_{2}$.
For each region, the central position
($l, b$), $r_{1}$, and $r_{2}$ are summarized in Table~A1.
We used a large radius for the region of high $T_{\rm d}$ located near 3C~454.3,
even though the origin of such high temperature is unknown.
Because 3C~454.3 is a very bright $\gamma$-ray source (see Figure~8),
the impact on the $\gamma$-ray data analysis was small.

\floattable
\begin{deluxetable}{ccccc}[ht!]
\tablecaption{
Infrared sources excised and interpolated across
in the \textit{Planck} dust maps}
\tablecolumns{5}
\tablewidth{0pt}
\tablehead{
\multicolumn{2}{c}{Position} & \colhead{$r_{1}$} & \colhead{$r_{2}$} &\colhead{Object name} \\
\cline{1-2}
\colhead{$l$ (deg)} & \colhead{$b$ (deg)} & \colhead{(deg)} & \colhead{(deg)} &\colhead{}
}
\startdata
79.61 & $-30.25$ & 0.12 & 0.15 & \\
82.85 & $-50.65$ & 0.12 & 0.15 & \\
83.10 & $-45.46$ & 0.12 & 0.15 & \\
86.30 & $-38.20$ & 0.60 & 0.65 & 3C~454.3 (Active galactic nucleus)\\
87.46 & $-29.73$ & 0.12 & 0.15 & NGC~7339 (Radio galaxy)\\
87.57 & $-39.12$ & 0.12 & 0.15 & \\
93.53 & $-40.35$ & 0.12 & 0.15 & RAFGL~3068 (Variable star)\\
93.91 & $-40.47$ & 0.12 & 0.15 & NGC~7625 (Interacting galaxies)\\
97.29 & $-32.52$ & 0.12 & 0.15 & IC~5298 (Seyfert2 galaxy)\\
98.88 & $-36.55$ & 0.12 & 0.15 & NGC~7678 (Active galactic nucleus)\\
104.26 & $-40.58$ & 0.12 & 0.15 & \\
104.46 & $-40.14$ & 0.12 & 0.15 & \\
111.37 & $-36.00$ & 0.12 & 0.15 & \\
\enddata
\end{deluxetable}

\clearpage 

\section{Intermediate-Velocity Clouds}

In our ROI, we have identified coherent structures of atomic hydrogen with velocities from $-80$ to $-30~{\rm km~s^{-1}}$
which correspond to some of the IVCs in the southern sky \citep{Wakker2001},
while the main $\HI$ clouds have velocities from $-30$ to $+20~{\rm km~s^{-1}}$, as shown by Figure~B1.
In panel~(b) of the figure, we can identify a strip-like structure which has an intensity peak at $(l,b) \sim (87\arcdeg, -38\arcdeg)$
and runs toward the lower-left corner of the image. We can also identify intense emissions in the upper-left corner of the ROI,
although some fraction of them are likely to be the contamination from the Galactic plane, not the IVCs.
In Section~4.2, we masked those structures [using green lines in panel~(b)] to examine the effect on the $\gamma$-ray data analysis.

The relative contribution of the clouds to the $\gamma$-ray flux (assuming uniform CR density) 
and the mass of the ISM gas (assuming the same distance) can be evaluated by
integrating $\WHI$ in the ROI. 
The relative contribution of the main clouds (defined as the $\HI$ clouds having velocities from $-30$ to $+20~{\rm km~s^{-1}}$)
to the whole emission of $\WHI$ (integrated from $-450$ to $+400~{\rm km~s^{-1}}$)
was found to be ${\sim}93.3\%$.
The contribution of IVCs were evaluated by integrating $\WHI$ having
velocities from $-80$ to $-30~{\rm km~s^{-1}}$ over the entire ROI, and were found to be ${\sim}5.5\%$.

\begin{figure}[ht!]
\gridline{
\fig{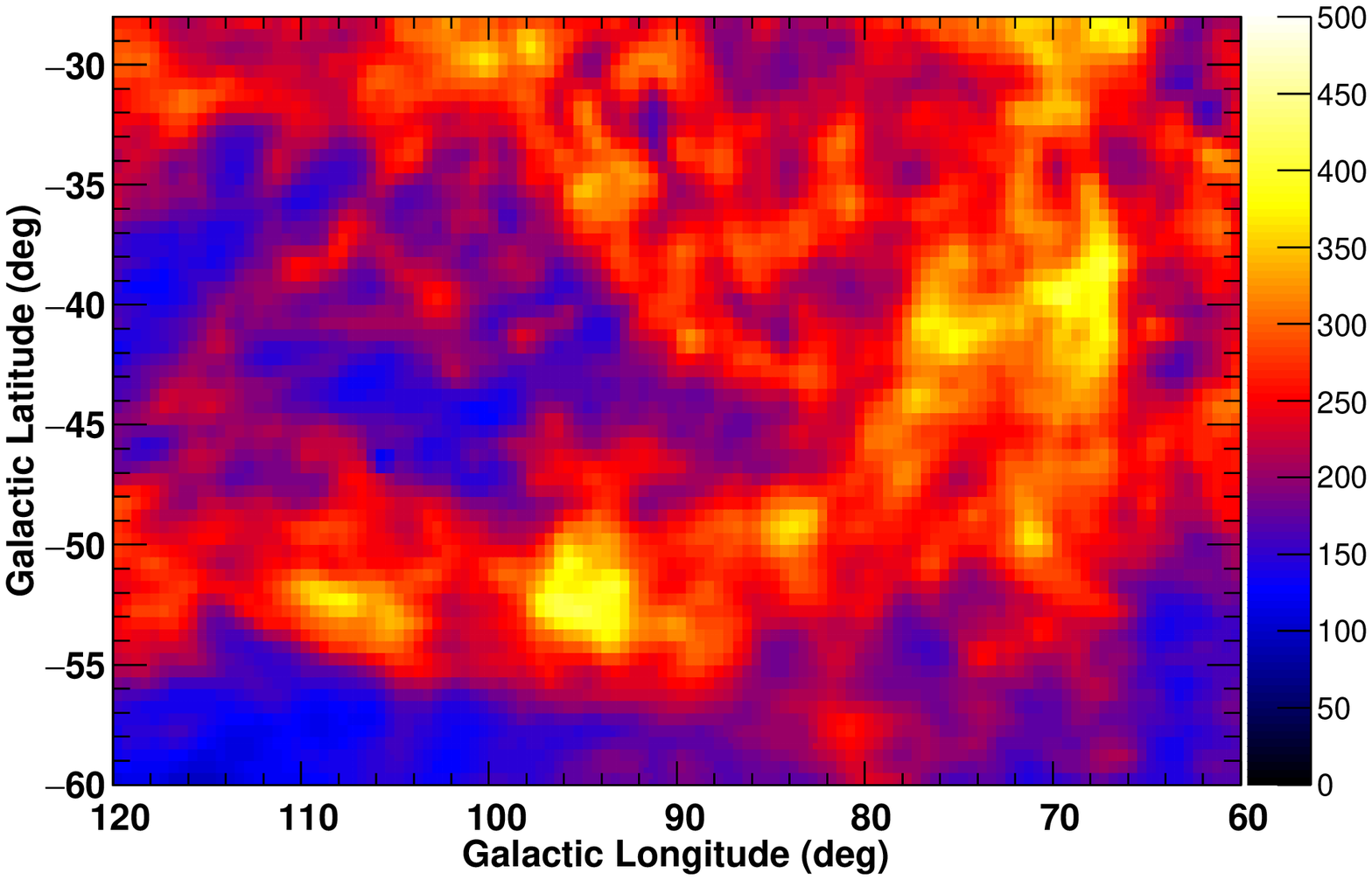}
{0.5\textwidth}{(a)}
\fig{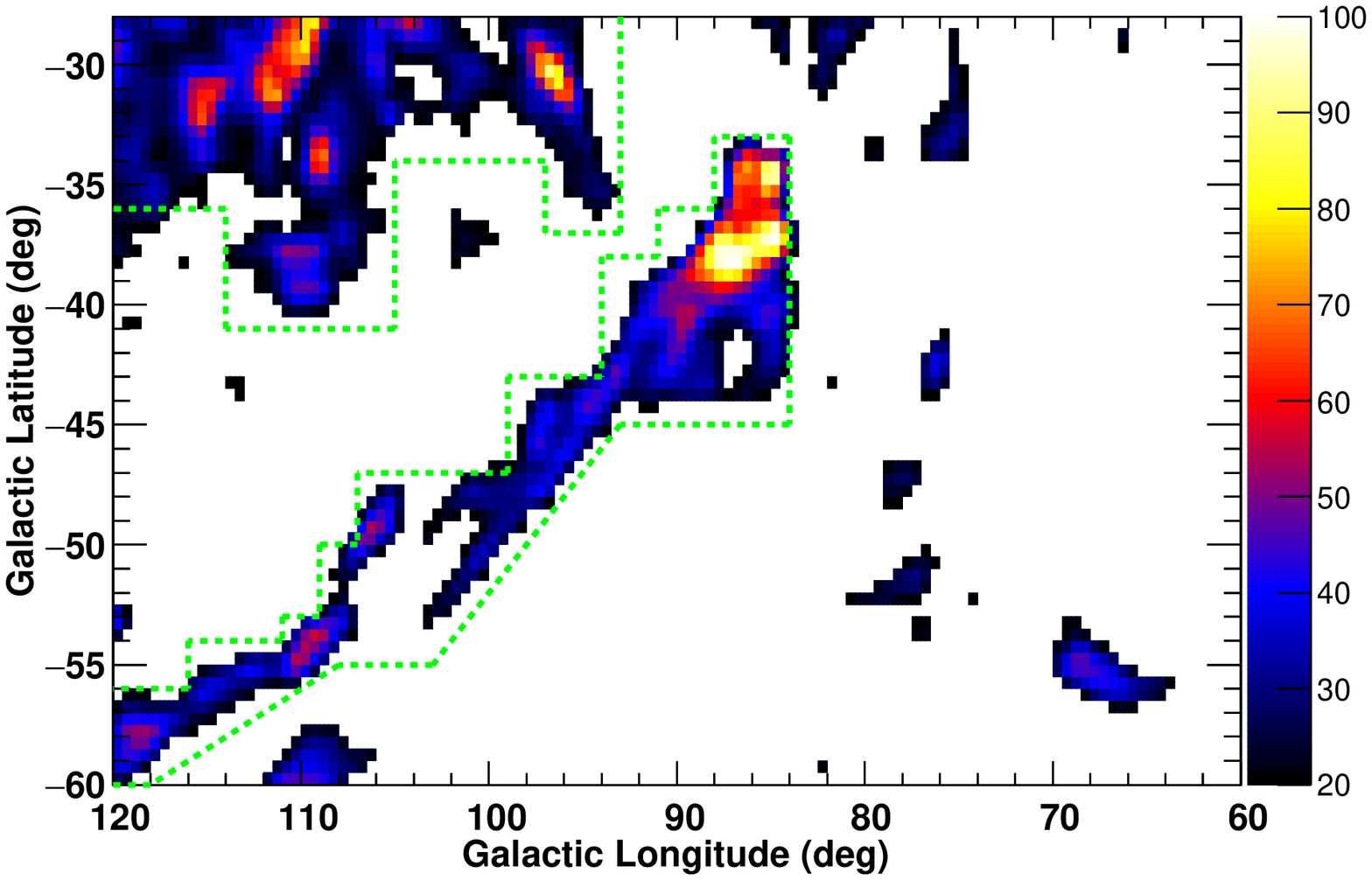}
{0.5\textwidth}{(b)}
}
\caption{
The $\WHI$ map in the velocities (a) from $-30$ to $+20~{\rm km~s^{-1}}$ and (b) from $-80$ to $-30~{\rm km~s^{-1}}$.
The former shows the structure of the main clouds, while the latter shows the distribution of IVCs.
The dotted lines in panel~(b) indicate the areas to be masked in Section~4.2.
}
\end{figure}

\clearpage

\section{Results with the Maps Sorted by Dust Temperature}

We split the $N({\rm H_{tot}})$ template map (constructed from $R$ or $\tau_{353}$)
into four based on $T_{\rm d}$, with $T_{\rm d} \le 18~{\rm K}$, $T_{\rm d}={\rm 18\mbox{--}19~{K}}$,
$T_{\rm d}={\rm 19\mbox{--}20~{K}}$ and $T_{\rm d} \ge 20~{\rm K}$,
and fit the $\gamma$-ray data with Equation~(1)
using the four template maps, with scaling factors ($c_{1}$ for each of the four templates)
free to individually vary instead of using a single $N({\rm H_{tot}})$ map. 
The obtained fit parameters and the spectrum of each component are summarized in Tables~C2 and C3 and Figure~C2.

\floattable
\begin{deluxetable}{ccccccccc}[ht!]
\tablecaption{Results with the $R$-based $N(\rm H_{tot})$ maps sorted by $T_{d}$ \label{tab:rad_Td}}
\tablecolumns{9}
\tablewidth{0pt}
\tablehead{
\colhead{Energy} & \colhead{$c_{1,1}$} & \colhead{$c_{1,2}$} & 
\colhead{$c_{1,3}$} & \colhead{$c_{1,4}$} & \colhead{$c_{\rm 2n}$} & 
\colhead{$c_{\rm 2i}$} & \colhead{$I_{\rm iso}$} & \colhead{$I_{\rm iso}$} \\
\colhead{(GeV)} & \colhead{($T_{\rm d} \le 18~{\rm K}$)} & \colhead{($18\mbox{--}19~{\rm K}$)} & 
\colhead{($19\mbox{--}20~{\rm K}$)} & \colhead{($T_{\rm d} \ge 20~{\rm K}$)} & & 
& \colhead{(norm\tablenotemark{a})} & \colhead{(index)} 
}
\startdata
0.3--0.9 & $0.84\pm0.02$ & $0.80\pm0.03$ & $0.75\pm0.04$ & $0.70\pm0.06$ & $0.70\pm0.18$ & $0.00\pm0.16$ & $3.94\pm0.06$ & $2.22\pm0.02$ \\
0.9--2.7 & $0.91\pm0.02$ & $0.81\pm0.03$ & $0.78\pm0.04$ & $0.63\pm0.05$ & $0.84\pm0.17$ & $0.39\pm0.25$ & $0.89\pm0.03$ & $2.36\pm0.05$ \\
2.7--8.1 & $1.00\pm0.05$ & $0.89\pm0.06$ & $0.76\pm0.07$ & $0.73\pm0.10$ & $1.17\pm0.27$ & $-1.0$\tablenotemark{b} & $0.17\pm0.01$ & $2.64\pm0.11$ \\
8.1--72.9 & $0.84\pm0.14$ & $0.54\pm0.18$ & $0.40\pm0.23$ & $0.16\pm0.32$ & $1.54\pm0.56$ & $-0.44\pm0.21$ & $0.05\pm0.01$ & $2.69\pm0.10$ \\
\enddata
\tablenotetext{a}{The integrated intensity ($10^{-6}~{\rm ph~s^{-1}~cm^{-2}~sr^{-1}}$) in each band.}
\tablenotetext{b}{Not well determined and reached at the smallest parameter boundary we set.}
\tablecomments{
The errors are 1-sigma statistical uncertainties.
Each of the four scale factors ($c_{1,1}$, $c_{1,2}$, $c_{1,3}$, and $c_{1,4}$) gives the normalization 
for a specified range of $T_{\rm d}$
of the gas-related component in each energy bin.
The scale factor for IC ($c_{2}$) is modeled by a power law in each energy bin (from $E_{\rm min}$ to $E_{\rm max}$) 
as $c_{2}(E)=c_{\rm 2n} \cdot (E/E_{0})^{c_{\rm 2i}}$ where $E_{0} = \sqrt{E_{\rm min} \cdot E_{\rm max}}$.
$I_{\rm iso}$ is modeled with a power law with the integrated intensity and the photon index as free parameters.
}
\end{deluxetable}

\floattable
\begin{deluxetable}{ccccccccc}[ht!]
\tablecaption{Results with the $\tau_{353}$-based $N({\rm H_{tot}})$ maps sorted by $T_{d}$ \label{tab:rad_Td}}
\tablecolumns{9}
\tablewidth{0pt}
\tablehead{
\colhead{Energy} & \colhead{$c_{1,1}$} & \colhead{$c_{1,2}$} & 
\colhead{$c_{1,3}$} & \colhead{$c_{1,4}$} & \colhead{$c_{\rm 2n}$} & 
\colhead{$c_{\rm 2i}$} & \colhead{$I_{\rm iso}$} & \colhead{$I_{\rm iso}$} \\
\colhead{(GeV)} & \colhead{($T_{\rm d} \le 18~{\rm K}$)} & \colhead{($18\mbox{--}19~{\rm K}$)} & 
\colhead{($19\mbox{--}20~{\rm K}$)} & \colhead{($T_{\rm d} \ge 20~{\rm K}$)} & & 
 & \colhead{(norm\tablenotemark{a})} & \colhead{(index)} 
}
\startdata
0.3--0.9 & $0.39\pm0.01$ & $0.47\pm0.01$ & $0.53\pm0.02$ & $0.63\pm0.03$ & $0.28\pm0.14$ & $0.07\pm0.42$ & $4.06\pm0.07$ & $2.22\pm0.02$ \\
0.9--2.7 & $0.41\pm0.01$ & $0.45\pm0.02$ & $0.52\pm0.02$ & $0.54\pm0.04$ & $0.59\pm0.18$ & $0.44\pm0.32$ & $0.91\pm0.03$ & $2.37\pm0.04$ \\
2.7--8.1 & $0.45\pm0.02$ & $0.50\pm0.03$ & $0.50\pm0.05$ & $0.60\pm0.08$ & $1.04\pm0.28$ & $-1.0$\tablenotemark{b} & $0.17\pm0.01$ & $2.58\pm0.11$ \\
8.1--72.9 & $0.39\pm0.06$ & $0.32\pm0.10$ & $0.30\pm0.14$ & $0.19\pm0.25$ & $1.41\pm0.58$ & $-0.47\pm0.23$ & $0.05\pm0.01$ & $2.67\pm0.10$  \\
\enddata
\tablenotetext{a}{The integrated intensity ($10^{-6}~{\rm ph~s^{-1}~cm^{-2}~sr^{-1}}$) in each band.}
\tablenotetext{b}{Not well determined and reached at the smallest parameter boundary we set.}
\tablecomments{
The errors are 1-sigma statistical uncertainties.
Each of the four scale factors ($c_{1,1}$, $c_{1,2}$, $c_{1,3}$, and $c_{1,4}$) gives the normalization 
for a specified range of $T_{\rm d}$
of the gas-related component in each energy bin.
The scale factor for IC ($c_{2}$) is modeled by a power law in each energy bin (from $E_{\rm min}$ to $E_{\rm max}$) 
as $c_{2}(E)=c_{\rm 2n} \cdot (E/E_{0})^{c_{\rm 2i}}$ where $E_{0} = \sqrt{E_{\rm min} \cdot E_{\rm max}}$.
$I_{\rm iso}$ is modeled with a power law with the integrated intensity and the photon index as free parameters.
}
\end{deluxetable}

\begin{figure}[ht!]
\gridline{
\fig{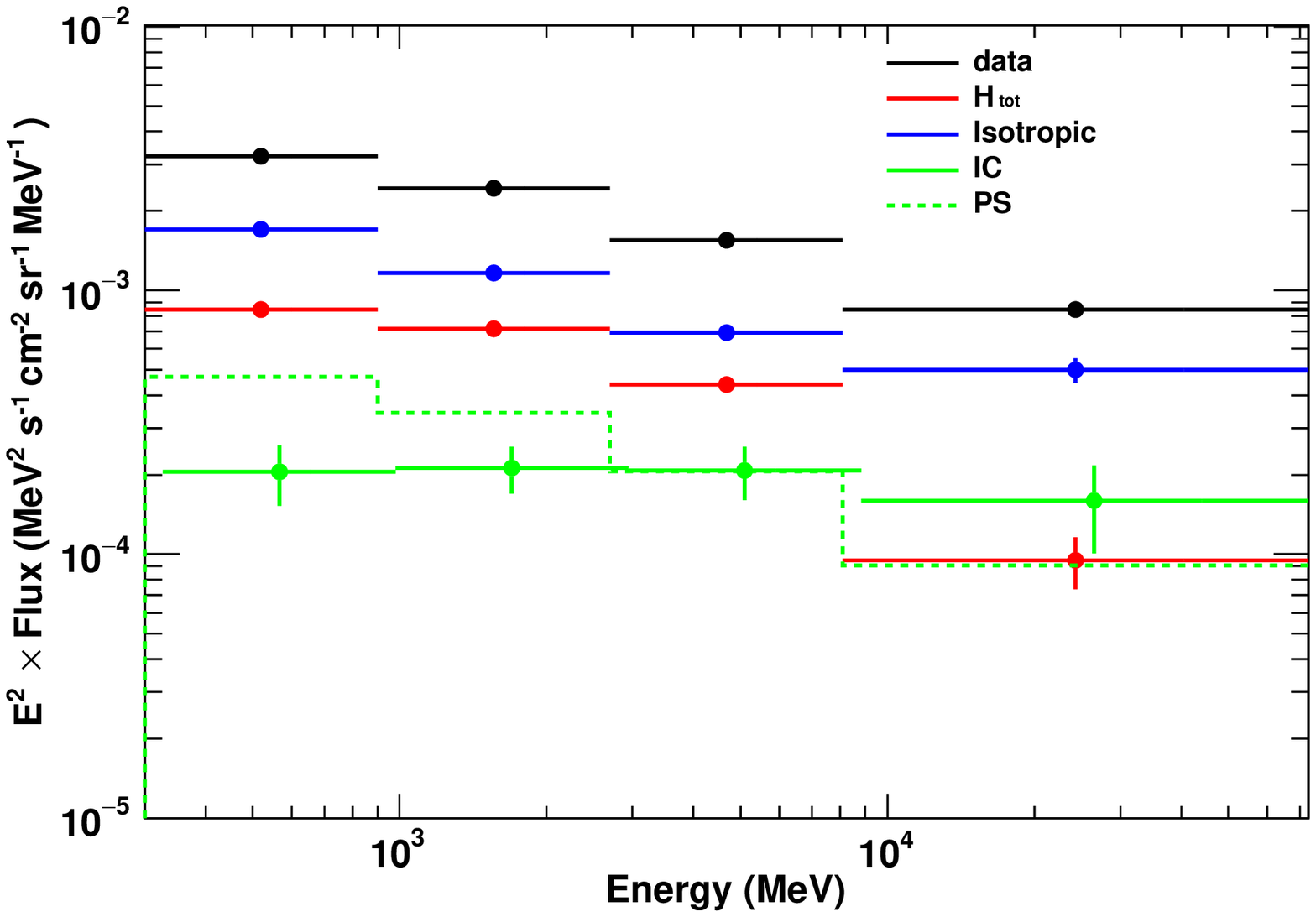}
{0.5\textwidth}{(a)}
\fig{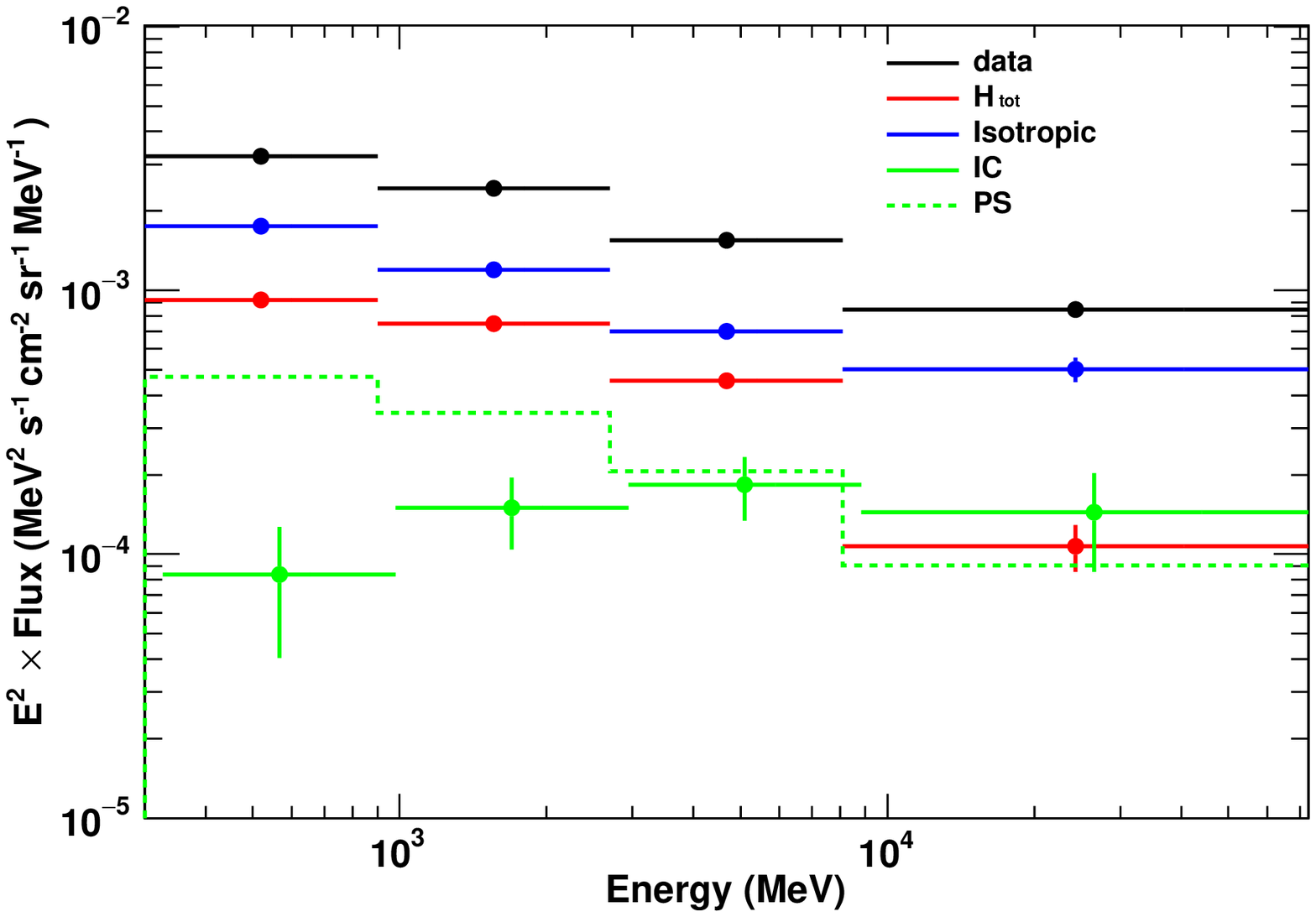}
{0.5\textwidth}{(b)}
}
\caption{
Spectrum of each component obtained by the four $N({\rm H_{tot}})$ maps sorted by $T_{\rm d}$,
for (a) $R$-based analysis and (b) $\tau_{353}$-based analysis.
}
\end{figure}

\clearpage

\section{Results by a Conventional Template-Fitting Method}

In order to prepare a template map of the dark gas, we fit the original $R$ map
with a linear combination of $N(\HI)$ map (Figure~1a) and the $W_{\rm CO}$ map (Figure~1b):
for simplicity we assumed the optically thin case to construct $N({\HI})$ map.
Then, the fit to the $R$ map is expressed as
\begin{equation}
R(l, b) = \yHI \cdot N(\HI_{\rm thin})(l, b) + y_{\rm CO} \cdot W_{\rm CO}(l,b)~~,
\end{equation}
where $\yHI$ and $y_{\rm CO}$ are coefficients for $N(\HI)$ and $W_{\rm CO}$ maps, respectively.
Through least-squares fitting, we obtained
$\yHI = (2.787\pm0.008) \times 10^{-28}~{\rm W~m^{-2}~sr^{-1}~cm^{2}}$ and
$y_{\rm CO} = (5.41\pm0.13) \times 10^{-8}~{\rm W~m^{-2}~sr^{-1}~(K~km~s^{-1})^{-1}}$.
We can convert $\yHI$ into the ratio of $N(\HI_{\rm thin})$ to $R$ as
$(0.3588\pm0.0010) \times 10^{28}~{\rm cm^{-2}~(W~m^{-2}~sr^{-1})^{-1}}$, and
calculate the dust-fit based $X_{\rm CO}$ from $\yHI$ and $y_{\rm CO}$ as
$X_{\rm CO,dust} = y_{\rm CO}/2\yHI = (0.974\pm0.023) \times 10^{21}~{\rm cm^{-2}~(K~km~s^{-1})^{-1}}$.
We used the positive residuals as a template map for dark gas ($R_{\rm res}^{\rm DG}$) as shown in Figure~D3.
Then, instead of Equation~(1), the $\gamma$-ray intensities can be modeled as

\begin{eqnarray}
I_{\gamma}(l, b, E) & = &
\left [
\cHI(E) \cdot N(\HI_{\rm thin})(l, b)
+ c_{\rm CO}(E) \cdot W_{\rm CO}(l, b)
+ c_{\rm DG}(E) \cdot R_{\rm res}^{\rm DG}(l, b)
 \right]
q_{\gamma}(E)  \nonumber \\
& & + c_{2}(E) \cdot I_{\rm IC}(l, b, E) +
I_{\rm iso}(E) + \sum_{j} {\rm PS}_{j}(l, b, E)~~,
\end{eqnarray}
where $\cHI(E)$, $c_{\rm CO}(E)$ and $c_{\rm DG}(E)$ are scale factors 
for the atomic gas, the molecular gas traced by CO, and the dark gas, respectively.
Since we employed three template gas maps instead of a single $N({\rm H_{tot}})$ map
(and therefore we have more free parameters), we used wider energy ranges as we did in Section~4.2.
We note that this analysis was simple and did not adopt detailed procedures
such as a denoising of $R_{\rm res}^{\rm DG}$ and iterative fittings in dust and $\gamma$-rays employed by, e.g., \citet[][]{Planck2015}.
In this analysis we do not aim to perform an optimized analysis in the framework of the conventional template-fitting technique,
but to compare the method we have developed in this paper with a conventional one semiquantitatively.

The obtained best-fit parameters are summarized in Table~D4. For comparison, we also tabulate the best-fit parameters
obtained by using the $N({\rm H_{tot,mod}})$ map with $T_{\rm bk}=20.5~{\rm K}$ and $C=2$ 
[see Equation~(3) in Section~4.3] in Table~D5.
The average of $\cHI$, $c_{\rm CO}$, and $c_{\rm DG}$ are $0.924\pm0.022$, 
$(1.658\pm0.070) \times 10^{20}~{\rm cm^{-2}~(K~km~s^{-1})^{-1}}$, 
and $(0.360\pm0.011) \times 10^{28}~{\rm cm^{-2}~(W~m^{-2}~sr^{-1})^{-1}}$, respectively.
(For comparison, the average of $c_{1}$ obtained by the analysis using the $N({\rm H_{tot,mod}})$ map is
$0.677\pm0.009$.)
From these scale factors, we can calculate $X_{\rm CO}$ based on $\gamma$-ray data analysis as
$X_{\rm CO,\gamma} = (0.897\pm0.043) \times 10^{20}~{\rm cm^{-2}~(K~km~s^{-1})^{-1}}$,
and the conversion factor from $R$ to the dark gas column density, $X_{\rm DG}$ as
$c_{\rm DG}/\cHI = (0.3911\pm0.0093) \times 10^{28}~{\rm cm^{-2}~(W~m^{-2}~sr^{-1})^{-1}}$.
They agree with the corresponding quantities obtained from the dust fit 
($X_{\rm CO,dust}$ and $1/\yHI$, respectively) described above within ${\le}10\%$.

The values of $\ln{L}$ obtained by the conventional template-fitting method and the analysis using the single $N({\rm H_{tot,mod}})$ map
summed over individual energy ranges in 0.3--72.9~GeV
are 1262809.3 and 1262815.2, respectively.
\footnote{
We note that we give the values of $\ln{L}$ for reference. Since our conventional template-fitting
analysis is not optimized as described in the text,
a statistical comparison based on the values of $\ln{L}$ is not appropriate.
}
The spectra of each component from the two analyses are summarized in Figure~D4 and fit residuals are compared in Figure~D5,
in which the ratio of the $\gamma$-ray model maps is also presented.
We also show in Figure~D6 the integrated gas column densities of each phase (dotted histograms)
and the integrated total gas column density (thick solid histogram) 
as a function of $T_{\rm d}$. The integrals of $N({\HI_{\rm thin}})$,
$N({\rm H_{DG}})$, and $2 X_{\rm CO} W_{\rm CO}$ are 60.9, 7.4 and 1.9
in units of $10^{22}~{\rm cm^{-2}~deg^{2}}$, respectively,
where $10^{22}~{\rm cm^{-2}~deg^{2}}$ corresponds to {$\sim$}740~${\rm M_{\sun}}$ for
$d=150~{\rm pc}$ (see Section~5 for details).
In the same plot, the integral of $N({\rm H_{tot,mod}})$ (already shown in Figure~9b)
is also presented for comparing the inferred total gas column density distributions
between the two analyses.

\floattable
\begin{deluxetable}{cccccccc}[ht!]
\tablecaption{Best-fit parameters with 1-sigma statistical uncertainties, obtained by the conventional template-fitting method}
\tablecolumns{8}
\tablewidth{0pt}
\tablehead{
\colhead{Energy} & \colhead{$\cHI$} & \colhead{$c_{\rm CO}$\tablenotemark{a}} & 
\colhead{$c_{\rm DG}$\tablenotemark{b}} & \colhead{$c_{\rm 2n}$\tablenotemark{c}} & 
\colhead{$c_{\rm 2i}$\tablenotemark{c}} & \colhead{$I_{\rm iso}$} & \colhead{$I_{\rm iso}$} \\
\colhead{(GeV)} & & & & & 
 & \colhead{(norm\tablenotemark{d})} & \colhead{(index)} 
}
\startdata
0.3--0.9 & $0.92\pm0.03$ & $1.57\pm0.10$ & $0.34\pm0.02$ & $0.35\pm0.18$ & $0.06\pm0.21$ & $3.72\pm0.07$ & $2.24\pm0.02$ \\
0.9--2.7 & $0.93\pm0.04$ & $1.62\pm0.12$ & $0.39\pm0.02$ & $0.33\pm0.12$ & $0.53\pm0.48$ & $0.85\pm0.03$ & $2.38\pm0.04$ \\
2.7--8.1 & $1.03\pm0.09$ & $2.54\pm0.27$ & $0.36\pm0.04$ & $0.13\pm0.06$ & $3.51\pm0.70$ & $0.17\pm0.01$ & $2.01\pm0.08$ \\
8.1--72.9 & $0.56\pm0.25$ & $1.86\pm0.69$ & $0.46\pm0.11$ & $0.89\pm0.45$ & $0.63\pm0.30$ & $0.06\pm0.01$ & $2.61\pm0.09$ \\
\enddata
\tablenotetext{a}{$10^{20}~{\rm cm^{-2}~(K~km~s^{-1})^{-1}}$}
\tablenotetext{b}{$10^{28}~{\rm cm^{-2}~(W~m^{-2}~sr^{-1})^{-1}}$}
\tablenotetext{c}{Since the IC model is not a structured component and its intensity is lower
than that of the isotropic component, and the energy bands analyzed are relatively narrow, 
the obtained values for the normalization and index ($c_{\rm 2n}$ and $c_{\rm 2i}$) are uncertain and quite correlated
(e.g., smaller normalization and larger index in 2.7--8.1~GeV than those in other energy ranges).}
\tablenotetext{d}{The integrated intensity ($10^{-6}~{\rm ph~s^{-1}~cm^{-2}~sr^{-1}}$) in each band.}
\tablecomments{
In each energy bin, $\cHI(E)$, $c_{\rm CO}(E)$ and $c_{\rm DG}(E)$ give scale factors for
the atomic gas, the molecular gas, and the dark gas, respectively.
The scale factor for IC ($c_{2}$) is modeled by a power law in each energy bin (from $E_{\rm min}$ to $E_{\rm max}$) 
as $c_{2}(E)=c_{\rm 2n} \cdot (E/E_{0})^{c_{\rm 2i}}$ where $E_{0} = \sqrt{E_{\rm min} \cdot E_{\rm max}}$.
$I_{\rm iso}$ is modeled with a power law with the integrated intensity and the photon index as free parameters.
}
\end{deluxetable}

\floattable
\begin{deluxetable}{cccccc}[ht!]
\tablecaption{Best-fit parameters with 1-sigma statistical uncertainties, obtained by the fit using the $N({\rm H_{tot,mod}})$ map}
\tablecolumns{6}
\tablewidth{0pt}
\tablehead{
\colhead{Energy} & \colhead{$c_{1}$} 
& \colhead{$c_{\rm 2n}$\tablenotemark{a}} & \colhead{$c_{\rm 2i}$\tablenotemark{a}} & \colhead{$I_{\rm iso}$} & \colhead{$I_{\rm iso}$} \\
\colhead{(GeV)} & & & 
 & \colhead{(norm\tablenotemark{b})} & \colhead{(index)} 
}
\startdata
0.3--0.9 & $0.65\pm0.01$ & $0.85\pm0.15$ & $-0.02\pm0.12$ & $3.92\pm0.08$ & $2.22\pm0.02$ \\
0.9--2.7 & $0.70\pm0.01$ & $0.77\pm0.11$ & $0.40\pm0.26$ & $0.89\pm0.02$ & $2.36\pm0.05$ \\
2.7--8.1 & $0.77\pm0.03$ & $0.45\pm0.21$ & $2.06\pm0.66$ & $0.18\pm0.01$ & $1.98\pm0.08$ \\
8.1--72.9 & $0.64\pm0.09$ & $0.97\pm0.43$ & $-0.60\pm0.28$ & $0.06\pm0.01$ & $2.61\pm0.09$ \\
\enddata
\tablenotetext{a}{Since the IC model is not a structured component and its intensity is lower
than that of the isotropic component, and the energy bands analyzed are relatively narrow, 
the obtained values for the normalization and index ($c_{\rm 2n}$ and $c_{\rm 2i}$) are uncertain and quite correlated
(e.g., smaller normalization and larger index in 2.7--8.1~GeV than those in other energy ranges).}
\tablenotetext{b}{The integrated intensity ($10^{-6}~{\rm ph~s^{-1}~cm^{-2}~sr^{-1}}$) in each band.}
\tablecomments{
In each energy bin, $c_{1}$ gives the scale factor of the gas-related component.
The scale factor for IC ($c_{2}$) is modeled by a power law in each energy bin (from $E_{\rm min}$ to $E_{\rm max}$) 
as $c_{2}(E)=c_{\rm 2n} \cdot (E/E_{0})^{c_{\rm 2i}}$ where $E_{0} = \sqrt{E_{\rm min} \cdot E_{\rm max}}$.
$I_{\rm iso}$ is modeled with a power law with the integrated intensity and the photon index as free parameters.
}
\end{deluxetable}

\begin{figure}[ht!]
\gridline{
\fig{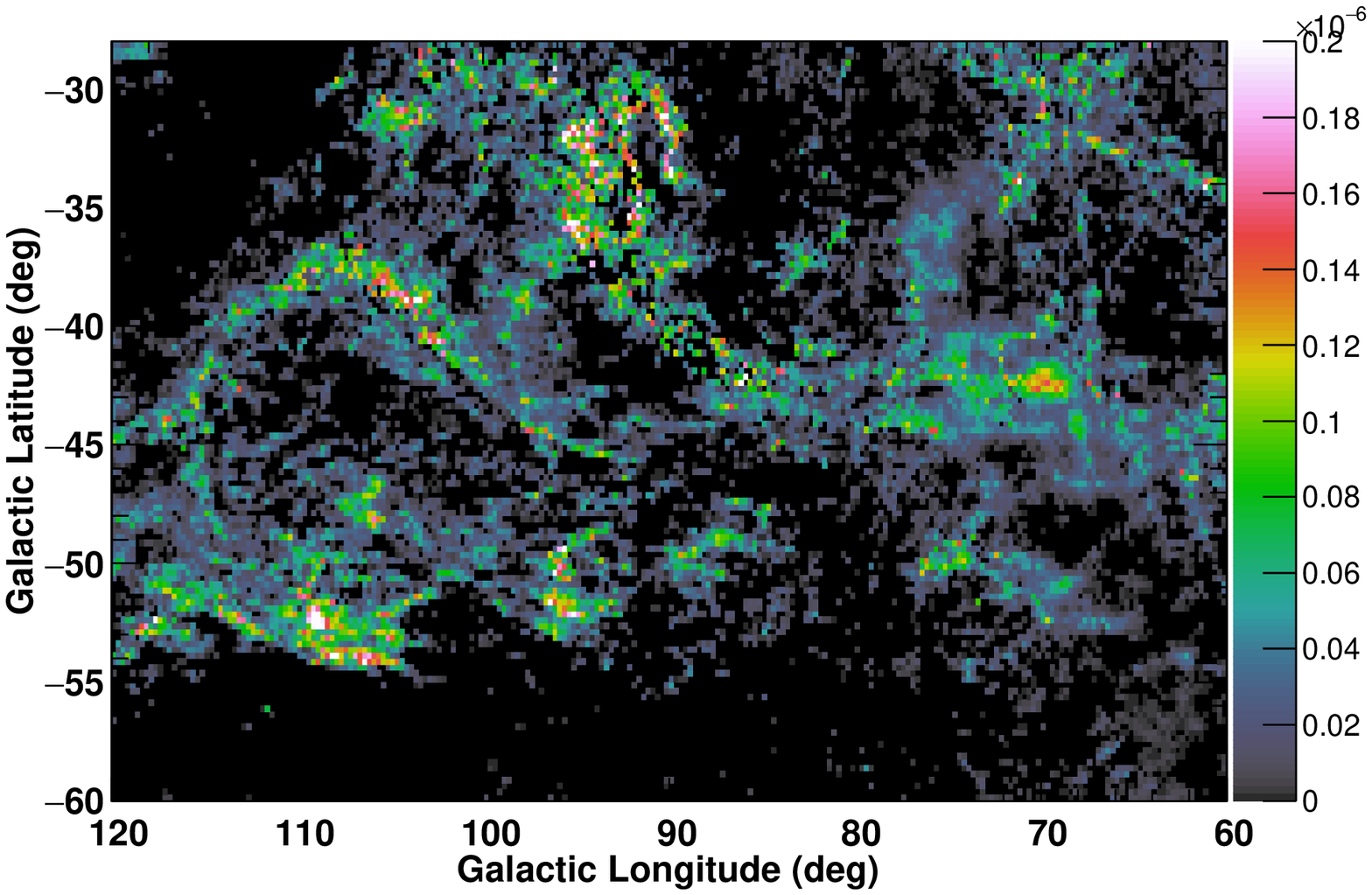}
{0.5\textwidth}{}
}
\caption{
The template dark gas map based on the \textit{Planck} $R$ map in units of ${\rm W~m^{-2}~sr^{-1}}$.
}
\end{figure}

\begin{figure}[ht!]
\gridline{
\fig{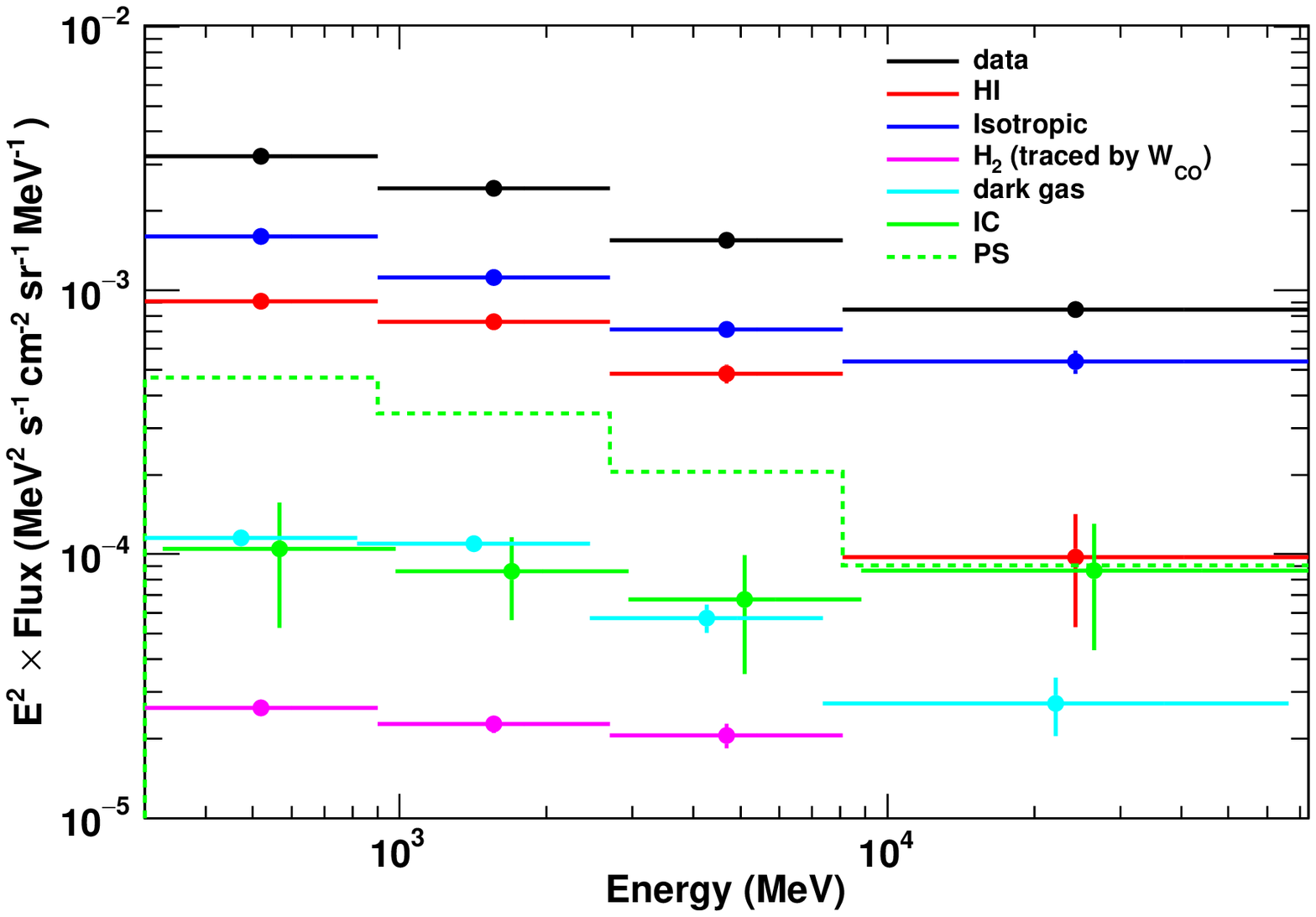}
{0.5\textwidth}{(a)}
\fig{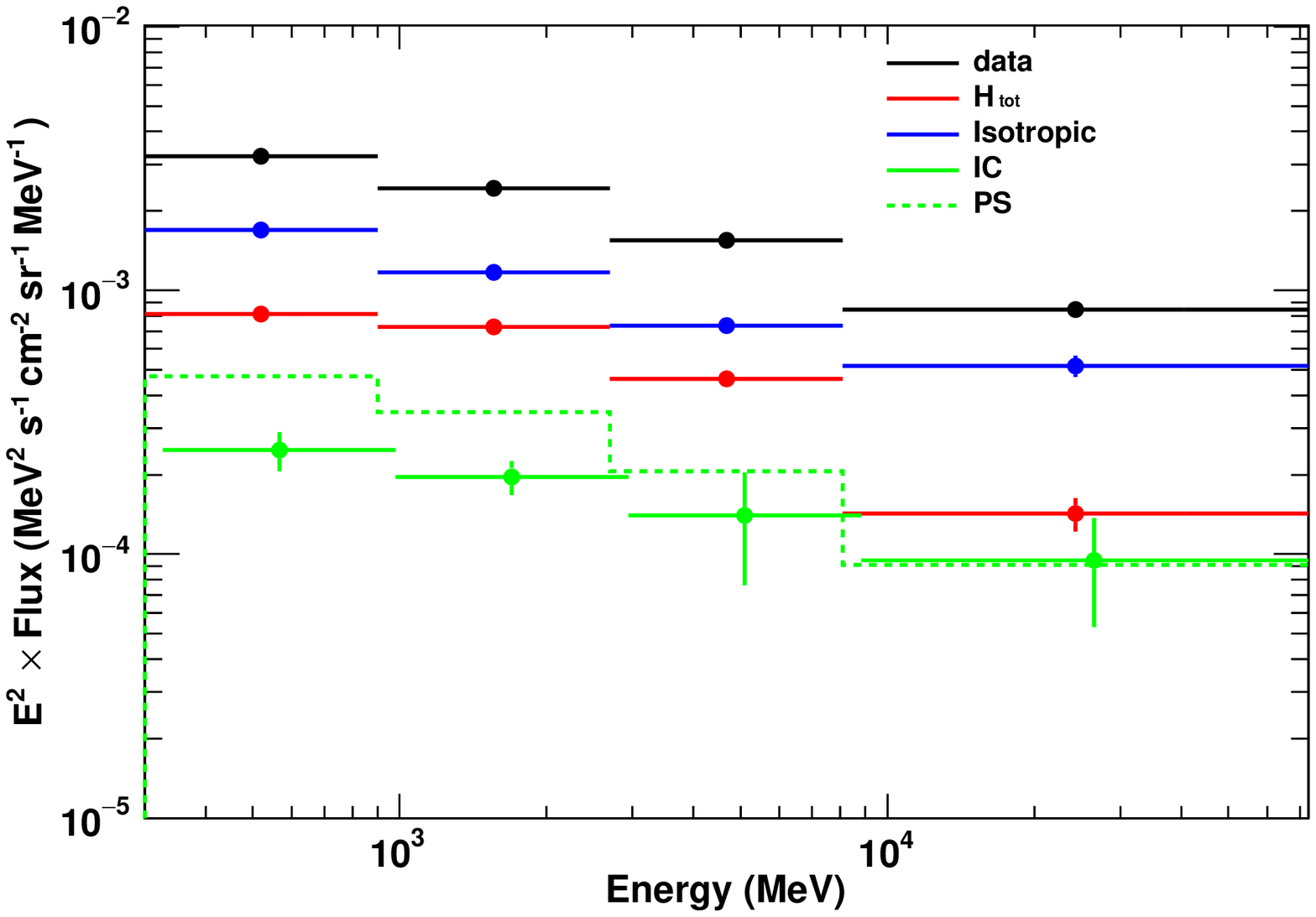}
{0.5\textwidth}{(b)}
}
\caption{
Spectrum of each component obtained by (a) the conventional template-fitting method and
(b) the analysis using the $N({\rm H_{tot,mod}})$ map.
}
\end{figure}

\begin{figure}[ht!]
\gridline{
\fig{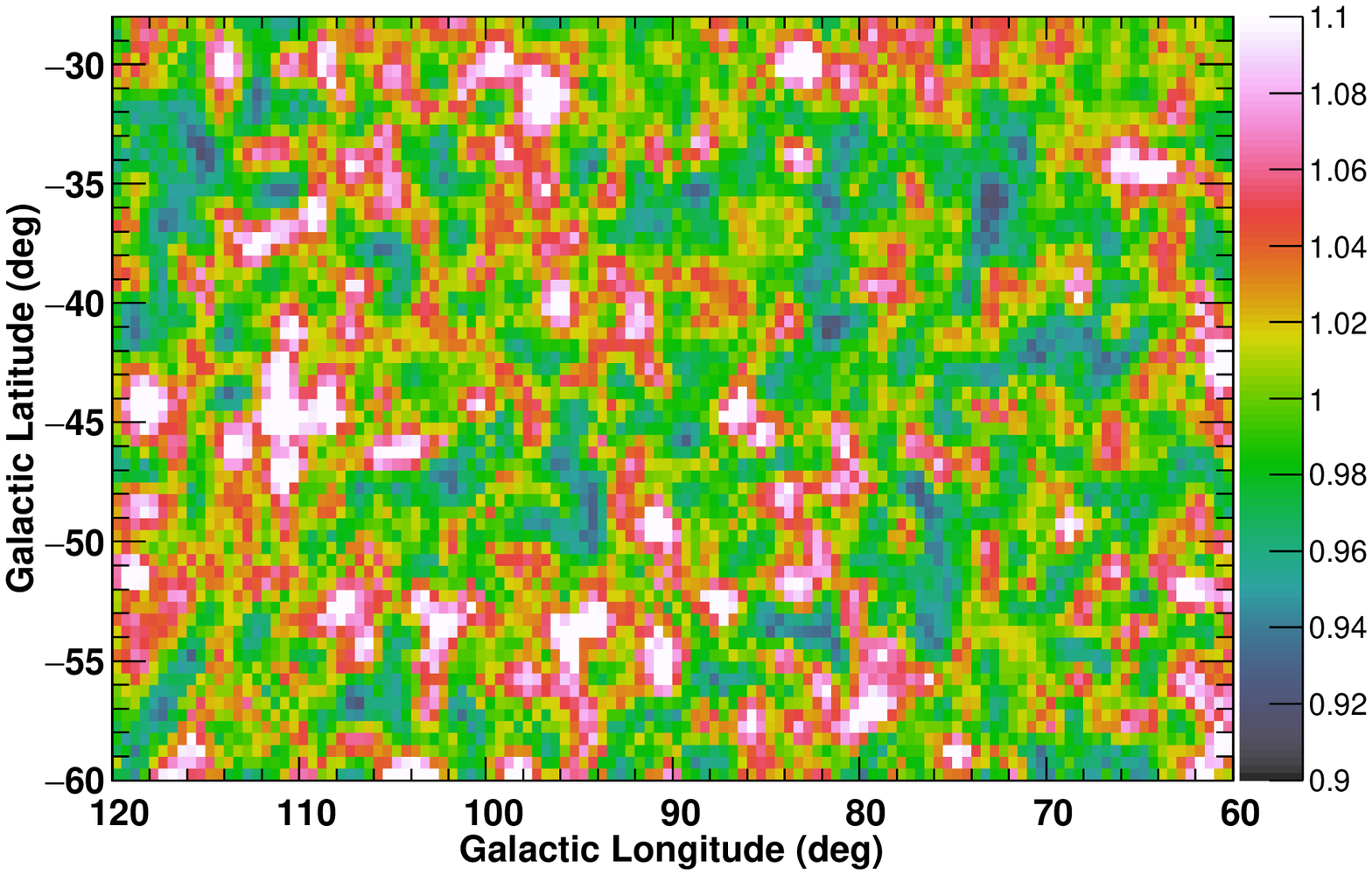}
{0.5\textwidth}{(a)}
\fig{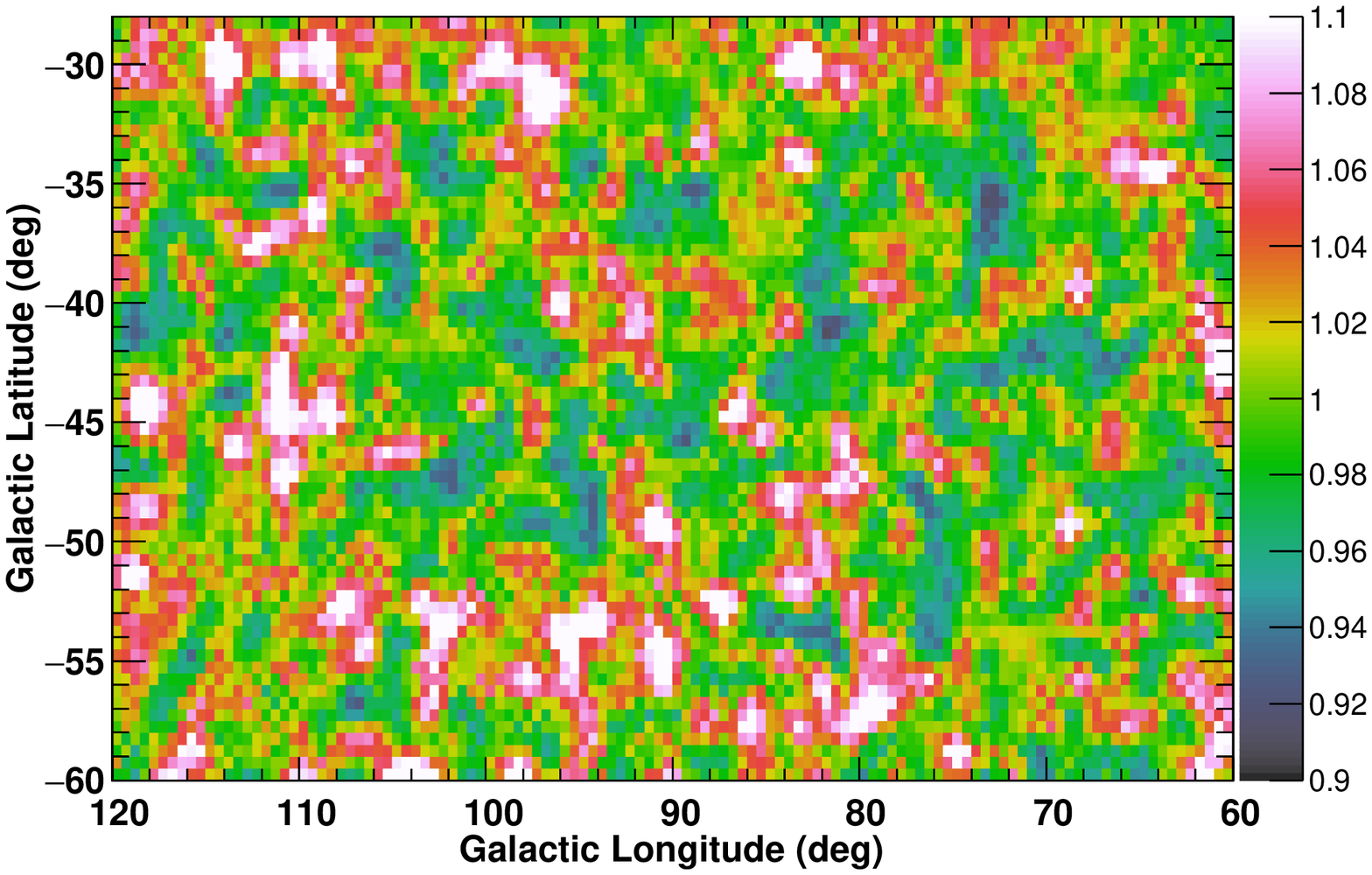}
{0.5\textwidth}{(b)}
}
\gridline{
\fig{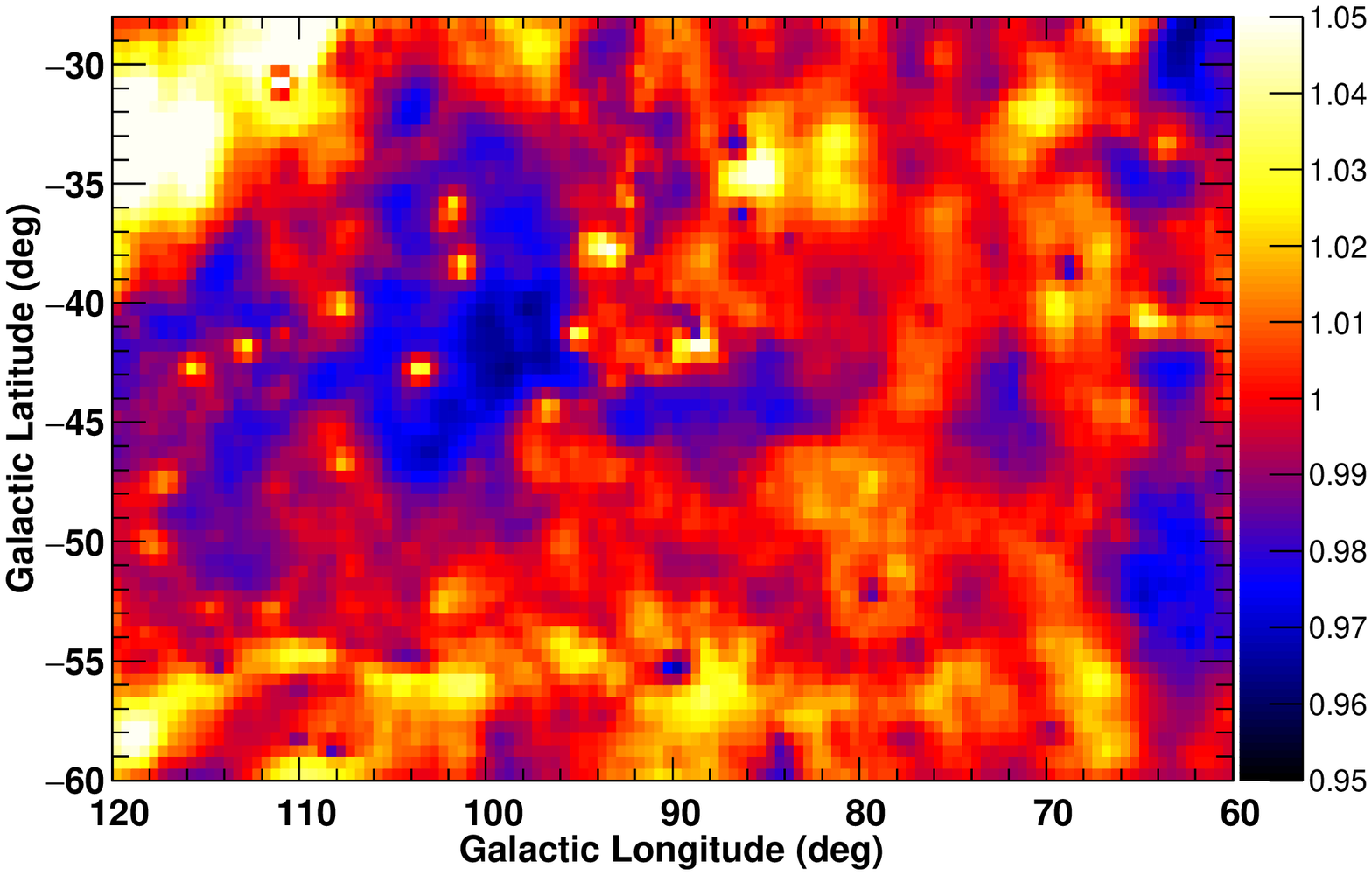}
{0.5\textwidth}{(c)}
}
\caption{
(a) The data/model ratio map obtained by the conventional template-fitting method.
(b) The same map, but obtained by the analysis using the $N({\rm H_{tot,mod}})$ map.
(c) The ratio of $\gamma$-ray model maps of the conventional template-fitting method to the
analysis using the $N({\rm H_{tot,mod}})$ map. In panels~(a) and (b), 
a smoothing with a k5a kernel (1-2-5-2-1 two-dimensional boxcar smoothing) in ROOT framework
(\url{https://root.cern.ch}) was applied. 
Positive/negative spikes in panel~(c) correcpond to point sources considered.
}
\end{figure}

\begin{figure}[ht!]
\gridline{
\fig{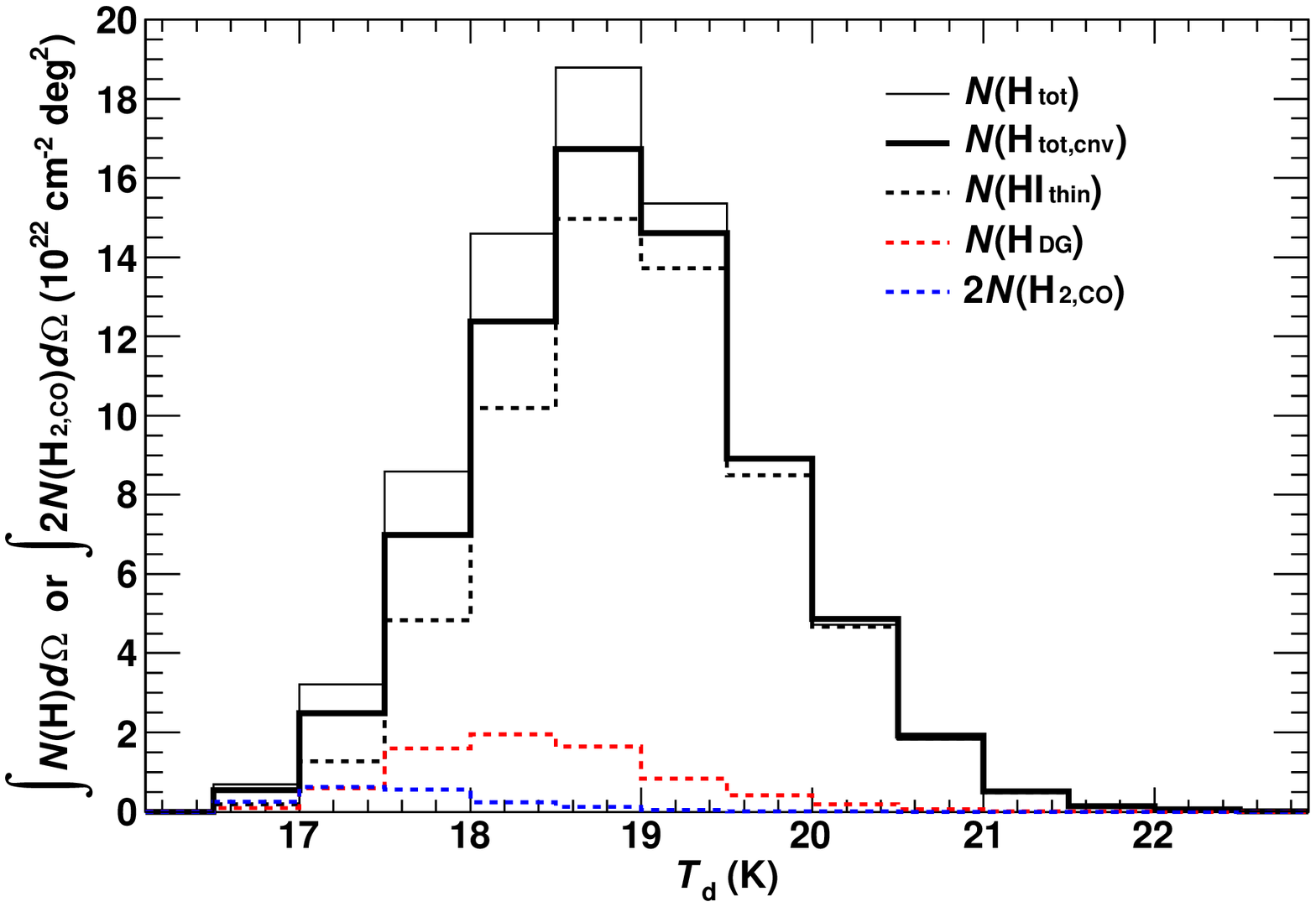}
{0.5\textwidth}{}
}
\caption{
The distributions of the integrated hydrogen column density inferred from $\WHI$ 
for the optically thin case ($\int N(\HI_{\rm thin}) \,d\Omega$), 
that inferred from the dark gas template ($\int X_{\rm DG}R_{\rm res}^{\rm DG} \,d\Omega$),
and that inferred from CO emission ($\int 2X_{\rm CO}W_{\rm CO} \,d\Omega$ as a measure of CO-bright $\Htwo$)
obtained by the conventional template-fitting method (dotted histograms).
The integral of the total gas column density is
also shown as the thick solid histogram. For comparison, the integral of the
$N({\rm H_{tot,mod}})$ (already shown in Figure~9b) is also presented as the thin solid histogram.
}
\end{figure}

\clearpage

\color{black}
\listofchanges

\end{document}